\documentclass[12pt]{article}
\usepackage{latexsym}
\usepackage[mathscr]{eucal}
\usepackage{amsfonts}
\usepackage{amscd}
\usepackage{cite}
\usepackage{amssymb}
\usepackage[centertags]{amsmath}
\usepackage{dsfont}
\usepackage{enumerate}
\usepackage{graphicx}
\usepackage{float}
\usepackage{epsfig}

\usepackage{multirow}
\usepackage{multicol}
\usepackage{amsmath}

\usepackage{array}
\newcolumntype{P}[1]{>{\centering\arraybackslash}p{#1}}
\newcolumntype{M}[1]{>{\centering\arraybackslash}m{#1}}

\restylefloat{figure}

\usepackage{array} 
\usepackage{amssymb}
\usepackage{graphics,graphpap}
\usepackage{graphicx}
\usepackage{color}
\usepackage{graphicx}
\usepackage{dcolumn}
\usepackage{epsfig}
\usepackage{epstopdf}
\usepackage{bm}
\usepackage{amsmath}
\usepackage{amsfonts}
\usepackage{textcomp}
\usepackage{bbm}
\usepackage{setspace}
\usepackage[centertags]{amsmath}

\usepackage[usenames,dvipsnames]{xcolor}

\def\simge{\mathrel{%
   \rlap{\raise 0.511ex \hbox{$>$}}{\lower 0.511ex \hbox{$\sim$}}}}

\def\simle{\mathrel{
   \rlap{\raise 0.511ex \hbox{$<$}}{\lower 0.511ex \hbox{$\sim$}}}}

\def\s#1{\setbox0=\hbox{$#1$}%
\rlap{\ifdim\wd0>.7em\kern.22\wd0\else\kern.1\wd0\fi /}#1}

\newcommand{\cleqn}{\setcounter{equation}{0}}

\newcommand{\newc}{\newcommand}
\newc{\be}{\begin{equation}}
\newc{\ee}{\end{equation}}
\newc{\bea}{\begin{eqnarray}}
\newc{\eea}{\end{eqnarray}}
\newc{\ben}{\begin{equation*}}
\newc{\een}{\end{equation*}}
\newc{\bean}{\begin{eqnarray*}}
\newc{\eean}{\end{eqnarray*}}
\newc{\ol}{\overline}
\newc{\wt}{\widetilde}
\newc{\bs}{\boldsymbol}
\newc{\m}{\mathcal}
\newc{\la}{\lambda}
\newc{\lra}{\longrightarrow}
\newc{\vp}{\varphi}
\newc{\ti}{\tilde}

\newcommand{\Red}[1]{\textcolor[rgb]{1,0,0}{#1}}
\newcommand{\Blue}[1]{\textcolor[rgb]{0,0,1}{#1}}

\begin{document}

\title{\hfill ~\\[-40mm]
          \hfill\mbox{\small  SI-HEP-2015-26}\\[-3.5mm]
          \hfill\mbox{\small  QFET-2015-33}\\[13mm]
      \textbf{Phenomenological Implications of an   \\
  $\bs{SU(5) \times S_4\times U(1)}$ SUSY GUT of Flavour}}

\date{}

\author{\\Maria Dimou\footnote{E-mail: {\tt md1e10@soton.ac.uk}}$\;\;^a$, Stephen F. King\footnote{E-mail: {\tt king@soton.ac.uk}}$\;\;^a$,
Christoph Luhn\footnote{E-mail: {\tt  christoph.luhn@uni-siegen.de}}$\;\;^b$\\[10mm]
 $^a$ \emph{\small{}School of Physics and Astronomy, University of Southampton,}\\
  \emph{\small Southampton, SO17 1BJ, United Kingdom}\\[2mm]
 $^b$ \emph{\small Theoretische Physik 1, Naturwissenschaftlich-Technische Fakult\"at,}\\
  \emph{\small Universit\"at Siegen, Walter-Flex-Stra{\ss}e 3, 57068 Siegen, Germany}}

\maketitle

\begin{abstract}
\noindent 
We discuss the characteristic low energy phenomenological implications of 
an $SU(5)$ Supersymmetric Grand Unified Theory (SUSY GUT) whose flavour
structure is controlled by the family symmetry $S_4 \times U(1)$, which 
provides a good description of all quark and lepton masses, mixings as
well as CP violation.  
Although the model closely mimics Minimal Flavour Violation (MFV)
as shown in~\cite{companion}, here we focus on the differences.
We first present numerical estimates of the low energy mass insertion parameters,
including canonical normalisation and renormalisation group running, 
for well-defined ranges of SUSY parameters and compare the
naive model expectations to the numerical scans and the experimental bounds.
Our results are then used to estimate the model-specific predictions
for Electric Dipole Moments (EDMs), Lepton Flavour Violation (LFV), $B$ and
$K$ meson mixing as well as rare $B$ decays. The largest observable deviations
from MFV come from the LFV process $\mu\rightarrow e \gamma$ and the electron
EDM. 
\end{abstract}
\thispagestyle{empty}
\vfill

\newpage


\tableofcontents

\vspace{1mm}

\hrulefill

\vspace{-3mm}



\setcounter{page}{1}

\section{Introduction}
\cleqn

The flavour problem has been around for a long time, but only relatively recently has
new information been provided in the form of neutrino mass and lepton mixing.
Subsequently, 
a lot of effort has been put into trying to formulate a theory of flavour
(for reviews see e.g.~\cite{King:2015aea}) which can account for
the observed pattern of fermion masses and mixing, while providing more
accurate predictions for the less well measured (or unmeasured) flavour
parameters in the neutrino sector, see e.g.~\cite{Ballett:2013wya}. 

A possible
additional source of experimental information which could shed light on the
flavour puzzle would be the observation of rare flavour changing processes at
rates beyond that predicted by the Standard Model (SM). Such observations
could in principle 
provide insight into the nature of the theory of flavour beyond the SM. So
far, experiment has unfortunately not measured any flavour or CP violation
beyond SM expectations.  Indeed all data are consistent with the concept of
Minimal Flavour Violation (MFV)~\cite{Buras:2000dm}, in which all flavour and
CP-violating transitions are governed by the CKM matrix and the only relevant
local operators are the ones that are relevant in the SM. Although the
formulation of MFV in an effective field theory, involving an approximate $SU(3)^5$ 
symmetry\footnote{In the framework of Grand Unified Theories (GUTs) it is not
possible to implement $SU(3)^5$ symmetry at the GUT scale. However, in GUTs
based on $SU(5)$~\cite{Georgi:1974sy} or Pati-Salam~\cite{Pati:1973uk}, it is
certainly possible to introduce an $SU(3)^2$ flavour symmetry, and this has
been shown to be sufficient~\cite{Barbieri:2015bda}.} broken by the Yukawa
matrices, allows some new operators which can in principle give significant
contributions~\cite{D'Ambrosio:2002ex,Bobeth:2002ch},  in all cases, MFV
predicts very SM-like flavour and CP violation consistent with observation.

The absence of flavour violation is consistent with the absence of any new
physics beyond the SM, such as Supersymmetry (SUSY) which, if softly broken at
the TeV scale, would in general imply large deviations from SM flavour and CP
violation~\cite{Chung:2003fi}. For example, SUSY models involve one-loop
diagrams that induce Flavour Changing Neutral Current (FCNC) processes such
as $b\to s\gamma$ and $\mu \to e \gamma$ at rates which are proportional to
the mass insertion parameters, i.e. the off-diagonal elements of the scalar
mass matrices in the super-CKM (SCKM) basis where the Yukawa matrices are
diagonal~\cite{Chung:2003fi,Gabbiani:1996hi}. Such SUSY contributions are very
small in the the Constrained Minimal Supersymmetric Standard Model (CMSSM)
where the squark and slepton mass squared matrices are proportional to the unit
matrix at the high energy scale and the trilinear $A$-terms are aligned with
the Yukawa matrices, resulting in an (approximate) MFV-like structure at low
energy~\cite{Chung:2003fi}. But there is no convincing theoretical basis for
either the CMSSM or MFV. Moreover, in SUSY GUTs, the CMSSM framework while
providing suppressed flavour violation, cannot easily control CP violation in
the form of Electric Dipole Moments (EDMs) which remains a
challenge~\cite{Chung:2003fi}. However, the real challenge is
to justify the assumptions of MFV or the CMSSM, while at the same time
providing a realistic explanation of quark and lepton (including neutrino)
masses, mixing and CP violation.

Following the discovery of neutrino mass and mixing, there has been an impetus to 
revisit the favour problem using a family symmetry of some kind, in particular
discrete non-Abelian family symmetry~\cite{King:2015aea}. It was realised that
in such models, spontaneous CP and flavour violation could solve the CP and
flavour problems of the SM~\cite{Ross:2002mr,Ross:2004qn} without any {\it ad
  hoc} assumptions about MFV or the CMSSM. The family symmetry that is
responsible for the structure of the Yukawa sector will automatically control
the soft SUSY breaking sector as long as the SUSY breaking hidden sector
respects the family symmetry.  This is realised for instance in supergravity
induced SUSY breaking.

Considering a SUSY framework, the choice of an $SU(3)$ family
symmetry~\cite{Ross:2004qn,Antusch:2007re} provides a benchmark scenario where
flavour and CP violation is controlled by family symmetry. The spontaneous
breaking of  family and CP symmetry by Vacuum Expectation Values (VEVs) of the
so-called flavon fields perturbs the SUSY breaking sector, thereby
generating distinct deviations from MFV or the CMSSM. Unfortunately, these
signatures which were expected to appear in Run 1 of the
LHC~\cite{AnatomyandPhenomenology} did not in fact materialise, and the allowed
parameter space has been much reduced~\cite{Buras:2012ts}.
At leading order, the CMSSM is enforced by the $SU(3)$ family symmetry 
acting on the squark and slepton mass squared matrices. 
When $SU(3)$ is broken by flavon VEVs, to generate quark and lepton flavour,
those flavons appearing in the K\"ahler potential give important
contributions to the kinetic terms, requiring extra canonical
normalisation~\cite{King:2003xq}. Since SUSY breaking also originates from the
K\"ahler potential, the flavons also  modify the couplings of squarks and
sleptons to the fields with SUSY breaking $F$-terms, where the corrections
have a different form to the flavon corrections appearing in the
superpotential. All of this occurs at the high scale. Additional flavour
violation is generated by renormalisation group (RG) running down to low energy,
taking into account the seesaw mechanism~\cite{seesaw} and threshold
corrections~\cite{Yamada:1992kv}. 

In this paper we discuss the characteristic low energy
phenomenological implications of an $SU(5)$ Supersymmetric Grand Unified
Theory (SUSY GUT) whose flavour structure is controlled by the family symmetry
$S_4 \times U(1)$, which provides a good description of all quark and lepton
masses, mixings as well as CP violation. In a recent paper we showed how MFV
emerges approximately in this setup~\cite{companion}. Assuming a SUSY
breaking mechanism which respects the family symmetry, we calculated in
full explicit detail the low energy mass insertion parameters in the SCKM
basis, including the effects of canonical normalisation and renormalisation
group running, showing that the peculiar flavour structure of the model, defined by the
small family symmetry $S_4 \times U(1)$, is sufficient to approximately mimic 
MFV.\footnote{Depending on the implementation of a particular family
symmetry, SUSY GUTs of flavour typically realise some approximation of
MFV at high as well as low scales~\cite{Paradisi:2008qh}.} 
However there are important phenomenological 
differences which can provide tell-tale signatures of the model,
and it is the main purpose of this paper to discuss these in detail.
In other words, we exploit the low energy mass insertion parameters of the model
calculated in~\cite{companion} to analyse a panoply of rare and flavour
changing processes as well as EDMs in both the lepton and quark sectors.
The results are quite illuminating: while we find only small new effects 
in $B$ physics, very large effects arise for Lepton Flavour Violation (LFV)
and the electron EDM which are therefore predicted to be observed soon. 

The layout of the remainder of the paper is as follows.
In Section~\ref{HighScaleMatrices} we give a succinct summary of the analytic 
Yukawa matrices and mass insertion parameters calculated 
in~\cite{companion}. In Section~\ref{numerical} 
we discuss numerical estimates of the low energy mass
insertion parameters for ranges of SUSY parameters which are
consistent with the bounds from direct searches for squarks and sleptons at
LHC Run~1. We compare the naive model expectations to the numerical scans
and the experimental bounds. 
In Section~\ref{sec:pheno} these results are then used to estimate the
predictions for EDMs, LFV, $B$ and $K$ meson mixing
as well as rare $B$ decays. The largest observable deviations from MFV come
from the LFV process $\mu\rightarrow e \gamma$ and the electron EDM. 
Section~\ref{conclusions} concludes the paper.




\section{Yukawa matrices and SUSY breaking parameters} 
\label{HighScaleMatrices}
\cleqn

In this section, we briefly summarise the GUT scale Yukawa matrices and soft
SUSY breaking parameters constructed within the framework of the family
symmetry model in~\cite{companion}. Working in a power expansion of the
Wolfenstein parameter $\lambda\approx 0.225$~\cite{Wolfenstein:1983yz}, we
present all expressions to Leading Order (LO). The entries of the flavour
matrices are generally complex, where the phases are given in terms of two free
parameters $\theta^d_2$, $\theta^d_3$, with the exception of the soft
trilinear terms whose phases are not identified with the corresponding Yukawa
phases but are kept as free parameters, even though their flavour structure is
the same as that of the Yukawas. Details on this aspect can be found
in~\cite{companion}. In the present work, we will comment on the consequences
of this generalisation where relevant.


\subsection{Yukawa sector}
\label{YukawaSector}


The fermion structure was already scrutinised in~\cite {M1}, and we have
completed this analysis by including the effects of canonical
normalisation. In the basis with canonical kinetic terms, that is after
redefining the superfields such that the K\"ahler metrics are identified with
the unit matrix, the Yukawa matrix for the up-type quarks reads
\begin{eqnarray}
Y^{u}_{\text{GUT}}&\approx&\left(
\begin{array}{ccc}
 y_u \,\lambda^8 & -\frac{1}{2}k_2\,y_c \,\lambda^8  &-\frac{1}{2}k_{4}\,y_te^{i(\theta^d_3-\theta^d_2)}\,\lambda ^6 \\
-\frac{1}{2}k_2\,y_c \lambda^8 & y_c\, \lambda ^4 & -\frac{1}{2}k_{3}\,y_te^{-i5\theta^d_2}\lambda^5 \\
\!\!-\frac{1}{2}k_{4}\,y_te^{-i(3\theta^d_2+2\theta^d_3)}\,\lambda ^6 &~~-\frac{1}{2}k_{3}\,y_te^{-i(7\theta^d_2+3\theta^d_3)}\lambda^5~~ &y_t
\end{array}
\right)\ , ~~~~
\label{YuC}
\end{eqnarray}
where $y_f$ and $k_i$ are real order one coefficients, with the former
stemming from the Yukawa part of the superpotential of the theory and the
latter from the K\"ahler potential. In particular, $k_2,k_3$ and $k_4$ appear
in the non-canonical K\"ahler metric of the $SU(5)$ ${\bf{10}}$-plets, in the
(12), (23) and (13) elements, respectively.


The Yukawa matrices for the down-type quarks and charged leptons take the form
{\small{
\begin{eqnarray}
 Y^{d}_{{\text{GUT}}}& \!\approx\!&\left(
\begin{array}{ccc}
 z^d_1e^{-i\theta^d_2}\lambda^8&\tilde x_2\lambda^5&\!-\tilde{x}_2e^{i(3\theta^d_2+2\theta^d_3)} \lambda^5\\
 -\tilde{x}_2\lambda^5&~~y_se^{-i\theta^d_2}\lambda^4&\!-y_se^{2i(\theta^d_2+\theta^d_3)}\lambda^4\\
\!\left(z^d_{3}\!-\!\frac{K_{3}}{2}y_b\right)e^{-i(3\theta^d_2+2\theta^d_3)}\lambda^6\!&~\left(z^d_{2}\!-\!\frac{K_{3}}{2}y_b\right)e^{-i(3\theta^d_2+2\theta^d_3)}\lambda^6\!\!&y_b \lambda^2
\end{array}
\right) \!, ~~~~~~~~\label{YdC}\\
 Y^{e}_{{\text{GUT}}}&\!\approx\!&\left(
\begin{array}{ccc}
 -3z^d_1e^{-i\theta^d_2}\lambda^8&-\tilde{x}_2\lambda^5&~~\left(z^d_{3}-\frac{K_{3}}{2}y_b\right)\lambda^6 \\
 \tilde{x}_2\lambda^5&-3\,e^{-i\theta^d_2}y_s\lambda^4&~~\left(z^d_{2}-\frac{K_{3}}{2}y_b\right)\lambda^6\\
-\tilde{x}_2\lambda^5&~~3\,e^{-i\theta^d_2}y_s\lambda^4&~~y_b \lambda^2
\end{array}
\right) \!.\label{YeC}
\end{eqnarray}}}%
Again, these expressions are given in the canonical basis and all coefficients
are real and of order one. $\tilde x_2$, $y_f$ and $z^d_i$  arise from the superpotential
operators and $K_3$ from the K\"ahler potential, where it enters symmetrically
in all off-diagonal elements of the non-canonical K\"ahler metric of the
$SU(5)$ $\bf{\bar{5}}$-plets.

Finally, the Dirac neutrino Yukawa matrix in the canonical basis is given by
\begin{eqnarray}\label{YnuC}
Y^{\nu}
&\!\approx\!&\left(
\begin{array}{ccc}
y_D&-\frac{y_D(K_{3}+K^N_3)}{2}\lambda^4&\left(z^D_1-\frac{y_D(K_{3}+K^N_3)}{2}\right)\lambda^4\\
-\frac{y_D(K_{3}+K^N_3)}{2}\lambda^4&\left(z^D_1-\frac{y_D(K_{3}+K^N_3)}{2}\right)\lambda^4\!\!&y_D\\
\!\left(z^D_1-\frac{y_D(K_{3}+K^N_3)}{2}\right)\lambda^4&y_D&-\frac{y_D(K_{3}+K^N_3)}{2}\lambda^4
\end{array}
\right), ~~~~~~~~~ 
\end{eqnarray}
which is real up to LO in $\lambda$. The parameters $y_D$ and $z^D_1$
originate from the superpotential, while $K^N_3$ is associated to the K\"ahler
metric of the right-handed neutrinos. Note that this metric is identical to
that of the $SU(5)$ $\bf{\bar{5}}$-plets, 
up to renaming the order one coefficients, see~\cite{companion} for details.

Transforming the left- and right-handed superfields $f_{L,R}$ by unitary
matrices $U^f_{L,R}$, we obtain the canonically normalised diagonal and
positive Yukawas in the SCKM basis  
\begin{equation} 
\tilde{Y}^u_{\text{GUT}}\approx
\left(
\begin{array}{ccc}
y_u\lambda ^8 & 0 & 0 \\
 0 &y_c \lambda ^4& 0 \\
 0 & 0 &y_t
\end{array}
\right)  , \qquad
\tilde{Y}^d_{\text{GUT}}\approx
\left(
\begin{array}{ccc}
\frac{\tilde{x}_2^2}{y_s}\lambda^6& 0 & 0 \\
 0 &y_s\lambda ^4& 0 \\
 0 & 0 &y_b\lambda ^2
\end{array}
\right) , \label{Yudt}
\end{equation}

\begin{equation}
\tilde{Y}^e_{\text{GUT}}\approx
\left(
\begin{array}{ccc}
\frac{\tilde{x}_2^2}{3y_s}\lambda^6& 0 & 0 \\
 0 &3y_s\lambda ^4& 0 \\
 0 & 0 &y_b\lambda^2
\end{array}
\right)  . \label{Yet}
\end{equation}%
Up to phase convention, the CKM matrix is given by
$V_{\text{CKM}_{\text{GUT}}}=(U^u_L)^TU^{d*}_L$, leading to the mixing angles
\begin{equation}
 \sin(\theta^q_{13})_{\text{GUT}}\approx
\frac{\tilde{x}_2}{y_b}\lambda^3\ ,\qquad
\tan (\theta^q_{23})_{\text{GUT}}\approx
\frac{y_s}{y_b}\lambda^2\ ,\qquad
\tan(\theta^q_{12})_{\text{GUT}}\approx
\frac{\tilde{x}_2}{y_s}\lambda \ .
\end{equation}%
The mixing arises purely from the down-type quark sector and incorporates the
Gatto-Sartori-Tonin relation~\cite{Gatto:1968ss}
$\theta^q_{12}\approx\sqrt{m_d/m_s}$.  The amount of CP violation is given by
the Jarlskog invariant~\cite{Jarlskog:1985ht}\\[-8mm] 
\begin{eqnarray}
 J^q_{\text{CP}_{\text{GUT}}}
&\approx&\lambda^7\frac{\tilde{x}_2^3}{y_b^2y_s}\sin \theta^d_2
 \ .
\label{JCPq}
\end{eqnarray}%

These results are in agreement with the LO expressions derived in~\cite{M1}, where
canonical normalisation effects were ignored. As discussed
in~\cite{companion}, the LO results for the quark and charged lepton masses
and mixing angles remain unaffected by the process of canonicalising the
kinetic terms. 
We point out that these 13 observables of the charged fermion
sector are given in terms of only 8 input parameters ($\lambda$, $y_{u,c,t}$,
$y_{s,b}$, $\tilde x_2$ and $\theta_2^d$) at LO.


\subsection{Soft SUSY breaking sector}

The soft trilinear $A$-terms and the Yukawa couplings originate in the same
superpotential terms. Hence, they have a similar flavour structure and, in the
basis of canonical kinetic terms, the soft flavour matrices
$A^f_{\text{GUT}}/A_0$, where $A_0$ denotes the scale of the trilinear
terms, can be deduced from Eqs.~(\ref{YuC}-\ref{YnuC}) by simply replacing
$y_u\to a_u\, e^{i(\theta^a_u-\theta^y_u)}$, 
$y_c\to a_c\,e^{i(\theta^a_c-\theta^y_c)}$, 
$y_t\to a_t$, 
$y_s\to a_s\,e^{i(\theta^a_s-\theta^y_s)}$, 
$y_b\to a_b\,e^{i(\theta^a_b-\theta^y_b)}$, 
$\tilde{x}_2\to
\tilde{x}^a_2\,e^{i(\theta^{\tilde{x}_a}_2-\theta^{\tilde{x}}_2)}$, 
$z^f_i\to z^{f_a}_i\,e^{i(\theta^{z_{f_a}}_i-\theta^{z_{f}}_i)}$ and 
$y_D\to\alpha_D$.
Here, the Yukawa phases are all given in terms of $\theta^d_2,~\theta^d_3$ as
follows: 
$\theta^y_u=\theta^y_c=\theta^y_s=\theta^{z_d}_1=2\theta^d_2+3\theta^d_3$, 
$\theta^y_b=\theta^{z_d}_2=\theta^{z_d}_3=\theta^d_3$ and 
$\theta^{\tilde{x}}_2=3(\theta^d_2+\theta^d_3)$. 
On the other hand, the trilinear phases
$\theta^a_f,~\theta^{\tilde{x}_a}_2,~\theta^{z_{f_a}}_i$ are kept free.

Turning to the soft scalar mass squared matrices in the canonical basis, we
find
\begin{eqnarray}
\frac{M^2_{T_{\text{GUT}}}}{m_0^2}&\approx&
\left(
\begin{array}{ccc}
 b_{01} & ~~(b_2-b_{01}k_2)\lambda^4 & ~~~e^{i(\theta^d_2-\theta^d_3)}(b_4-\frac{k_4(b_{01}+b_{02})}{2})\lambda^6 \\
\cdot & ~~b_{01} &~~~e^{5i\theta^d_2}(b_3-\frac{k_3(b_{01}+b_{02})}{2})\lambda^5 \\
\cdot & \cdot &~~ b_{02}
\end{array}
\right) \ , \label{MTC}
\end{eqnarray}
for the $SU(5)$ ${\bf{10}}$-plets as well as
\begin{eqnarray}
 \frac{M^2_{F(N)_{\text{GUT}}}}{m_0^2}&\approx&
\left(
\begin{array}{ccc}
 B^{(N)}_0 & ~~~(B^{(N)}_{3}
-K^{(N)}_3)\lambda^4&~~~(B^{(N)}_{3}
-K^{(N)}_3)\lambda^4 \\
 \cdot& ~~~B^{(N)}_0 &~~~ (B^{(N)}_{3}
-K^{(N)}_3)\lambda^4\\
 \cdot & \cdot & B^{(N)}_0
\end{array}
\right), \label{MFC}
\end{eqnarray}%
for the $SU(5)$ ${\bf{5}}$-plets and the right-handed neutrinos, 
with the latter being associated to the coefficients with index $N$.
For convenience, we absorb the universal order one parameter $B_0$ on the
diagonal into the soft SUSY breaking mass $m_0$, so that the leading
contribution to the diagonal entries of $M^2_{F_{\text{GUT}}}/m_0^2$ is one.


\subsection{Mass insertion parameters}

In order to study the phenomenological implications of the soft SUSY breaking
sector,  it is useful to rotate all quantities into the physical basis where
the Yukawa matrices are diagonal and positive, i.e. the SCKM basis. Any
misalignment between the fermion and sfermion flavour matrices constitutes a
source of flavour violation, with the off-diagonal entries of the sfermionic
mass matrices contributing to FCNCs. The sfermion mass matrices are given as
\begin{equation}
m^2_{\tilde{f}_{LL}}\!\!=(\tilde{m}_f^2)_{LL}+\tilde{Y}_f\tilde{Y}_f^\dagger \upsilon^2_{u,d}\ ,~\quad
m^2_{\tilde{f}_{RR}}\!\!=(\tilde{m}_f^2)_{RR}+\tilde{Y}_f^\dagger\tilde{Y}_f
\upsilon^2_{u,d}\ , ~\quad
m^2_{\tilde{f}_{LR}}\!\!=\tilde{A}_f \upsilon_{u,d}-\mu \tilde{Y}_f
\upsilon_{d,u}\ ,
\label{fullmasses}
\end{equation}%
where $\tilde m^2_f$ and $\tilde A_f$ denote the soft flavour matrices in the
SCKM basis, and $\tilde Y_f$ are the diagonal Yukawa matrices.  $\mu$ is the
(real) higgsino mass parameter, and the VEVs of the two neutral Higgses are
defined as
\begin{eqnarray}
\upsilon_u&=&\frac{\upsilon}{\sqrt{1+t_\beta^2}}\,t_\beta\ ,~~~~~~\upsilon_d=\frac{\upsilon}{\sqrt{1+t_\beta^2}}\ ,
\end{eqnarray}
where $t_\beta\equiv \tan\beta=\frac{\upsilon_u}{\upsilon_d}$
and $\upsilon = \sqrt{\upsilon_u^2+\upsilon_d^2}=174$~GeV. 
The indices $L$ and $R$ refer to the chirality of the corresponding SM
fermions and $m^2_{\tilde{f}_{RL}}\equiv (m^2_{\tilde{f}_{LR}})^\dagger$. 
With these definitions, the amount of flavour violation can be measured in
terms of the dimensionless mass insertion parameters~\cite{Gabbiani:1996hi}
\begin{eqnarray}
(\delta^f_{LL})_{ij}=\frac{(m^2_{\tilde{f}_{LL}})_{ij}}{\langle
    m_{\tilde{f}}\rangle ^2_{LL}}\ ,\qquad
(\delta^f_{RR})_{ij}=\frac{(m^2_{\tilde{f}_{RR}})_{ij}}{\langle
    m_{\tilde{f}}\rangle ^2_{RR}}\ ,\qquad
(\delta^f_{LR})_{ij}=\frac{(m^2_{\tilde{f}_{LR}})_{ij}}{\langle
    m_{\tilde{f}}\rangle ^2_{LR}}\ ,
\label{ins}
\end{eqnarray}%
where the average masses in the denominators are defined by
\begin{equation}
\langle m_{\tilde{f}}\rangle
^2_{AB}=\sqrt{(m^2_{\tilde{f}_{AA}})_{ii}(m^2_{\tilde{f}_{BB}})_{jj}}\ .\label{mav}
\end{equation}%
We mention in passing that the phase structure of the mass insertion
parameters depends on the choice of the phase conventions of the CKM and PMNS
matrices. In~\cite{companion}, we have worked out the expressions in
Eq.~\eqref{ins} explicitly for our model at the GUT scale, choosing a phase
convention in which $V_\text{CKM}$ and $U_\text{PMNS}$ take their standard form.

The effects of RG running down to the low energy scales where experiments are
performed were also estimated, using the leading logarithmic
approximation. Introducing the parameters 
\begin{eqnarray}
\eta&=&\frac{1}{16\pi^2}\ln\left(\frac{M_{\text{GUT}}}{M_{\text{low}}}\right),~~~
\eta_N=\frac{1}{16\pi^2}\ln\left(\frac{M_{\text{GUT}}}{M_{\text{R}}}\right),
\label{etas}
\end{eqnarray}
we performed a two-stage running ($i$)~from
$M_{\text{GUT}}$ to $M_R$, where the right-handed neutrinos are integrated
out, and ($ii$)~from $M_R$ to $M_{\text{SUSY}}\sim M_\text{W}\equiv
M_{\text{low}}$.
For $M_{\text{GUT}}\approx 2\times 10^{16}$~GeV, $M_{\text{R}}\approx
10^{14}$~GeV and $M_{\text{low}}\approx 10^{3}$~GeV, $\eta\approx0.19$ is of
the order of our expansion parameter $\lambda\approx0.22$  and
$\eta_N\approx0.03$.  In terms of their $\lambda$-suppression, the resulting flavour
structures of the low energy mass insertion parameters~$\delta$ read
\begin{eqnarray}\label{eq:deltau}
&&\delta^u_{LL}\sim\left(
\begin{array}{ccc}
1  & \lambda^4 & \lambda^6 \\
\cdot &1 &\lambda^5 \\
\cdot &\cdot  &1
\end{array}
\right),~~~
\delta^u_{RR}\sim\left(
\begin{array}{ccc}
1  & \lambda^4 & \lambda^6 \\
\cdot &1 &\lambda^5 \\
\cdot &\cdot  &1
\end{array}
\right),~~~
\delta^u_{LR}\sim\left(
\begin{array}{ccc}
\lambda^8 & 0& \lambda^7 \\
0&\lambda^4&\lambda^6 \\
0 &\lambda^7&1
\end{array}
\right),~~~~\\[-2mm]
\label{eq:deltad}
&&\delta^d_{LL}\sim\left(
\begin{array}{ccc}
1  & \lambda^3 & \lambda^4 \\
\cdot &1 &\lambda^2 \\
\cdot &\cdot  &1
\end{array}
\right),~~~
\delta^d_{RR}\sim\left(
\begin{array}{ccc}
1  & \lambda^4 & \lambda^4 \\
\cdot &1 &\lambda^4 \\
\cdot &\cdot  &1
\end{array}
\right),~~~
\delta^d_{LR}\sim\left(
\begin{array}{ccc}
\lambda^6 &\lambda^5&\lambda^5 \\
\lambda^5&\lambda^4&\lambda^4\\
\lambda^6&\lambda^6&\lambda^2
\end{array}
\right),~~~~\\[-2mm]
\label{eq:deltae}
&&\delta^e_{LL}\sim\left(
\begin{array}{ccc}
1  & \lambda^4 & \lambda^4 \\
\cdot &1 &\lambda^4 \\
\cdot &\cdot  &1
\end{array}
\right),~~~
\delta^e_{RR}\sim\left(
\begin{array}{ccc}
1  & \lambda^3 & \lambda^4 \\
\cdot &1 &\lambda^2 \\
\cdot &\cdot  &1
\end{array}
\right),~~~
\delta^e_{LR}\sim\left(
\begin{array}{ccc}
\lambda^6 &\lambda^5&\lambda^6 \\
\lambda^5&\lambda^4&\lambda^6\\
\lambda^5&\lambda^4&\lambda^2
\end{array}
\right).~~~~
\end{eqnarray}
Appendix~\ref{App:LowMIs} provides the explicit expressions for each entry
in terms of the parameters of the model.



\section{Numerical analysis}
\label{numerical}
\cleqn


\subsection{Parameter range}

Numerical results for the running quark and charged lepton masses as well as
for the quark mixing angles at the GUT scale can be found
in~\cite{GUTvalues}. The matching conditions from the SM to the Minimal
Supersymmetric Standard Model (MSSM), 
imposed at the SUSY scale, take the form
\begin{eqnarray}
\nonumber y_{u,c,t}^\text{SM}&\approx&\, y_{u,c,t}^\text{MSSM}\sin\bar{\beta},\\
\nonumber y_{d,s}^\text{SM}&\approx&
(1+\bar{\eta}_q)\,y_{d,s}^\text{MSSM}\cos\bar{\beta},\\
\nonumber y_b^\text{SM}&\approx& (1+\bar{\eta}_b)\,y_b^\text{MSSM}\cos\bar{\beta},\\
\nonumber y_{e,\mu}^\text{SM}&\approx& (1+\bar{\eta}_l)\,y_{e,\mu}^\text{MSSM}\cos\bar{\beta},\\
          y_\tau^\text{SM}&\approx& y_\tau^\text{MSSM}\cos\bar{\beta},\label{matchingconditions}
\end{eqnarray}
for the singular values of the Yukawa matrices. Similarly, we have for the CKM mixing
\begin{eqnarray}
 \theta^{q,\text{SM}}_{i3}\approx\frac{1+\bar{\eta}_q}{1+\bar{\eta}_b}\theta^{q,\text{MSSM}}_{i3},\qquad
 \theta^{q,\text{SM}}_{12}\approx\theta^{q,\text{MSSM}}_{12},\qquad
          \delta^{q,\text{SM}}\approx\delta^{q,\text{MSSM}}.
\end{eqnarray}
Here
\begin{eqnarray}
\bar{\eta}_q=\eta_q-\eta_l',\qquad
\bar{\eta}_b=\eta_q'+\eta_A-\eta_l',\qquad 
\bar{\eta}_l=\eta_l-\eta_l'\,,
\end{eqnarray}
represent SUSY radiative threshold corrections that are parametrised by
$\eta_i=\epsilon_i\,\tan\beta$, with explicit expressions for $\epsilon_i$
available in~\cite{GUTvalues2}. The unprimed $\eta$ parameters correspond to
corrections to the first two generations, the primed ones to the third
generation, and the one with index ``$A$" to a correction due to the soft SUSY breaking trilinear
terms. The parameter~$\bar{\beta}$ follows from the absorption of $\eta_l'$ into $\beta$,
\begin{eqnarray}
\cos\bar{\beta}\equiv(1+\eta_l')\cos\beta,~~~~~\sin\bar{\beta}\approx \sin\beta,
\end{eqnarray}
with the approximation being valid for $\tan\beta\gtrsim 5$. In the limit where
threshold effects for the charged leptons are neglected, $\tan\bar{\beta}$ simply reduces to $\tan\beta$.

Our model predicts $\hat y_{{b,\tau}} =  y_b\,\lambda^2$, where the hat
indicates the diagonalised Yukawa sector at the GUT scale. As a consequence,
very large values of  $\tan{\beta}$ are excluded, and we only study the
parameter space in which $\tan{\beta}\in[5,25]$, keeping the value of $y_b$ below
four. In order to obtain viable ranges for our Yukawa input parameters, we plot
$y_{u,c,t,b}$, $(\tilde{x}_2/y_s)^2$ and $(1+\bar{\eta}_l)y_s$ against
$\tan\bar{\beta}$ using the results for the diagonalised Yukawa sector at the
GUT scale provided in~\cite{GUTvalues}. We remark that $y_b$, $y_s$ and
$\tilde{x}_2$ are extracted from the lepton sector. We fit the resulting
curves using the relative uncertainties 
$\sigma(y_u)/y_u=31\%$, 
$\sigma(y_c)/y_c=3.5\%$, 
$\sigma(y_t)/y_t=10\%$, 
$\sigma(y_b)/y_b=0.6\%$, see~\cite{GUTvalues}.
Concerning $y_s$ and $\tilde{x}_2$, we take $\sigma(y_s)/y_s=10\%$ and 
$\sigma(\tilde{x}_2)/\tilde{x}_2=10\%$, allowing for higher order corrections
to the mass ratios that would reduce the discrepancy between the 
values of $\tilde{x}_2/y_s$ predicted from the lepton and the quark sectors and maximise
the GUT scale value of $(\hat y_\mu\,\hat y_d)/(\hat y_s\,\hat y_e)$. Due to
the implementation of the Georgi-Jarlskog relation~\cite{GJ}, it is 
equal to 9 in our model at LO, while its preferred range is
$10.7^{+1.8}_{-0.8}$~\cite{GUTvalues}, which is independent of  threshold
corrections and also not sensitive to a change of the SUSY scale.

We estimate the low energy Yukawa couplings using the leading logarithmic
approximation as described in~\cite{companion}. Clearly, the resulting low
energy Yukawa matrices are only valid up to that approximation. Mindful of
such limitations, we obtain
\begin{eqnarray}
\tilde{Y}^u_{{\text{low}}}&\approx&\text{Diag}\,\Big[(1+R^y_u)\,y_u\,\lambda^8,(1+R^y_u)\,y_c\,\lambda^4,(1+R^y_t)\,y_t\Big],\label{YuLow}\\
\tilde{Y}^d_{{\text{low}}}&\approx&\text{Diag}\,\Big[(1+R^y_d)\frac{\tilde{x}_2^2}{y_s}\lambda^6,(1+R^y_d)\,y_s\,\lambda^4,(1+R^y_b)\,y_b\,\lambda^2\Big],\label{YdLow}\\
\tilde{Y}^e_{{\text{low}}}&\approx&\text{Diag}\,\Big[(1+R^y_e)\frac{\tilde{x}_2^2}{3\,y_s}\lambda^6,(1+R^y_e)3y_s\,\lambda^4,(1+R^y_e)\,y_b\,\lambda^2\Big],\label{YeLow}
\end{eqnarray}
where the corrections from the RG running are encoded in the parameters $R^y_f$
\begin{eqnarray}
  R^y_u&=&\eta\left(\frac{46}{5}g_U^2-3y_t^2\right)-3\eta_N\,y_D^2,\qquad
R^y_t=R^y_u-3\,\eta \,y_t^2,\label{Rysx} \\
R^y_d&=&\eta\frac{44}{5}g_U^2,\qquad
R^y_b=R^y_d-\eta\,y_t^2,\qquad
R^y_e=\eta\frac{24}{5}g_U^2-\eta_N\,y_D^2.
\label{Rys}
\end{eqnarray}
Here, $g_U\approx \sqrt{0.52}$ denotes the universal gauge coupling constant at the GUT scale.
Our scan produces the following values for the right-hand sides of Eq.~\eqref{matchingconditions}
\begin{eqnarray}
\nonumber &\!\!&\tilde{Y}^u_{\text{low}_{11}}\sin\bar{\beta}\in[3.4,6.9]\times 10^{-6}\!,~
\tilde{Y}^u_{\text{low}_{22}}\sin\bar{\beta}\in[2.34,2.65]\times 10^{-3}\!,~
\tilde{Y}^u_{\text{low}_{33}}\sin\bar{\beta}\in[0.77,0.89],\\
\nonumber
&&\tilde{Y}^d_{\text{low}_{11}}\cos\bar{\beta}(1+\bar{\eta}_q)\in[0.9,1.6]\times10^{-5},~~~~~
\tilde{Y}^d_{\text{low}_{22}}\cos\bar{\beta}(1+\bar{\eta}_q)\in[2.2,3.5]\times 10^{-4},\\
\nonumber
&&\tilde{Y}^d_{\text{low}_{33}}\cos\bar{\beta}(1+\bar{\eta}_b)\in[1.17,1.6]\times
10^{-2},\\
\nonumber&&
\tilde{Y}^e_{\text{low}_{11}}\cos\bar{\beta}(1+\bar{\eta}_l)\in[2.4,3.8]\times 10^{-6},~~~~~
\tilde{Y}^e_{\text{low}_{22}}\cos\bar{\beta}(1+\bar{\eta}_l)\in[5.6,7.7]\times 10^{-4},\\
&&\tilde{Y}^e_{\text{low}_{33}}\cos\bar{\beta}\in[1.06,1.14]\times 10^{-2},
\end{eqnarray}
which have to be compared to the SM values, taken from Table 2
of~\cite{GUTvalues},
\begin{eqnarray}
 y_u^\text{SM}&\in&[3.40,7.60]\times 10^{-6},~~~y_c^\text{SM}\in[2.69,3.20]\times 10^{-3},~~~y_t^\text{SM}\in[0.78,0.88],\label{SMYukawas}\\
\nonumber y_d^\text{SM}&\in&[1.15,1.56]\times 10^{-5},~~~y_s^\text{SM}\in[2.29,2.84]\times 10^{-4},~~~y_b^\text{SM}\in[1.21,1.42]\times 10^{-2},\\
y_e^\text{SM}&\in&[2.85,2.88]\times 10^{-6},~~~y_\mu^\text{SM}\in[6.01,6.08]\times 10^{-4},~~~y_\tau^\text{SM}\in[1.02,1.03]\times 10^{-2}.~~~~~~\nonumber
\end{eqnarray}
The corresponding ranges of the order one input parameters of the Yukawa sector
are listed in the first five rows of the first column of Table~\ref{Ranges}. 
All other coefficients that are not fixed by this fit, are scanned over the
interval $\pm[0.5,2]$, with the following exceptions: we allow the absolute
value of the Dirac neutrino Yukawa coupling $y_D$ to be as small as $0.2$ but
not larger than $0.6$, such that it does not exceed the maximum allowed
value of $y_t$. 
We also relax the lower bounds on $|\tilde{x}^a_2|$, $|a_s|$ and $|a_u|$ and
extend the upper bound on $|a_b|$, such that they are allowed to get the same
values as the corresponding Yukawa coefficients.
The coefficients $c_{H_u}$ and $c_{H_d}$ of the soft Higgs mass squares, 
\begin{eqnarray}
m^2_{{H_u}_\text{GUT}}&=&c_{H_u}\,m_0^2,\qquad 
m^2_{{H_d}_\text{GUT}}=c_{H_d}\,m_0^2\ ,~~\label{cHud}
\end{eqnarray}
are taken to be positive, just like the coefficients $b_{01}$, $b_{02}$ and
$B_0^{(N)}$ of the leading order diagonal elements of the soft scalar mass
squared matrices. Phases are generally allowed to take arbitrary values within $[0,2\pi]$.
As mentioned earlier, $\tan\beta$ is varied between~5~and~25. Concerning the
CMSSM parameters, we define
\begin{eqnarray}
\alpha_0&\equiv&A_0/m_0,\qquad 
x\equiv(M_{1/2}/m_0)^2,\label{x_alpha0}
\end{eqnarray}
and scan over $M_{1/2}\in[0.3,5]$~TeV, $m_0\in[0.05,5]$~TeV as well as
$\alpha_0\in[-3,3]$ in order to avoid charge and colour breaking
minima.\footnote{In our numerical scan, we have checked that the potentials
are always bounded from below and that the corresponding minima do not break
charge or colour~\cite{CCB}.}

\begin{table}[t]
\begin{center}
 \begin{tabular}{ |c|c||c |c|}
    \hline
    Yukawa terms & Range & Soft trilinear terms & Range\\ \hline
    $\tilde{x}_2,y_s$      &$[0.2,1.6]$&   $\tilde{x}^a_2,a_s$&$\pm[0.2,2]$ \\ \hline
    $y_b$                  &$[0.7,3.8]$&   $a_b$&$\pm[0.5,4] $             \\\hline
    $y_u$                  &$[0.3,0.6]$&   $a_u$&$\pm[0.3,2] $              \\\hline
    $y_c$                  &$[0.5,0.6]$&   $a_c$&\multirow{ 4}{*}{$\pm[0.5,2]$ } \\ \cline{1-3}
    $y_t$                  &$[0.46,0.6]$&  $a_t$&       \\ \cline{1-3}
    $y_D$                  &$\pm[0.2,0.6]$&$\alpha_D$&     \\ \cline{1-3}
    $z^f_i$                &$\pm[0.5,2]$&$z^{f_a}_i$&  \\ \cline{1-3}\hline\hline
    K\"ahler metric   &  Range  & Soft mass terms & Range       \\ \hline
     $k_2,k_3,k_4,K_3^{(N)}$  & $\pm[0.5,2]$   &
$b_2,b_3,b_4,B^{(N)}_3$     & $\pm[0.5,2]$ 
\\\hline
     \multicolumn{2}{|c||}{} 
& $b_{01},b_{02},B^{(N)}_0,c_{H_{u}},c_{H_{d}}$      &$[0.5,2]$
     \\ \hline\hline
   SUSY masses   &  Range  & SUSY ratios & Range       \\ \hline
$M_{1/2}$ & $[0.3,5]$\,TeV &   $\tan \beta$ & $[5,25]$ \\ \hline
 $m_0$ &$[0.05,5]$\,TeV & $\alpha_0$ &  $[-3,3]$  \\\hline
  \end{tabular}
\end{center}
\caption{Ranges of the input parameters used in our scan.}\label{Ranges}
\end{table}

The $\mu$ parameter, which we take as real, is given at the electroweak scale
by the relation\footnote{The lack of any evidence for
low energy supersymmetry requires a certain amount of cancellation between the
terms of Eq.~\eqref{eq:MzMh1}, see e.g.~\cite{Kane:1998im}.}
\begin{eqnarray}\label{eq:MzMh1}
\frac{M_Z^2}{2}&=&\frac{m_{H_d}^2+\Sigma^d_d-(m_{H_u}^2+\Sigma^u_u)t_\beta^2}{t_\beta^2-1}-\mu^2,
\end{eqnarray}
where $M_Z$ denotes the $Z$ boson mass~\cite{Baer-RNS-13}. $\Sigma^u_u$ and
$\Sigma^d_d$ are radiative corrections, with the most important contributions
coming from the stops 
\begin{eqnarray}
\Sigma^u_u\left(\tilde{t}_{1,2}\right)&=&\frac{3}{16\pi^2}F(m^2_{\tilde{t}_{1,2}})\left(Y_t^2-g_Z^2\mp\frac{A_t^2-8g_Z^2\left(\frac{1}{4}-\frac{2}{3}x_W\right)\Delta_t}{m^2_{\tilde{t}_{2}}-m^2_{\tilde{t}_{1}}}\right),\\
\Sigma^d_d\left(\tilde{t}_{1,2}\right)&=&\frac{3}{16\pi^2}F(m^2_{\tilde{t}_{1,2}})\left(g_Z^2\mp\frac{Y_t^2 \mu^2+8g_Z^2\left(\frac{1}{4}-\frac{2}{3}x_W\right)\Delta_t}{m^2_{\tilde{t}_{2}}-m^2_{\tilde{t}_{1}}}\right).
\end{eqnarray}
In these expressions, $Y_t$, $A_t$ and $\mu$ denote the low energy Yukawa and
trilinear couplings and the low energy $\mu$ parameter, respectively. Moreover
\begin{eqnarray}
\nonumber m^2_{\tilde{t}_{1,2}}&=&\frac{1}{2}\left(m^2_{\tilde{t}_{LL}}+m^2_{\tilde{t}_{RR}}\mp\sqrt{4\,m^2_{\tilde{t}_{LR}}+(m^2_{\tilde{t}_{LL}}-m^2_{\tilde{t}_{RR}})^2}\right),\\
\nonumber F(m^2)&=&m^2\left(\log\left(\frac{m^2}{M_S^2}\right)-1\right)\!,~~~~~~
\Delta_t=\frac{1}{2}\left(m^2_{\tilde{t}_{LL}}-m^2_{\tilde{t}_{RR}}\right)+M_Z^2\cos(2\beta)\left(\frac{1}{4}-\frac{2}{3}x_W\right)\!,\\
x_W&=&\sin^2\theta_W,\qquad
g_Z^2=\frac{M_Z^2}{4\upsilon^2},\qquad
M_S=\sqrt{m_{\tilde{t}_1}m_{\tilde{t}_2}},
\end{eqnarray}
with $\theta_W$ denoting the Weinberg angle. $m^2_{\tilde{t}_{LL}}$,
$m^2_{\tilde{t}_{RR}}$ and $m^2_{\tilde{t}_{LR}}$ are the low energy (33)
elements of the squark mass matrices defined in Eq.~\eqref{fullmasses}. 
The so-determined $\mu$ parameter can then be used to calculate the physical
Higgs mass. Adopting the approximate formulas of Section~2.4 of~\cite{MHiggs},
we demand that the resulting Higgs mass lies within the interval
$[110,135]$~GeV. 
Additionally, we impose cuts on the SUSY parameters from direct searches
by requiring that the first and the second generation squark masses are larger than
$1.4$~TeV.


\subsection{Estimates of the low energy mass insertion parameters }

In this section, we analyse the predictions for the low energy mass insertion
parameters~$\delta$ whose explicit expressions are given in
Appendix~\ref{App:LowMIs}.  
Tables~\ref{up limits}-\ref{electron limits} provide naive expectations for
the individual $\delta$s, where we take into account the $\lambda$-suppression
and the main effects of the RG running, while setting any order one
coefficients to one. 
Clearly, we still expect to see a spread within a few orders of magnitude due
to the variation of the SUSY scale and the order one coefficients. 
The third columns of Tables~\ref{up limits}-\ref{electron limits} list existing
experimental bounds. The full ranges of our $\delta$s arising from scanning
over the input parameters, given in Table~\ref{Ranges}, are depicted in
Figures~\ref{Fig:up MIs}-\ref{Fig:electron MIs}.

\subsubsection{Up-type quark sector}

\begin{table}[t]
\begin{center}
  \begin{tabular}{ |M{3cm} || M{7.85cm} | M{3.3cm} |}
    \hline
    Parameter & Naive expectation & Exp. bound \\ \hline
    $\sqrt{|\mathrm{Im}[(\delta^u_{LL,RR})^2_{12}]|}$&$\mathcal{O}\left(\frac{\sqrt{\sin(2\theta^d_2)}\lambda^4}{1+6.3\,x}\approx
    4\times{10^{-4}}\sqrt{\sin(2\theta^d_2)}\right)~~~~~~$ &
    $2.85\times10^{-2}$ \cite{Dmixing}\newline$\left(1.65\times10^{-3}\right)\!|_{_{LL=RR}}$ \\ \hline
    $\sqrt{|\mathrm{Im}[(\delta^u_{LR,RL})^2_{12}]|}$&$ 0 $ & $3.75\times10^{-3}$\cite{Dmixing}\\ \hline
    $|(\delta^u_{LL})_{13}|$&$\mathcal{O}\left( \frac{1+\eta\left(\frac{R_q}{1+6.5\,x}-y_t^2\right)}{1+6.5\,x}\lambda^6\approx  2\times 10^{-5}    \right)$ &$ \mathcal{O}(10^{-1})$ \cite{Khalil}\\ \hline
    $|(\delta^u_{RR})_{13}|$&$\mathcal{O}\left( \frac{1+2\eta\left(\frac{R_q}{1+6.15\,x}-y_t^2\right)}{1+6.15\,x}\lambda^6\approx 2\times 10^{-5}   \right)$ &\multirow{ 9}{*}{}\\\hline
    $|(\delta^u_{LL})_{23}|$&$\mathcal{O}\left( \frac{1+\eta\left(\frac{R_q}{1+6.5\,x}-y_t^2\right)}{1+6.5\,x}\lambda^5 \approx 8\times 10^{-5}\right)$         & \\\hline
    $|(\delta^u_{RR})_{23}|$&$\mathcal{O}\left( \frac{1+2\eta\left(\frac{R_q}{1+6.15\,x}-y_t^2\right)}{1+6.15\,x}\lambda^5\approx 8\times 10^{-5}   \right)$ & \\ \hline
  $|(\delta^u_{LR})_{13}|$&$\mathcal{O}\left(\frac{\alpha_0\,\upsilon_u}{m_0}\frac{2\,\eta}{(1+6.3\,x)}\lambda^7\approx  10^{-7}  \right)$ & \\ \hline
    $|(\delta^u_{LR})_{23}|$&$\mathcal{O}\left(\frac{\alpha_0\,\upsilon_u}{m_0}\frac{2\,\eta}{(1+6.3\,x)}\lambda^6
\approx 5\times 10^{-7}  \right)$ &$\mathcal{O}(10^{-1})$ \cite{Schacht}\\ \hline
   $|(\delta^u_{RL})_{13}|$&$0$ & \\ \hline
   $|(\delta^u_{RL})_{23}|$&$\mathcal{O}\left(\frac{\alpha_0\,\upsilon_u}{m_0}\frac{1+\eta\left(\frac{46\,g_U^2}{5}-8y_t^2+\frac{R_q}{1+6.5\,x}\right)}{1+6.3\,x}\lambda^7\approx 5\times 10^{-7}   \right)$ & 
\\ \hline
  \end{tabular}
\end{center}
\caption{The naive numerical expectations for the  low energy up-type mass insertion
  parameters as extracted from our model (second column), to be
  compared with experimental bounds in the literature (third
  column). The full ranges of the $\delta$s are shown in Figure~\ref{Fig:up
    MIs}. Note that the (12), (21) and (31)  $\delta^u_{LR}$ parameters
  remain zero up to order $\lambda^8$.}\label{up limits} 
\end{table}

\begin{figure}[H]
\minipage{0.48 \textwidth}
  \includegraphics[width=\linewidth]{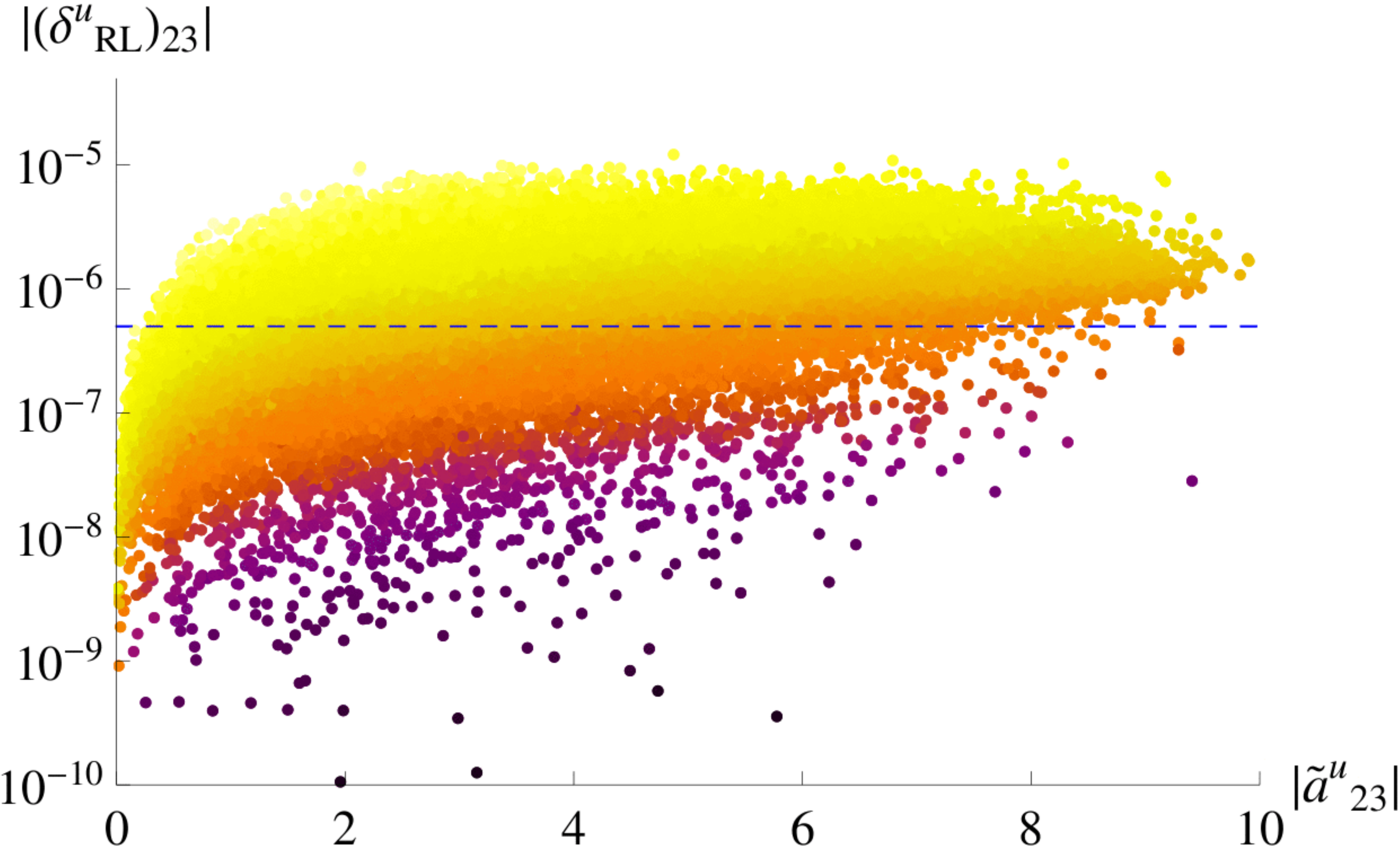}
\endminipage\hfill
\minipage{0.48 \textwidth}
  \includegraphics[width=\linewidth]{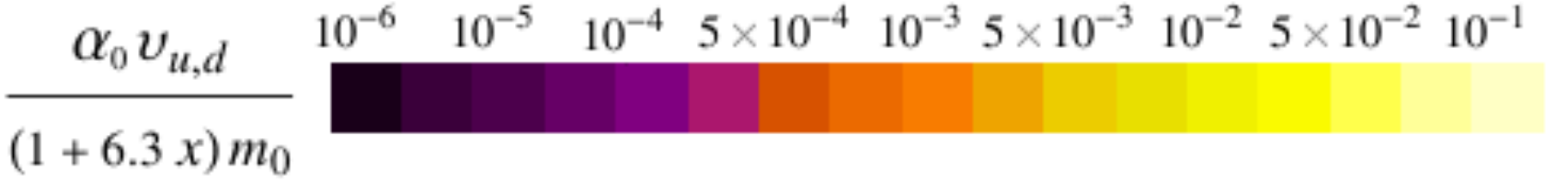}
\endminipage\hfill
\caption{The low energy up-type mass insertion parameters plotted against
  their GUT scale coefficients, defined in Eqs.~(\ref{Bt},\ref{adt}) [except
    for $(\delta^u_{LR})_{13,23}$ which are plotted against a coefficient
    multiplying the RG running contribution,
    cf. Eqs.~(\ref{deltauLR13Low},\ref{deltauLR23Low})]. The blue dashed
  lines represent our naive 
  numerical expectations according to the second column of Table~\ref{up
    limits}, while the red dotted lines (when available) represent their
  experimental limits, shown in the third column of Table~\ref{up limits}. 
Since $(\delta^u_{RR})_{12}\approx (\delta^u_{LL})_{12}$, only the $LL$
parameter is plotted. The plots have been produced by scanning over the input
parameters listed in Table~\ref{Ranges}.}
\label{Fig:up MIs}
\end{figure}

The strongest constraints on the up-type mass insertion parameters involve the
(12) sector and stem from $D^0-\bar{D}^0$ mixing. The SM contribution to this
amplitude conserves CP to a good approximation and provides significant
constraints on the imaginary parts of $(\delta^u_{AB})_{12}$, $A,B=L,R$. 
These limits were derived in~\cite{Dmixing}, assuming equal squark and gluino
masses of 1~TeV. We quote them in the third column of Table~\ref{up limits},
rescaled to masses of 1.5~TeV. The limits on the $RR$ and $RL$ parameters are
identical to the $LL$ and $LR$ ones due to the $L\leftrightarrow R$ symmetric
form of the gluino-squark box diagram. 
The index $LL=RR$ refers to the assumption that 
$(\delta^u_{LL})_{12}\approx(\delta^u_{RR})_{12}$, as is the case in our model.  
In the second column of Table~\ref{up limits}, we give a naive estimate for 
$\sqrt{|\mathrm{Im}[(\delta^u_{LL})^2_{12}]|}\approx 
\sqrt{|\mathrm{Im}[(\delta^u_{RR})^2_{12}]|} \approx 
\sqrt{|\mathrm{Im}[(\delta^u_{LL})_{12}(\delta^u_{RR})_{12}]|}$. 
For $\theta^d_2=\pi/2$, as suggested from maximising the Jarlskog
invariant of Eq.~\eqref{JCPq}, these quantities vanish to LO. 
Since $\sqrt{|\mathrm{Im}[(\delta^u_{LL,RR})^2_{12}]|}$ is at most 
$\sim |(\delta^u_{LL})_{12}|$, we only show the full range of the absolute
value of that parameter in Figure~\ref{Fig:up MIs}, plotted against the
corresponding GUT scale coefficient $\tilde{b}_{12}$, defined in Eq.~\eqref{Bt}. 
This coefficient quantifies the mismatch between the K\"ahler metric and the
soft mass matrix elements for the $SU(5)$ ${\bf{10}}$-plets and can be as
large as 6 when the associated parameters contribute constructively and
receive their maximum values in the scan. 
The effects of the RG running are trivial and depend only on $x=(M_{1/2}/m_0)^2$; 
for $x\approx 1$ and $\tilde{b}_{12}\approx 1$, we estimate a value of around
$4\times 10^{-4}$, shown by the blue dashed line in Figure~\ref{Fig:up MIs}. 
With increasing $x$, we obtain even smaller values, as the RG suppression is
increased. The red dotted line shows the experimental limit, adapted
from~\cite{Dmixing} and valid for $(\delta^u_{LL})_{12}\approx(\delta^u_{RR})_{12}$.

The $LL$ and $RR$ parameters of the $(i3)$ sector ($i=1,2$) have GUT scale
coefficients with the same range as the parameters of the (12) sector  but a different
RG suppression due to the milder running of the third generation sfermionic
masses.  
This is represented by the factor $\eta\,R_q$  appearing in Eq.~\eqref{pus},
where $\eta$ and $R_q$ are defined in Eqs.~(\ref{etas},\ref{Rq}), respectively. 
Approximating these $\delta$s as shown in Table~\ref{up limits} and taking
$x\approx 1$, $R_q\approx 3y_t^2+1$ 
as well as $y_t\approx 0.5$, we expect $|(\delta^u_{LL,RR})_{13}|\propto
\lambda^6$ and $|(\delta^u_{LL,RR})_{23}|\propto \lambda^5$ to vary around
$2\times 10^{-5}$ and $8\times 10^{-5}$, respectively. 
The existing bounds on these variables from flavour changing effects are very
weak, leaving them essentially unconstrained. 
$B_d$ mixing can place a bound on $|(\delta^u_{LL})_{13}|$ of the order of
$10^{-1}$ at most, as described in~\cite{Khalil}.

\begin{figure}[H]
\minipage{0.48 \textwidth}
  \includegraphics[width=\linewidth]{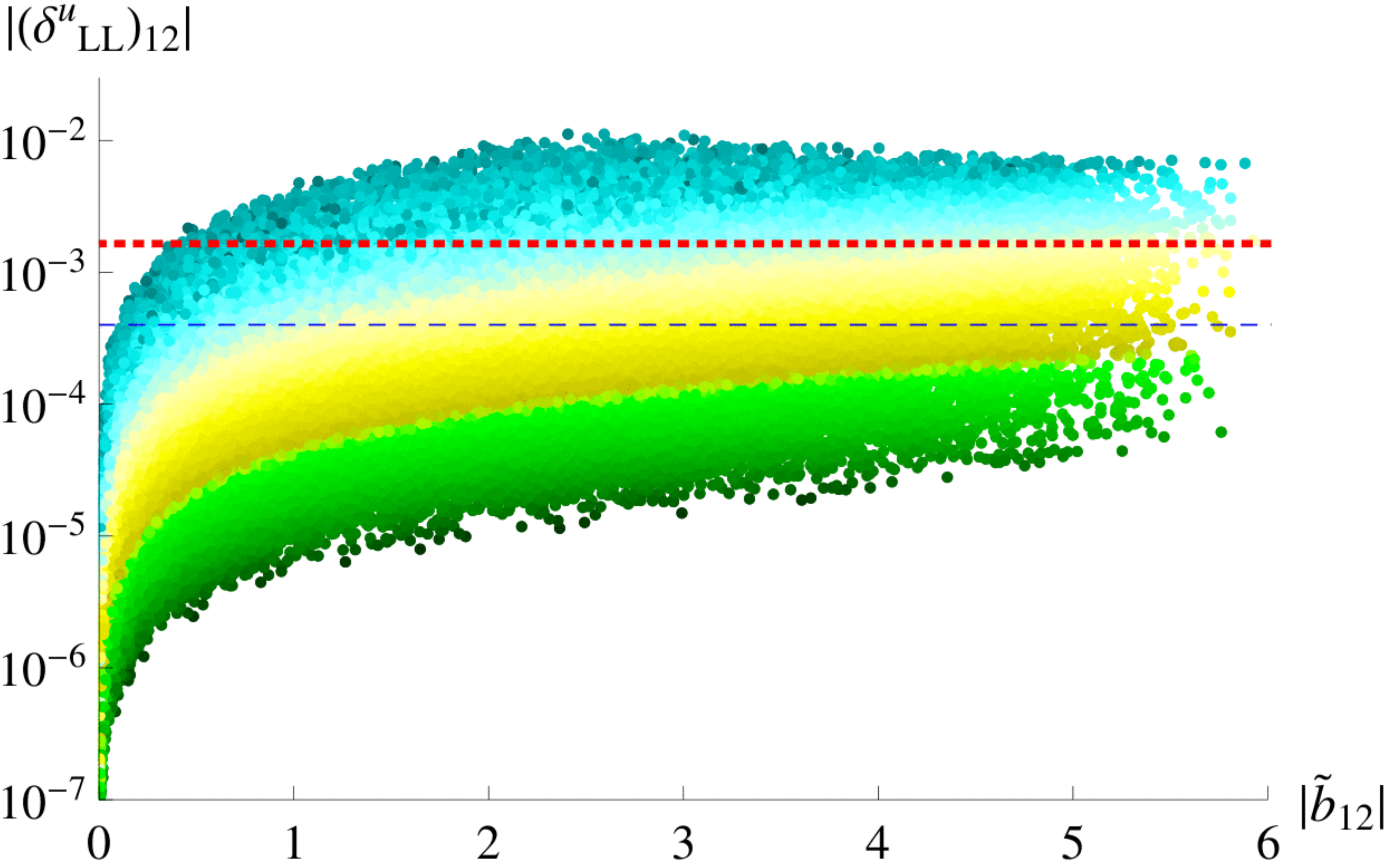}
\endminipage\hfill
\hspace{-6.00mm}
\minipage{0.4\linewidth}
  \includegraphics[width=70mm]{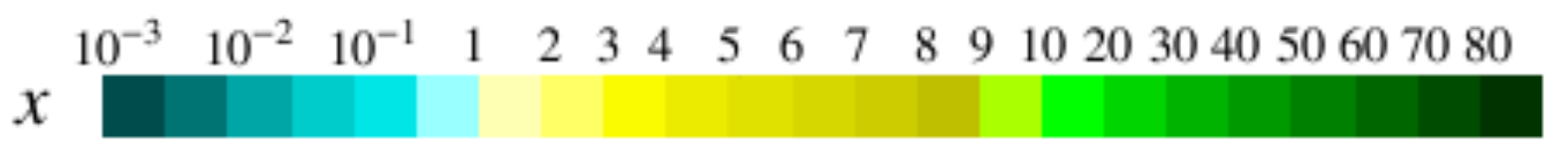}
\endminipage\hfill
\minipage{0.48 \textwidth}
  \includegraphics[width=\linewidth]{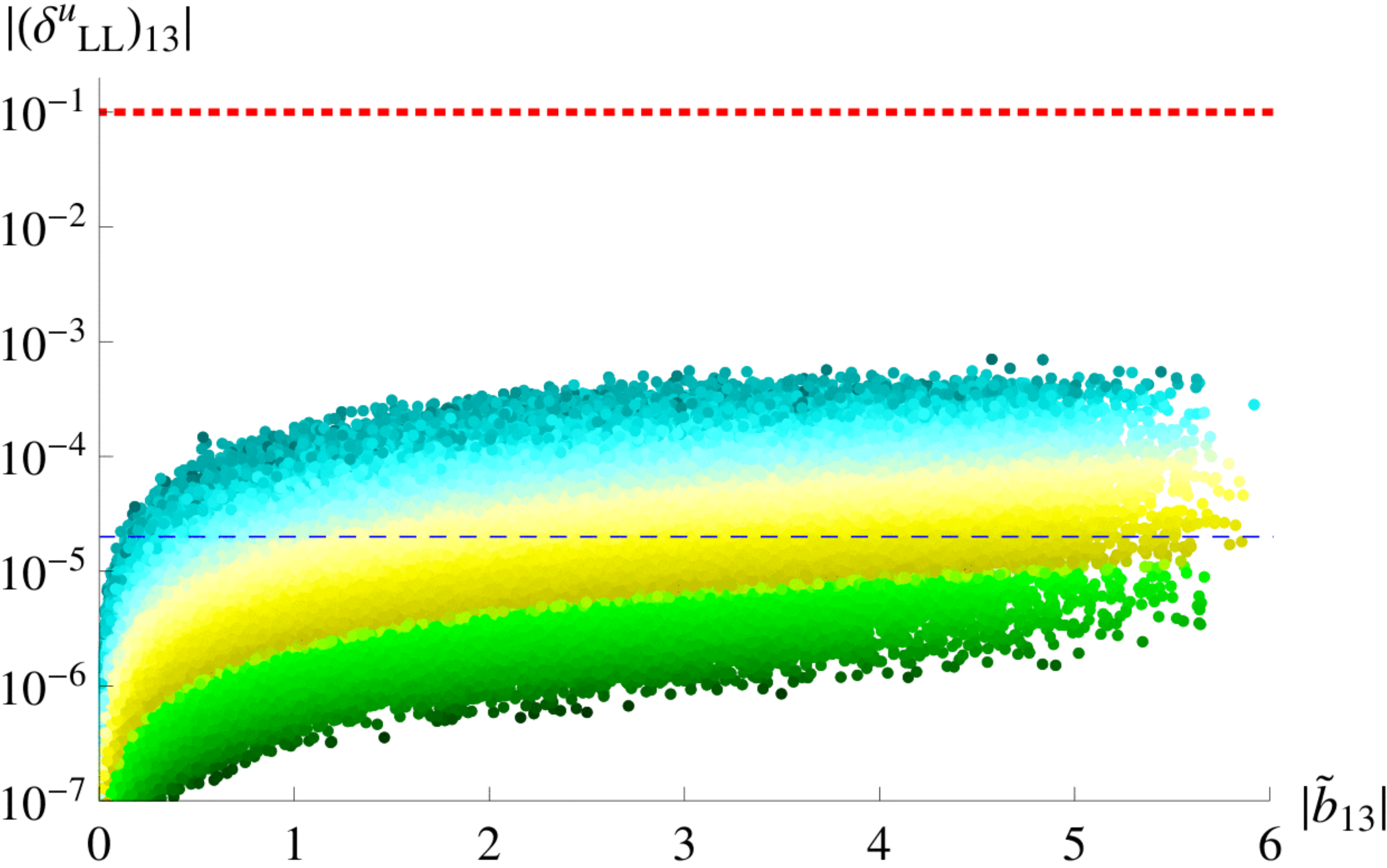}
\endminipage\hfill
\minipage{0.48 \textwidth}
  \includegraphics[width=\linewidth]{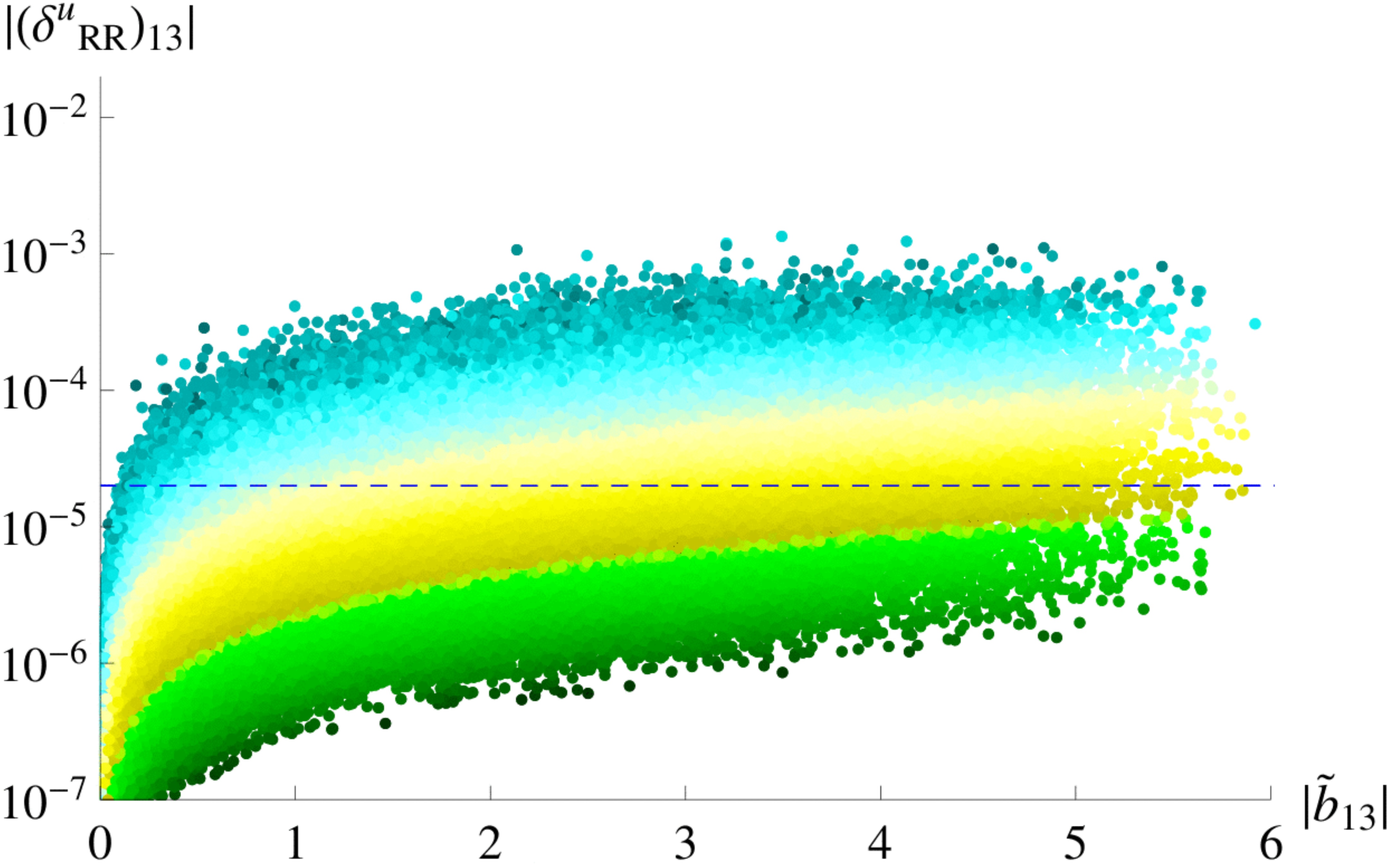}
\endminipage\hfill
\minipage{0.48 \textwidth}
  \includegraphics[width=\linewidth]{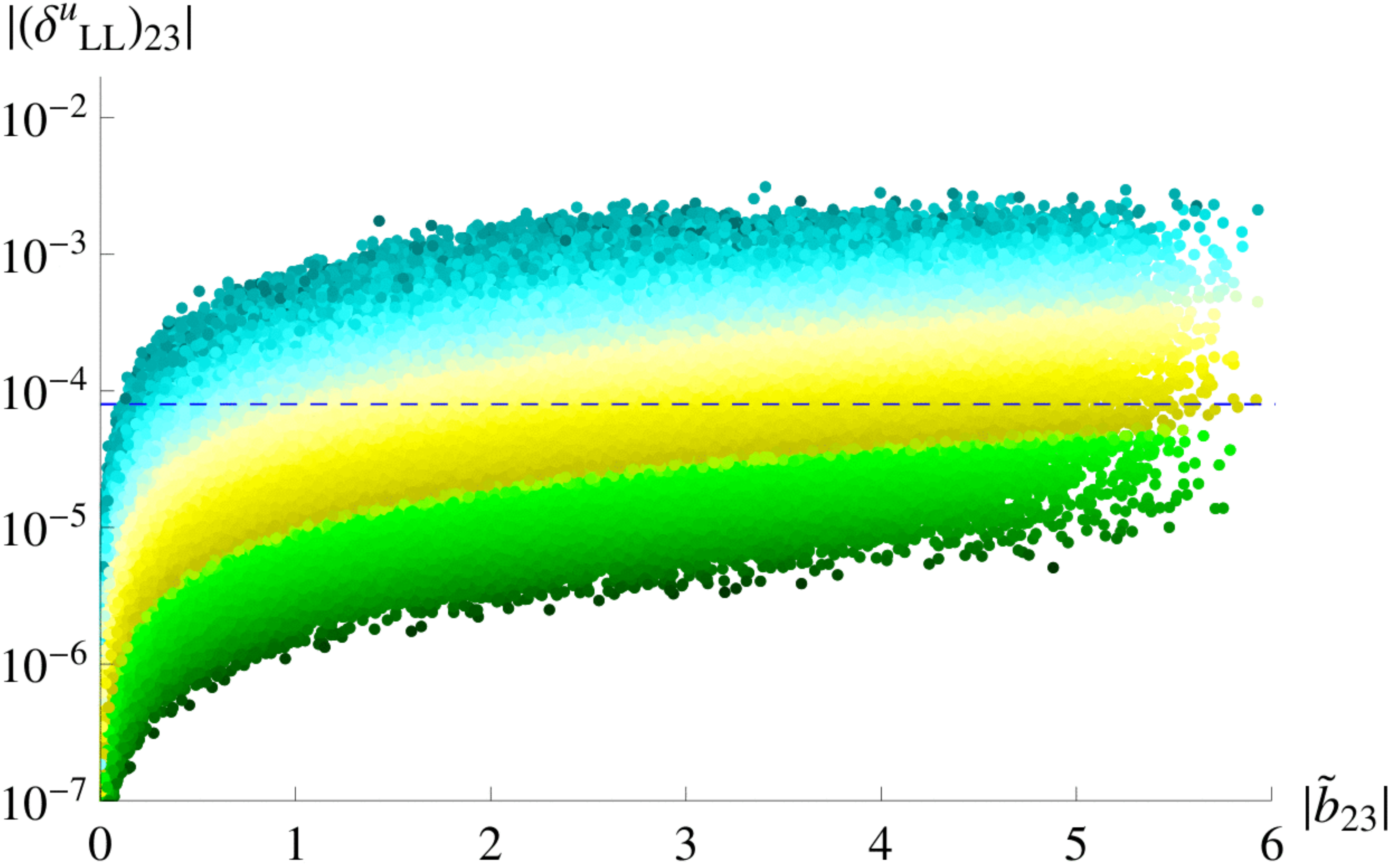}
\endminipage\hfill
\minipage{0.48 \textwidth}
  \includegraphics[width=\linewidth]{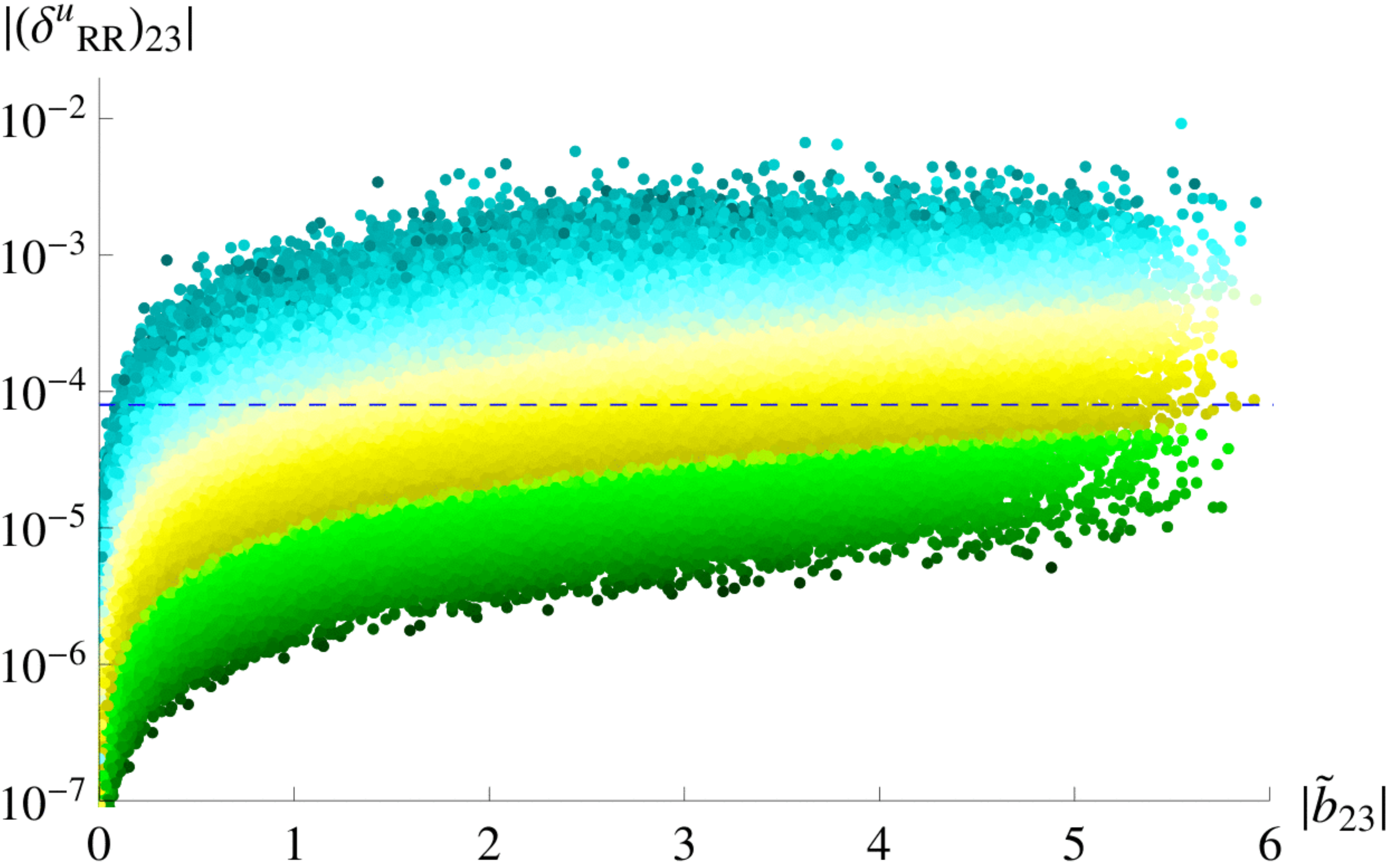}
\endminipage\hfill
\minipage{0.48 \textwidth}
  \includegraphics[width=\linewidth]{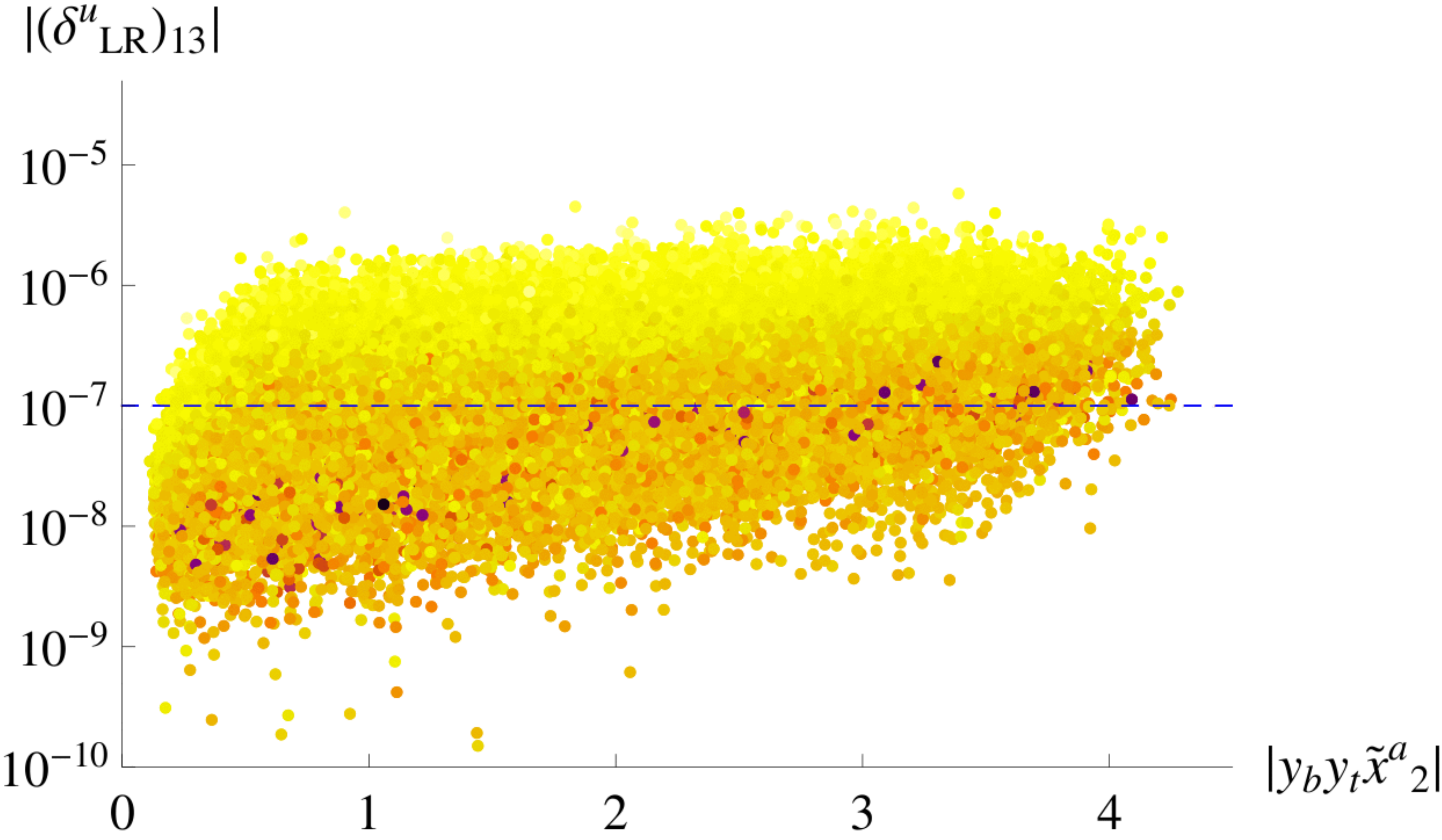}
\endminipage\hfill
\minipage{0.48 \textwidth}
  \includegraphics[width=\linewidth]{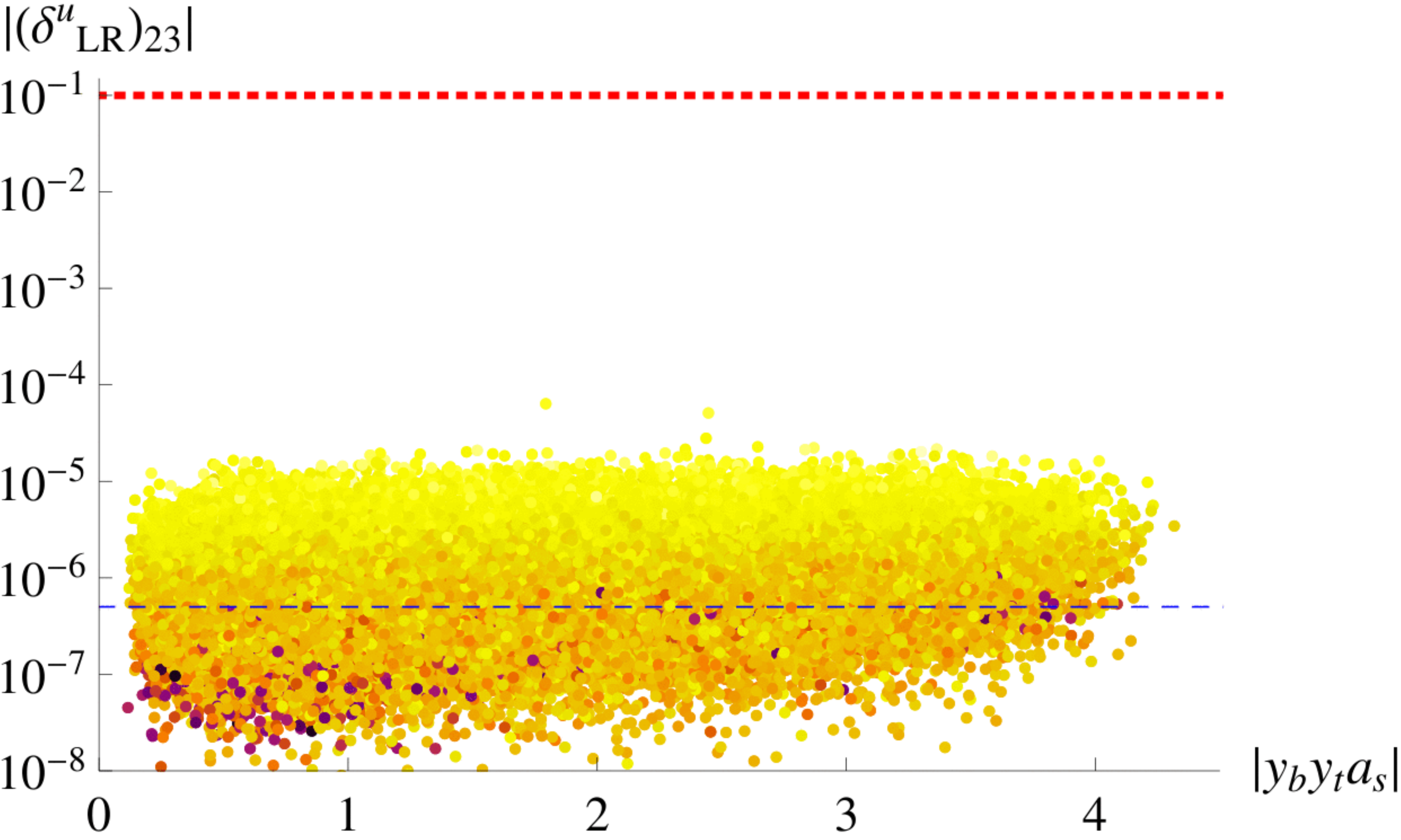}
\endminipage\hfill
\end{figure}

The parameters of $LR$ type have a slightly different behaviour. They are
proportional to the factor $(\alpha_0\,\upsilon_u/m_0)$ which, for
$|A_0|>0.5$~TeV, can cause an extra suppression of up to
$\mathcal{O}(10^{-3})$.  
Because of this factor, the $LR$ parameters show a dependence on the mass
scale, even at the GUT scale. 
$(\delta^f_{LR})_{ij}$ are also generally proportional to the mismatch
of the ratios of soft trilinear over Yukawa sector coefficients for
the $i$-th and the $j$-th generation and vanish, barring RG induced
corrections, if those are aligned.
To estimate the magnitude of these parameters in Table~\ref{up limits}, we
take $|\alpha_0|\,\upsilon_u/m_0\approx 10^{-1}$, $x\approx1$, $y_t\approx0.5$
and $R_q\approx 1.75$, while their full ranges are shown in 
Figure~\ref{Fig:up MIs}. 
The $(\delta^u_{LR})_{13}$ parameter was zero at the GUT scale but receives a
contribution through the RG running of the order of
$\eta\,\lambda^7$. Similarly, $(\delta^u_{LR})_{23}$, which was suppressed by
$\lambda^7$ at the GUT scale, receives a similar running contribution which
comes in at an even lower order, namely $\eta\,\lambda^6$. Such an effect is
not found in any other $\delta$ parameter.
Finally, we remark that $(\delta^u_{LR,RL})_{12}$ as well as
$(\delta^u_{RL})_{13}$ are zero up to order $\lambda^8$, where we truncate our
expansion.

The limits on the $LR$ parameters of the $(i3)$ sector ($i=1,2)$ originate
mainly from the requirement that the potential be bounded from below with a
vacuum that does not break charge or colour~\cite{CCB}. We have already
constrained the trilinear parameters accordingly and do not comment on
those effects any further.  
Other bounds on the $LR$ off-diagonal parameters can be deduced by demanding
that the supersymmetric radiative corrections to the CKM matrix elements do
not exceed their experimental values~\cite{Crivellin1}. The limit for
$|(\delta^u_{LR})_{23}|$ quoted in Table~\ref{up limits} has been obtained
in~\cite{Schacht} by considering chargino loop contributions to $b\to s
l^+l^-$. In our model, all up-type mass insertion parameters of the $LR$ type turn out
to be safely below any current bound.

\subsubsection{\label{Numerics:Down-quark sector}Down-type quark sector}

\begin{table}[t]
\begin{center}
  \begin{tabular}{ |M{4cm} || M{7.4cm} | M{2.7cm} |}
    \hline
    Parameter & Naive expectation & Exp. bound \\ \hline
    $\sqrt{\left|\mathrm{Re}\left[(\delta^d_{LL})^2_{12}\right]\right|}$&$\mathcal{O}\left(\frac{1}{1+6.5\,x}\lambda^3\approx 2\times10^{-3}\right)$ & \multirow{ 2}{*}{$[6.6\times10^{-2},$}\\ \cline{1-2}
    $\sqrt{\left|\mathrm{Re}\left[(\delta^d_{RR})^2_{12}\right]\right|}$&$\mathcal{O}\left(\frac{\sqrt{\cos(2\theta^d_2)}}{1+6.1\,x}\lambda^4\approx 4\times10^{-4}\sqrt{\cos(2\theta^d_2)}\right)$ & $3.3
\times 10^{-1}]$\\\hline
$\sqrt{\left|\mathrm{Im}\left[(\delta^d_{LL})^2_{12}\right]\right|}$&$\mathcal{O}\left(\frac{\sqrt{\sin(\theta^d_2)}}{1+6.5\,x}\lambda^{7/2}\approx 7\times 10^{-4}\sqrt{\sin(\theta^d_2)}\right)$ & \multirow{ 2}{*}{$[8.7\times10^{-3},$}\\ \cline{1-2}
    $\sqrt{\left|\mathrm{Im}\left[(\delta^d_{RR})^2_{12}\right]\right|}$&$\mathcal{O}\left(\frac{\sqrt{\sin(2\theta^d_2)}}{1+6.1\,x}\lambda^4\approx 4\times10^{-4}\sqrt{\sin(2\theta^d_2)}\right)$ &$4.2\times 10^{-2}]$ \\\hline
    $\sqrt{\left|\mathrm{Re}\left[(\delta^d_{LR(RL)})^2_{12}\right]\right|}$&$\mathcal{O}\Big(\frac{\alpha_0\,\upsilon_d}{m_0}\frac{1+\eta\,\frac{44\,g_U^2}{5}}{1+6.3\,x}\lambda^5\times$ & $[7.8,12]\times10^{-3}$\\ \cline{1-1}\cline{3-3}
    $\left|\mathrm{Im}\left[(\delta^d_{LR(RL)})_{12}\right]\right|$& $\mathrm{Re}(\mathrm{Im})\left[f(\theta^{\tilde{x}_a}_2-\theta^{\tilde{x}}_2,\theta^a_s-\theta^y_s)\right]\approx 7\times 10^{-7}\Big)$ & $[1,5.7]\times 10^{-4}$\\ \hline
  \end{tabular}
\end{center}
\caption{The naive expectation for the ranges of $(\delta^d_{AB})_{12}$,
  $A,B=L,R$, as extracted from our model (second column), to be compared with
  experimental bounds from~\cite{FlavourPhysicsandGrandUnification} for
  $m_{\tilde{q}}\approx1.5$ TeV and $(m_{\tilde{g}}/m_{\tilde{q}})^2\in[0.3,4]$
  (third column).  The full ranges of these $\delta$s as produced in our
  scan are shown in Figure~\ref{Fig:down MIs}.
}\label{down (12) limits}
\end{table}

We first consider the (12) elements of the down-type mass insertion parameters
$(\delta^d_{AB})_{12}$, where $A,B=L,R$. The corresponding bounds are derived 
from the results of~\cite{FlavourPhysicsandGrandUnification} which we have
rescaled to $m_{\tilde{q}}\approx1.5$~TeV and
$(m_{\tilde{g}}/m_{\tilde{q}})^2\in[0.3,4]$. These bounds are summarised in
the third column of Table~\ref{down (12) limits} and have been extracted
using observables related to Kaon mixing. They are given separately for the
real and imaginary parts due to a relative difference  of an order of magnitude.

\begin{figure}[H]
\minipage{0.48 \textwidth}
  \includegraphics[width=\linewidth]{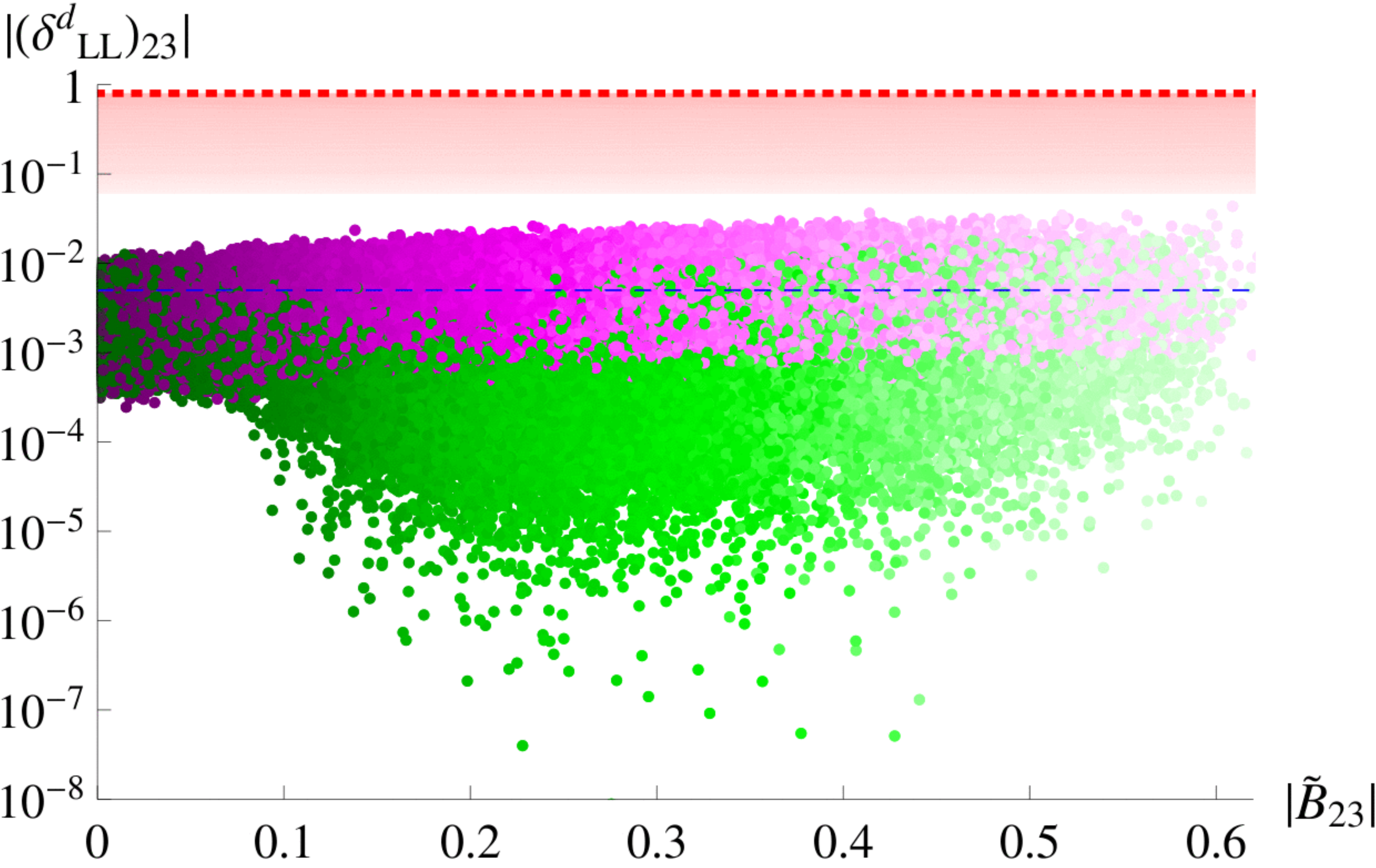}
\endminipage\hfill
\minipage{0.48 \textwidth}
  \includegraphics[width=\linewidth]{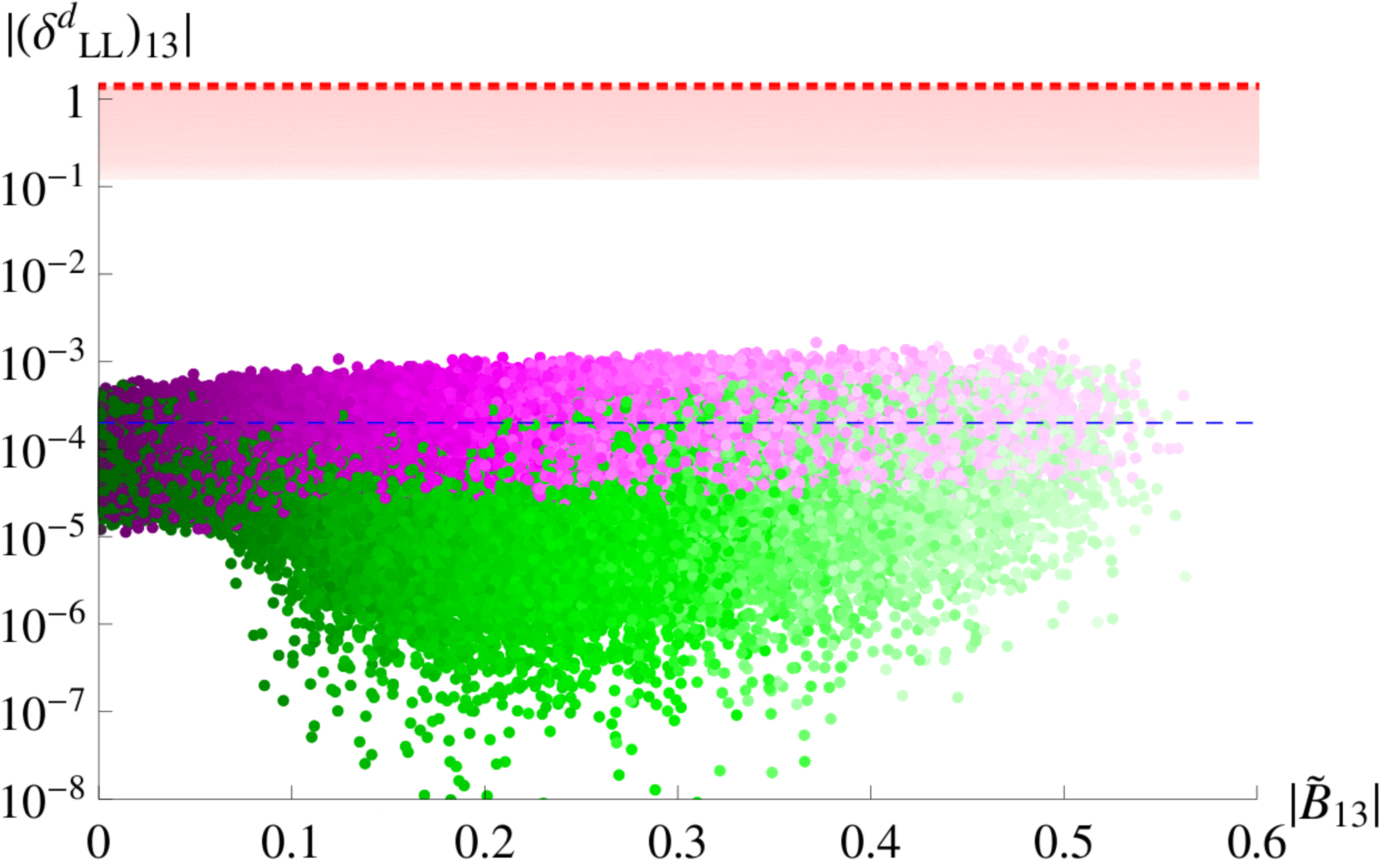}
\endminipage\hfill
\begin{center}
  \includegraphics[scale=0.48]{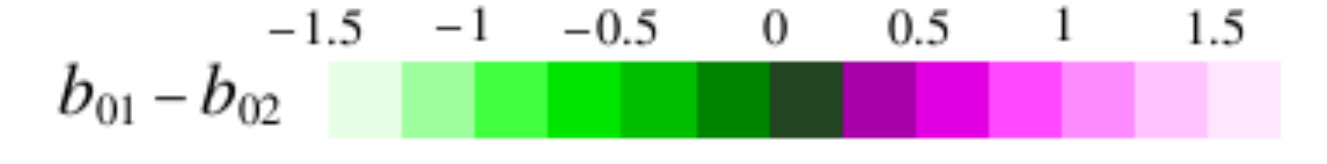}
\end{center}
\caption{The low energy down-type mass insertion parameters
  $(\delta^d_{AB})_{ij}$, $A,B=L,R$, $i=1,2,3$ plotted against their GUT scale
  coefficients, defined in Eqs.~(\ref{Bt},\ref{adt}).  
The blue dashed lines represent our naive numerical expectation according to
the second columns of 
Tables~\ref{down (12) limits}-\ref{down (13) limits}. The red shaded areas
cover the parameter space bounded by the limits shown in the third column of
the corresponding tables, with the red dotted lines denoting the weakest limit
in each case. The absolute values of $\delta^d_{RR}$ are equal in the
(12),(23) and (13) sectors and also
$|(\delta^d_{LR})_{12}|=|(\delta^d_{RL})_{12}|=|(\delta^d_{LR})_{13}|$.  
We therefore only show the bounds stemming from the (12) sector as they are
the strongest ones. All plots have been produced by scanning over the input
parameters shown in Table~\ref{Ranges}.} 
\label{Fig:down  MIs}
\end{figure}

In our model, $(\delta^d_{LL})_{{12}}\sim\lambda^3$ is real at LO, while the
next-to-leading order (NLO) contribution is a linear combination of
$e^{-i\theta^d_2}$ and $\cos(4\theta^d_2+\theta^d_3)$. Therefore,
$\sqrt{\left|\mathrm{Im}\left[(\delta^d_{LL})^2_{{12}_{\text{NLO}}}\right]\right|}$
is proportional to $\sqrt{\sin(\theta^d_2)}\lambda^{7/2}$. 
Setting $\theta^d_2=\pi/2$, i.e. the value preferred by the Jarlskog
invariant~$J^q_{CP}$, we expect
$\mathrm{Im}\left[(\delta^d_{LL})^2_{{12}_{\text{NLO}}}\right]$ to take its
maximum value.  
In Figure~\ref{Fig:down  MIs} we only plot the absolute value of this mass
insertion parameter versus its GUT scale coefficient~$\tilde B_{12}$, see
Eq.~\eqref{Bt}, which can take values between zero and twelve. 
Our naive numerical estimate of $|(\delta^d_{LL})_{12}|$, approximated as
shown in the second column of Table~\ref{down (12) limits}, is of the order of
$10^{-3}$ for $x\approx 1$, visualised by the blue dashed line in
Figure~\ref{Fig:down  MIs}. Since the experimental limits are given as ranges,
we depict them by the red shaded region.

\begin{figure}[H]
\minipage{0.48 \textwidth}
  \includegraphics[width=\linewidth]{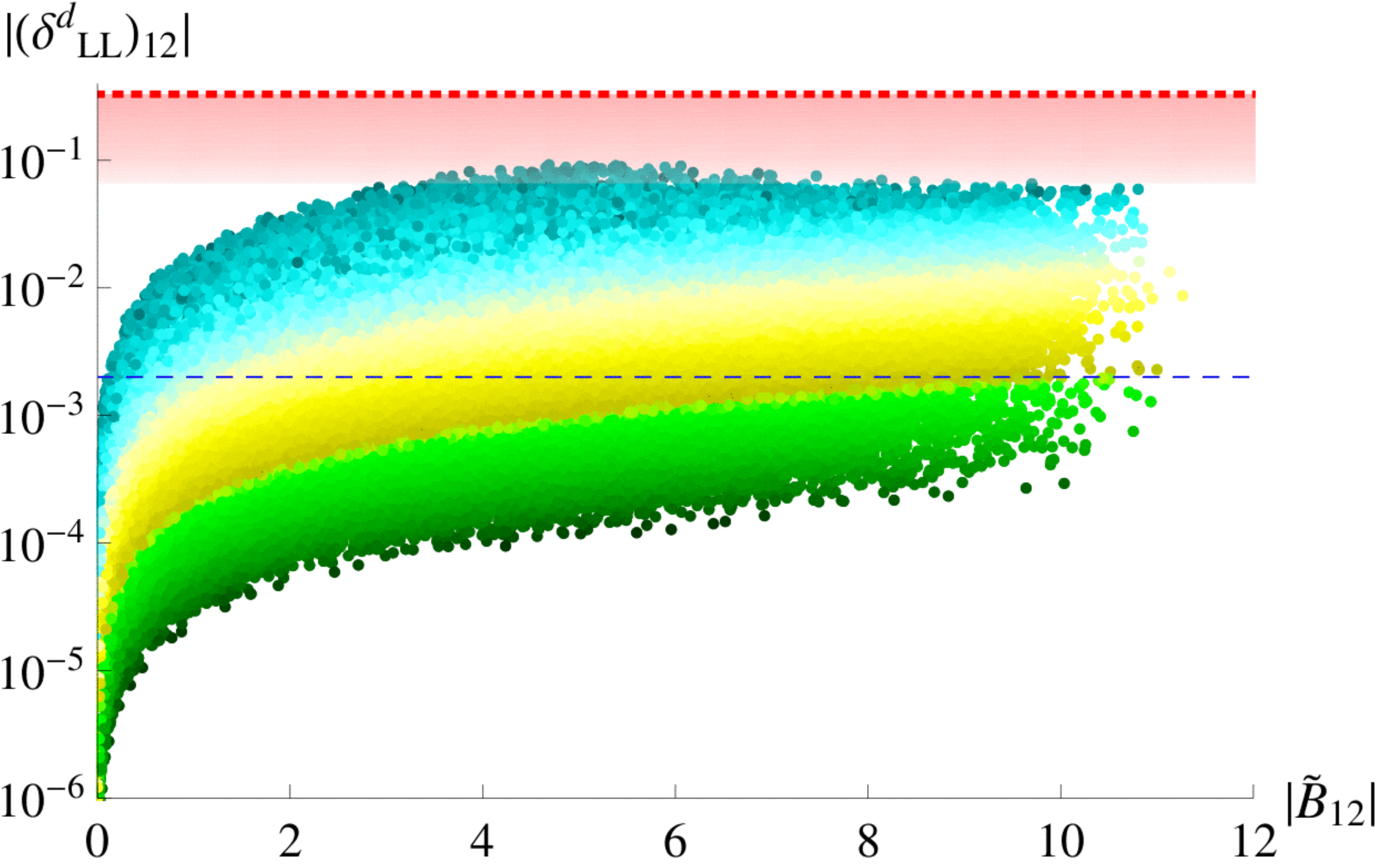}
\endminipage\hfill
\minipage{0.48 \textwidth}
  \includegraphics[width=\linewidth]{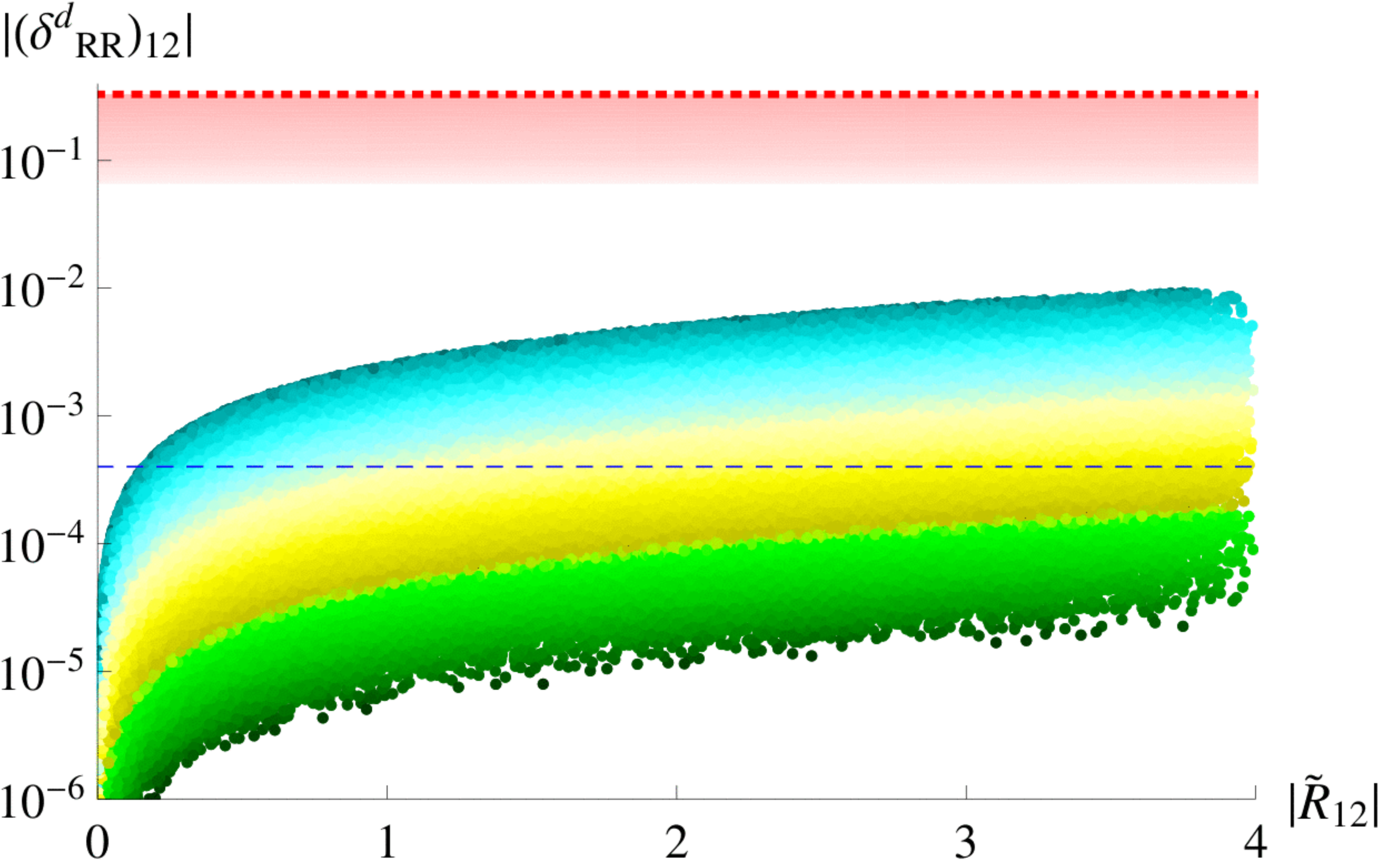}
\endminipage\hfill
\begin{center}
  \includegraphics[scale=0.48]{leg_x.pdf}
\end{center}
\minipage{0.48 \textwidth}
  \includegraphics[width=\linewidth]{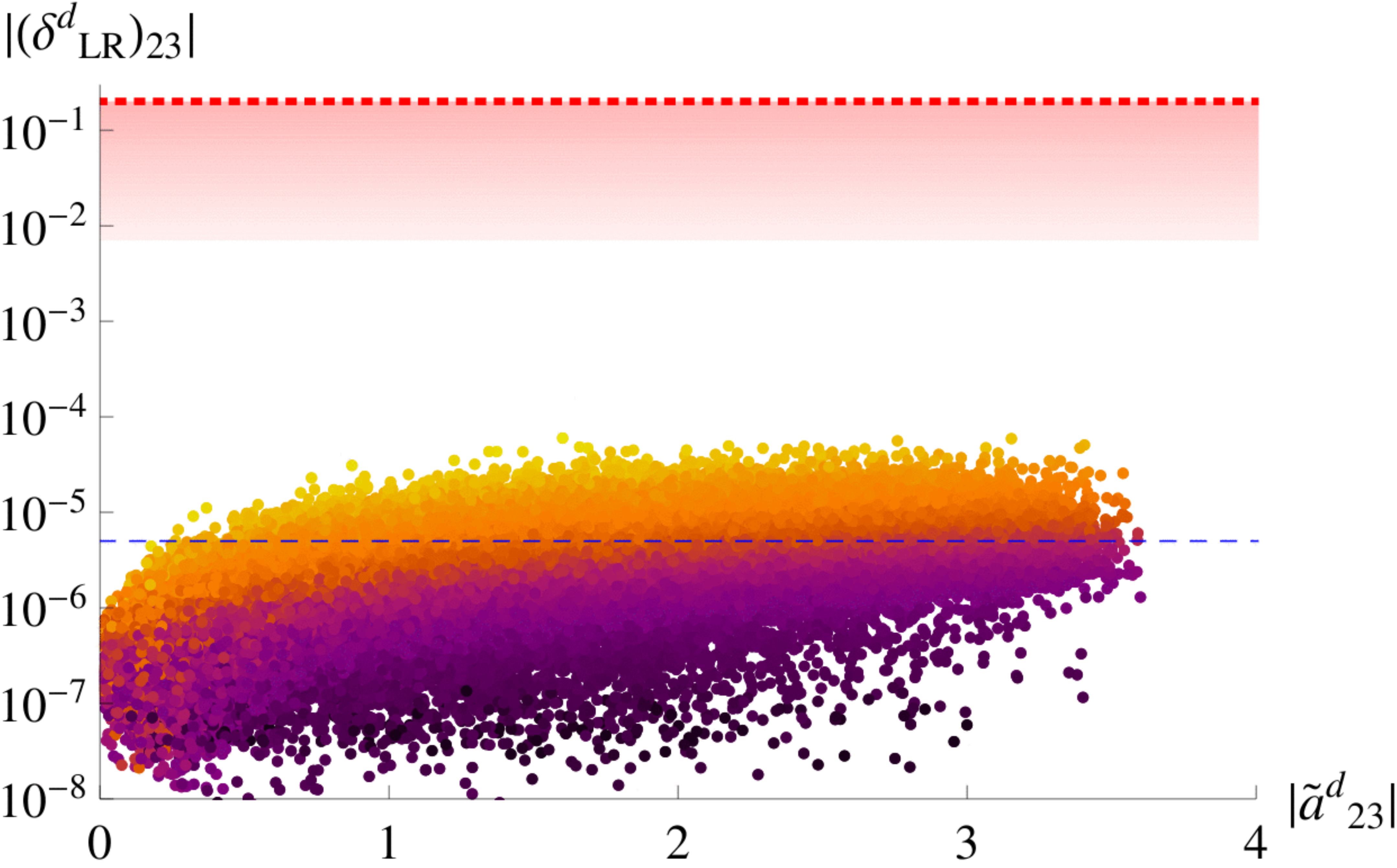}
\endminipage\hfill
\minipage{0.48 \textwidth}
  \includegraphics[width=\linewidth]{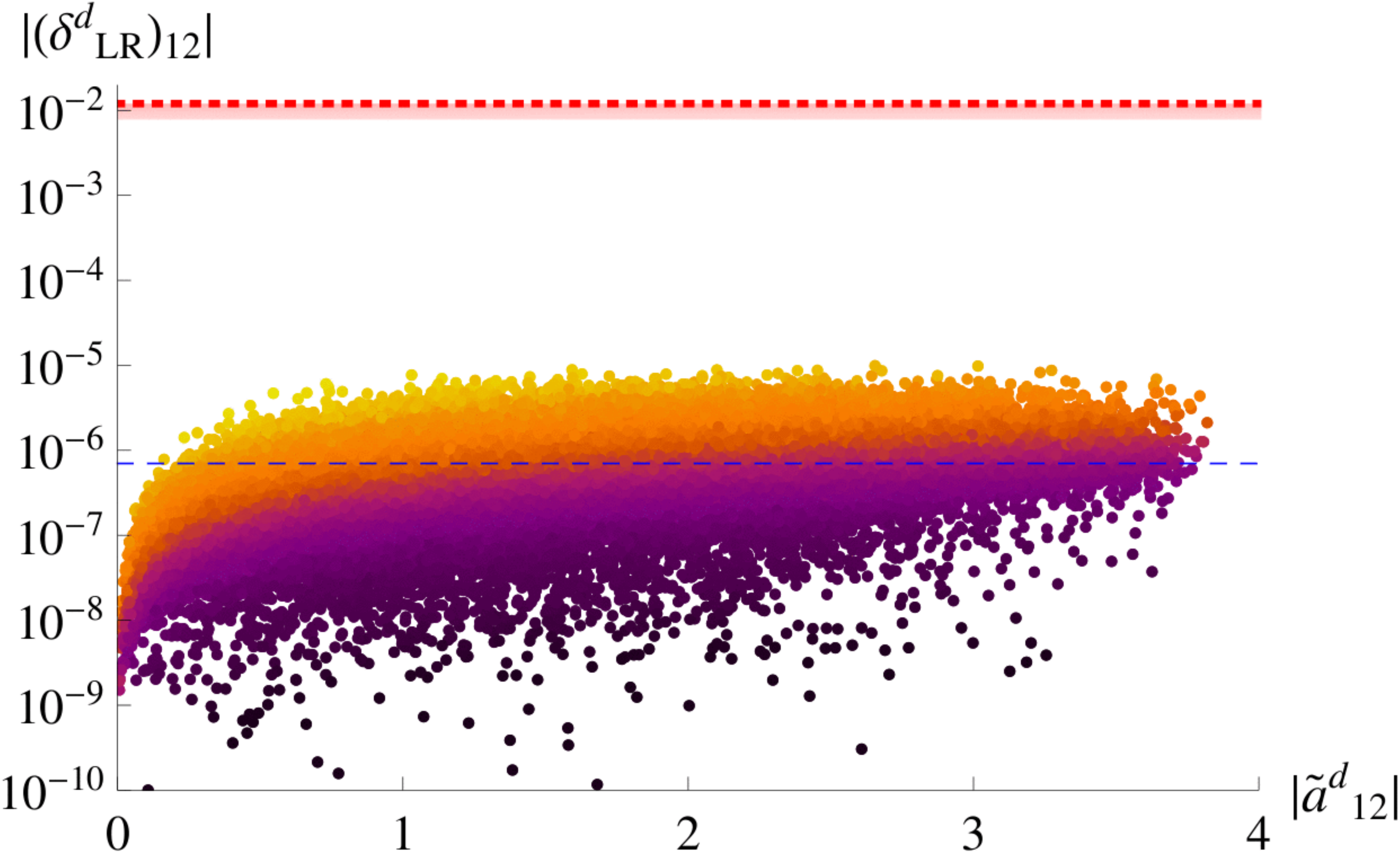}
\endminipage\hfill
\minipage{0.48 \textwidth}
  \includegraphics[width=\linewidth]{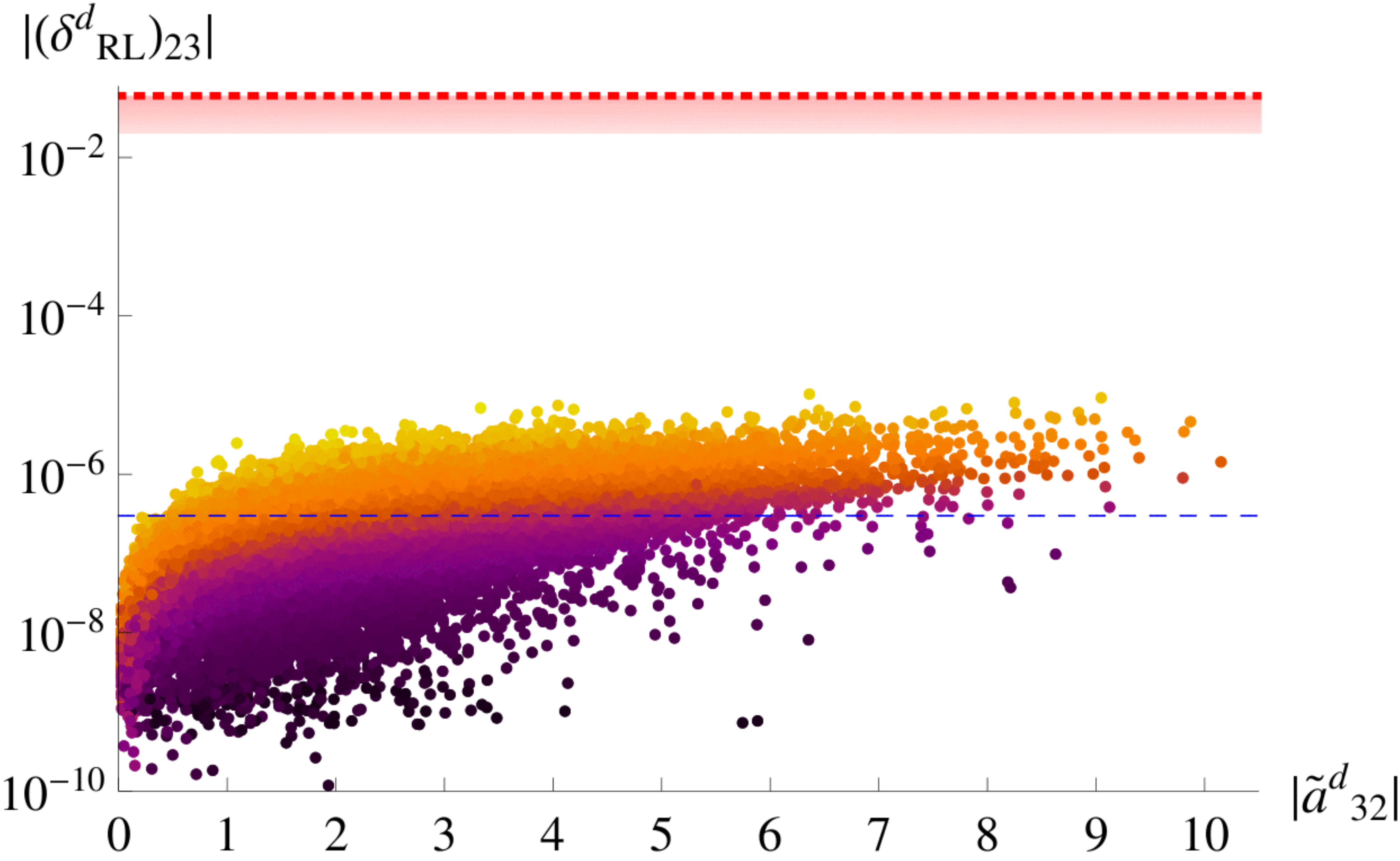}
\endminipage\hfill
\minipage{0.48 \textwidth}
  \includegraphics[width=\linewidth]{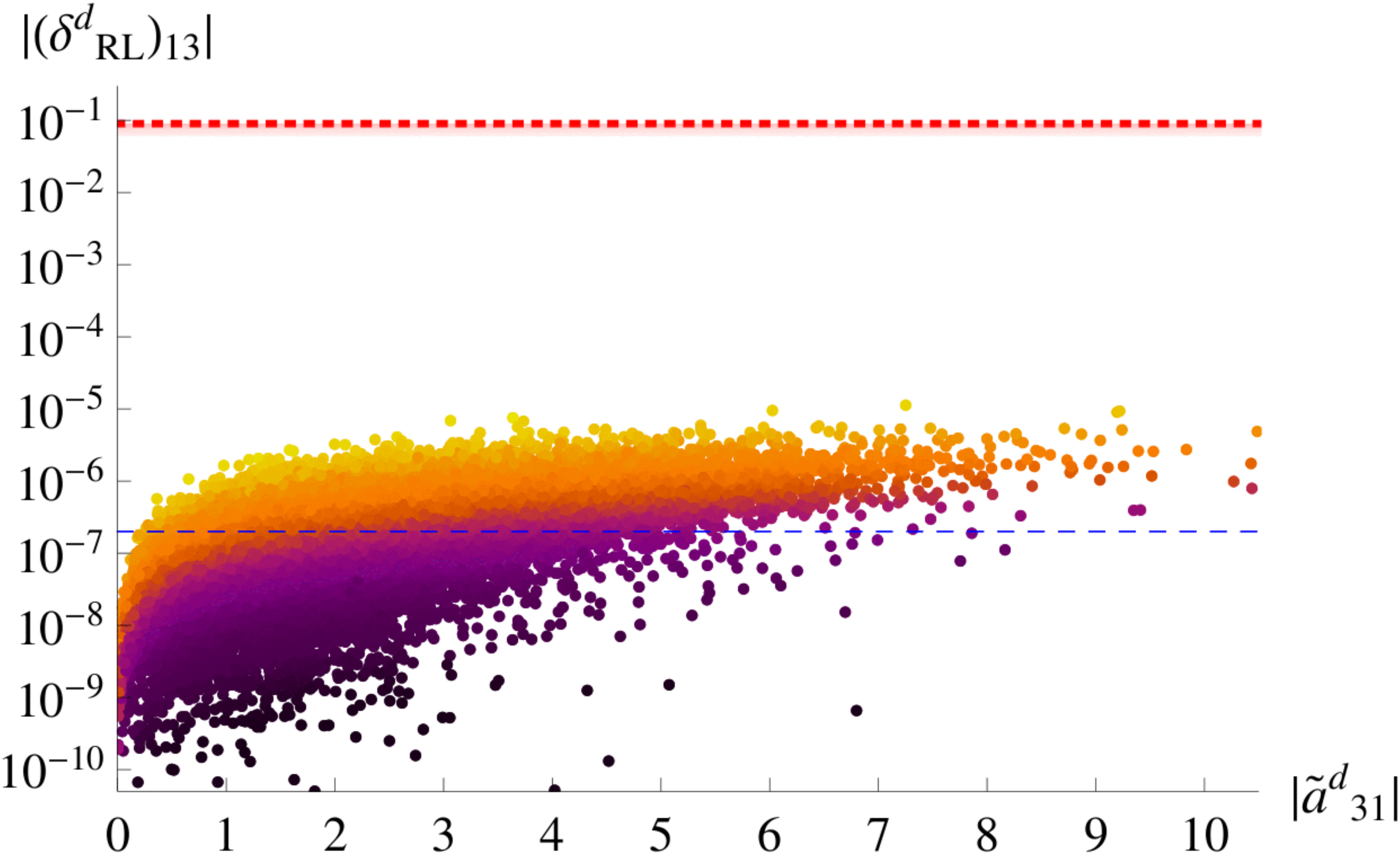}
\endminipage\hfill
\begin{center}
  \includegraphics[scale=0.48]{leg_LR.pdf}
\end{center}
\end{figure}

The parameter $(\delta^d_{RR})_{12}$ is proportional to $e^{i\theta^d_2}$, so that 
$\sqrt{\left|\mathrm{Im}\left[(\delta^d_{RR})^2_{12}\right]\right|}$ vanishes  
for $\theta^d_2=\pi/2$, while the corresponding real part is maximised. 
The RG suppression is again trivial, only depending on $x$, while the GUT scale 
$\delta$ parameter is proportional to $\tilde{R}_{12}=(B_3-K_3)$, see
Eq.~\eqref{Bt}.  
When $B_3=-K_3=2$ and $x\ll 1$, the absolute value of the mass insertion
reaches its maximum of $10^{-2}$, as can be seen in the associated plot in
Figure~\ref{Fig:down MIs}. 
On the other hand, for $B_3=0.5$, $K_3=1$ and $x \gg 1$, it can scale down to
about~$10^{-6}$. 
Note that $|(\delta^d_{RR})_{12}|=|(\delta^d_{RR})_{23}|=|(\delta^d_{RR})_{13}|$, as can be seen in Eqs.~(\ref{dRR13Low},\ref{dRR23Low}).

The mass insertion parameters $(\delta^d_{LR})_{12}=-(\delta^d_{RL})_{12}=(\delta^d_{LR})_{13}$
receive an extra suppression from the factor $\alpha_0\,\upsilon_d/m_0$, for
which we use the value of $5\times 10^{-3}$ in our naive numerical
estimates.
Then, for $x\approx 1$, we expect these $\delta$ parameters to vary around
$7\times 10^{-7}$, see the last two rows of Table~\ref{down (12) limits}. As
can be seen in Figure~\ref{Fig:down  MIs}, our model 
predictions lie well below the  limits. 
Furthermore, if the  Yukawa  and soft trilinear phase structures are aligned, 
the phases within $\tilde{a}^d_{12}$ cancel and $(\delta^d_{LR})_{12}$ becomes
real at the given order in $\lambda$. 

As parts of our parameter space place the down-type mass insertion parameter
$|(\delta^d_{LL})_{12}|$ within a region possibly excluded by Kaon mixing
observables, we study the relevant contributions in
Section~\ref{sec:pheno} in more detail. Due to additional strong constraints
on the product of $LL$ and $RR$ mass insertion parameters, we see that
actually a large fraction of the parameter space is excluded.

\begin{table}[t]
\begin{center}
  \begin{tabular}{ |M{2cm} || M{8.2cm} | M{3.5cm} |}
    \hline
    Parameter & Naive expectation & Exp. bound \\ \hline
    $|(\delta^d_{LL})_{23}|$&$\mathcal{O}\left(\frac{2\eta\,R_q}{1+6.5\,x}\lambda^2|_{b_{01}=b_{02}}\approx 5\times 10^{-3}\right)$ & $[6\times 10^{-2},8\times 10^{-1}]$\\ \hline
    $|(\delta^d_{RR})_{23}|$&$\mathcal{O}\left(\frac{1}{1+6.1\,x}\lambda^4\approx 4\times10^{-4}\right)$  & $[6.3,9.7]\times 10^{-1}$\\ \hline
    $|(\delta^d_{LR})_{23}|$&$\mathcal{O}\left(\frac{\alpha_0\,\upsilon_d}{m_0}\frac{1+\eta\left(\frac{44\,g_U^2}{5}+2a_t\,y_t\right)}{1+6.3\,x}\lambda^4\approx 5\times10^{-6}\right)$ & $[7\times10^{-3},2\times 10^{-1}]$ \\ \hline
  $|(\delta^d_{RL})_{23}|$&$\mathcal{O}\left(\frac{\alpha_0\,\upsilon_d}{m_0}\frac{1+\eta\left(\frac{44\,g_U^2}{5}+2a_t\,y_t+\frac{R_q}{1+6.5\,x}\right)}{1+6.3\,x}\lambda^6\approx 3\times10^{-7}\right)$   & $[2,6]\times 10^{-2}$ \\ \hline
  \end{tabular}
\end{center}
\caption{The naive expectation for the ranges of $(\delta^d_{AB})_{23}$,
  $A,B=L,R$, as extracted from our model (second column), to be compared with
  experimental bounds from~\cite{Catania1} (third column). The full ranges of
  each $\delta$ parameter, produced by scanning over the input parameters as
  shown in Table~\ref{Ranges}, are plotted in Figure~\ref{Fig:down MIs}.}\label{down (23) limits}
\end{table}

The bounds on $(\delta^d_{AB})_{23}$, $A,B=L,R$ are related to $b\to s$
transitions. They are taken from~\cite{Catania1} and were derived by demanding
that the contribution of each individual mass insertion parameter to the
flavour observables $\text{BR}(B\to X_s \gamma)$, 
$\text{BR}(B_s \to \mu^+\mu^-)$ and $\Delta M_{B_s}$ 
does not exceed the current experimental limits.
 The analysis was performed for six representative points of the MSSM
 parameter space which are compatible with LHC SUSY and Higgs searches as
 well as an explanation of the discrepancy of $(g-2)_\mu$ from its SM value in
 terms of one-loop SUSY contributions from charginos and neutralinos. 
We present the extracted bounds in the third column of 
Table~\ref{down (23) limits}, where the intervals arise due to the dependence
on the SUSY spectra. We note that, for simplicity, all $\delta$s were assumed
to be real in~\cite{Catania1}.

At the GUT scale, the parameter $(\delta^d_{LL})_{23}\sim \lambda^2$ is
proportional to $(b_{01}-b_{02})$; it can therefore vanish at that order if
$b_{02}\to b_{01}$. 
In that case, it would still receive a non-zero contribution through the
running, as can be seen in Eq.~\eqref{dLL23Low}, through the factor $R_q$,
defined in Eq.~\eqref{Rq}.  
To see this effect, we expand $(\delta^d_{LL})_{23}$ to first order in the
running parameter $\eta$, defined in Eq.~\eqref{etas}, taking the limit
$b_{02}\to b_{01}$. Then, for $R_q\approx 3y_t^2+1$, $y_t\approx 0.5$ 
and $x\approx1$, we expect the absolute value of $(\delta^d_{LL})_{23}$ to
vary around $5\times10^{-3}$ for $\tilde B_{23} \propto b_{01} - b_{02} \to
0$, as shown by the blue dashed line in Figure~\ref{Fig:down MIs}. 
The spread towards smaller values of $(\delta^d_{LL})_{23}$ as
$\tilde{B}_{23}$ deviates from zero, is mainly due to the parameter space where
$b_{01}-b_{02}$ is negative, thereby partly cancelling the $R_q$ contribution. 
As can be seen in Figure~\ref{Fig:down MIs}, all generated points lie below the
limits of the corresponding (23) sector.

\begin{table}[t]
\begin{center}
  \begin{tabular}{ |M{1.9cm} || M{8cm} | M{3.5cm} |}
    \hline
    Parameter & Naive expectation & Exp. bound \\ \hline
    $|(\delta^d_{LL})_{13}|$&$\mathcal{O}\left(\frac{2\eta\,R_q}{1+6.5\,x}\lambda^4|_{b_{01}=b_{02}}\approx 2\times 10^{-4}\right)$  &\multirow{ 2}{*}{$[1.2,14]\times 10^{-1}$}\\ \cline{1-2}
    $|(\delta^d_{RR})_{13}|$&$\mathcal{O}\left(\frac{1}{1+6.1\,x}\lambda^4\approx
    4\times10^{-4}\right)$ & \\\hline
    $|(\delta^d_{LR})_{13}|$&$\mathcal{O}\left(\frac{\alpha_0\,\upsilon_d}{m_0}\frac{1+\eta\frac{44\,g_U^2}{5}}{1+6.3\,x}\lambda^5\approx 7\times10^{-7}\right)$ & \multirow{4}{*}{$[6,9]\times 10^{-2}$} \\ \cline{1-2}
    $|(\delta^d_{RL})_{13}|$&$\mathcal{O}\left(\frac{\alpha_0\,\upsilon_d}{m_0}\frac{1+\eta\left(\frac{44\,g_U^2}{5}+\frac{R_q}{1+6.5\,x}-y_t^2\right)}{1+6.3\,x}\lambda^6\approx
    2\times 10^{-7}\right)$  & 
\\ \hline
  \end{tabular}
\end{center}
\caption{The naive expectation for the ranges of $(\delta^d_{AB})_{13}$,
  $A,B=L,R$, as extracted from our model (second column), to be compared with
  experimental bounds from~\cite{FlavourPhysicsandGrandUnification} for
  $m_{\tilde{q}}\approx1$~TeV and $(m_{\tilde g}/m_{\tilde q})^2\in [0.25,4]$
  (third column). The full ranges of the $\delta$s as produced in our scan are
  shown in Figure \ref{Fig:down MIs}.}\label{down (13) limits}
\end{table}

The experimental bounds for $(\delta^d_{AB})_{13}$ are taken
from~\cite{FlavourPhysicsandGrandUnification}, where they were extracted
from $B_d$ mixing related observables and given in terms of 
${|\mathrm{Re}\left[\delta^d_{AB}\right]|}$ and
${|\mathrm{Im}\left[\delta^d_{AB}\right]|}$. Their orders of magnitude are 
at most of the same order as $|\delta^d_{AB}|$, and for
$m_{\tilde{q}}\approx1$~TeV and $(m_{\tilde g}/m_{\tilde q})^2\in [0.25,4]$ 
they are summarised in the third column of Table~\ref{down (13) limits}. 
The limits for the $RR$ and $RL$ type $\delta$s are equal to the $LL$ and $LR$
type ones, respectively, as the gluino contribution to the box diagram for
meson mixing is symmetric under $L \leftrightarrow R$.

In our model, we expect $|(\delta^d_{LL})_{13}|$ to have a similar behaviour
as $|(\delta^d_{LL})_{23}|$ but with an extra suppression of $\lambda^2$.
Furthermore, $|(\delta^d_{LR})_{23}|$ mimics 
$|(\delta^d_{LR})_{12}|=|(\delta^d_{RL})_{12}|=|(\delta^d_{LR})_{13}|$
with an extra enhancement factor of $\lambda^{-1}$. 
The $RL$ parameters (13) and (23) sectors are of the same order in $\lambda$
and should therefore have a similar numerical range. All (13)  
sector mass insertion parameters $\delta^d_{AB}$ lie below the limits set by
$B_d$ mixing, as can be seen in Figure~\ref{Fig:down MIs}.

\subsubsection{Charged lepton sector}\label{Numerics:Charged-lepton sector}

\begin{table}[t]
\begin{center}
  \begin{tabular}{ |M{2cm} || M{8cm}  | M{3cm} |}
    \hline
    Parameter & Naive expectation & Exp. bound \\ \hline
    $|(\delta^e_{LL})_{12}|$&\multirow{ 2}{*}{$\mathcal{O}\left(\frac{2\,R_l\eta_N}{1+0.5\,x}\lambda^4|_{B_3=K_3}\approx2\times10^{-4}
\right)$}   & $[1.5,60]\times 10^{-5}$\\ \cline{3-3}\cline{1-1}
    $|(\delta^e_{LL})_{23,13}|$&    & $[0.7,35]\times 10^{-2}$\\ \hline
    $|(\delta^e_{RR})_{12}|$&$\mathcal{O}\left(\frac{\lambda^3}{1+0.15\,x}\approx10^{-2} \right)$  & $[0.35,25]\times 10^{-3}$\\\hline
    $|(\delta^e_{RR})_{23}|$&$\mathcal{O}\left(\frac{\lambda^2}{1+0.15\,x}\approx 4\times 10^{-2} \right)$  & \multirow{ 2}{*}{$[2,10]\times 10^{-1}$}\\\cline{1-2}
    $|(\delta^e_{RR})_{13}|$&$\mathcal{O}\left(\frac{\lambda^4}{1+0.15\,x}\approx2\times{10^{-3}} \right)$  & \\\hline
    $|(\delta^e_{LR(RL)})_{12}|$&\multirow{ 2}{*}{$\mathcal{O}\Big(\frac{\alpha_0\,\upsilon_d}{m_0}\frac{1+\eta\frac{24g_U^2}{5}+\eta_N\left(\frac{R_l}{1+0.5\,x}-y_D^2\right)}{1+0.3\,x}\lambda^5\approx 3\times 10^{-6}\Big)$}   & $[1.2,22]\times 10^{-6}$ \\  \cline{3-3}\cline{1-1}
    $|(\delta^e_{RL})_{13}|$&   & \multirow{ 4}{*}{$[1,22]\times 10^{-2}$} \\   \cline{1-2}
    $|(\delta^e_{LR})_{13}|$&\multirow{ 2}{*}{$\mathcal{O}\Big(\frac{\alpha_0\,\upsilon_d}{m_0}\frac{1+\eta\frac{24g_U^2}{5}+\eta_N\left(\frac{R_l}{1+0.5\,x}-y_D^2\right)}{1+0.3\,x}\lambda^6\approx 8\times 10^{-7}\Big)$} &   \\  \cline{1-1}
    $|(\delta^e_{LR})_{23}|$& &  \\ \cline{1-2}
    $|(\delta^e_{RL})_{23}|$&$\mathcal{O}\Big(\frac{\alpha_0\,\upsilon_d}{m_0}\frac{1+\eta\frac{24g_U^2}{5}+\eta_N\left(\frac{R_l}{1+0.5\,x}-y_D^2\right)}{1+0.3\,x}\lambda^4\approx 10^{-5}\Big)$   &  \\ \hline 
  \end{tabular}
\end{center}
\caption{The naive expectation for the ranges of $(\delta^e_{AB})_{ij}$,
  $A,B=L,R$, as extracted from our model (second column), to be compared with
  experimental bounds from~\cite{Catania2} (third column). The full ranges of
  the $\delta$ parameters produced in our scan are shown in
  Figure~\ref{Fig:electron MIs}.} 
\label{electron limits}
\end{table}

The bounds on the mass insertion parameters $(\delta^e_{AB})_{ij}$, $A,B=L,R$, of
the charged lepton sector are taken from~\cite{Catania2}. They were derived by
studying radiative, leptonic and semileptonic LFV decays as well as $\mu\to
e$ conversion in heavy nuclei. The analysis was performed for six
representative points in the MSSM parameter space, which are in agreement with
LHC SUSY and  Higgs searches as well as data on $(g-2)_\mu$. Moreover, four
additional, more general two-dimensional scenarios, characterised by universal
squark and slepton mass scales, were considered in~\cite{Catania2}.
The derived limits vary within an order of magnitude in all cases and are
summarised in the third column of Table~\ref{electron limits}.
We note that all $\delta$s were assumed to be real in~\cite{Catania2} for simplicity.

\begin{figure}[H]
\minipage{0.48 \textwidth}
  \includegraphics[width=\linewidth]{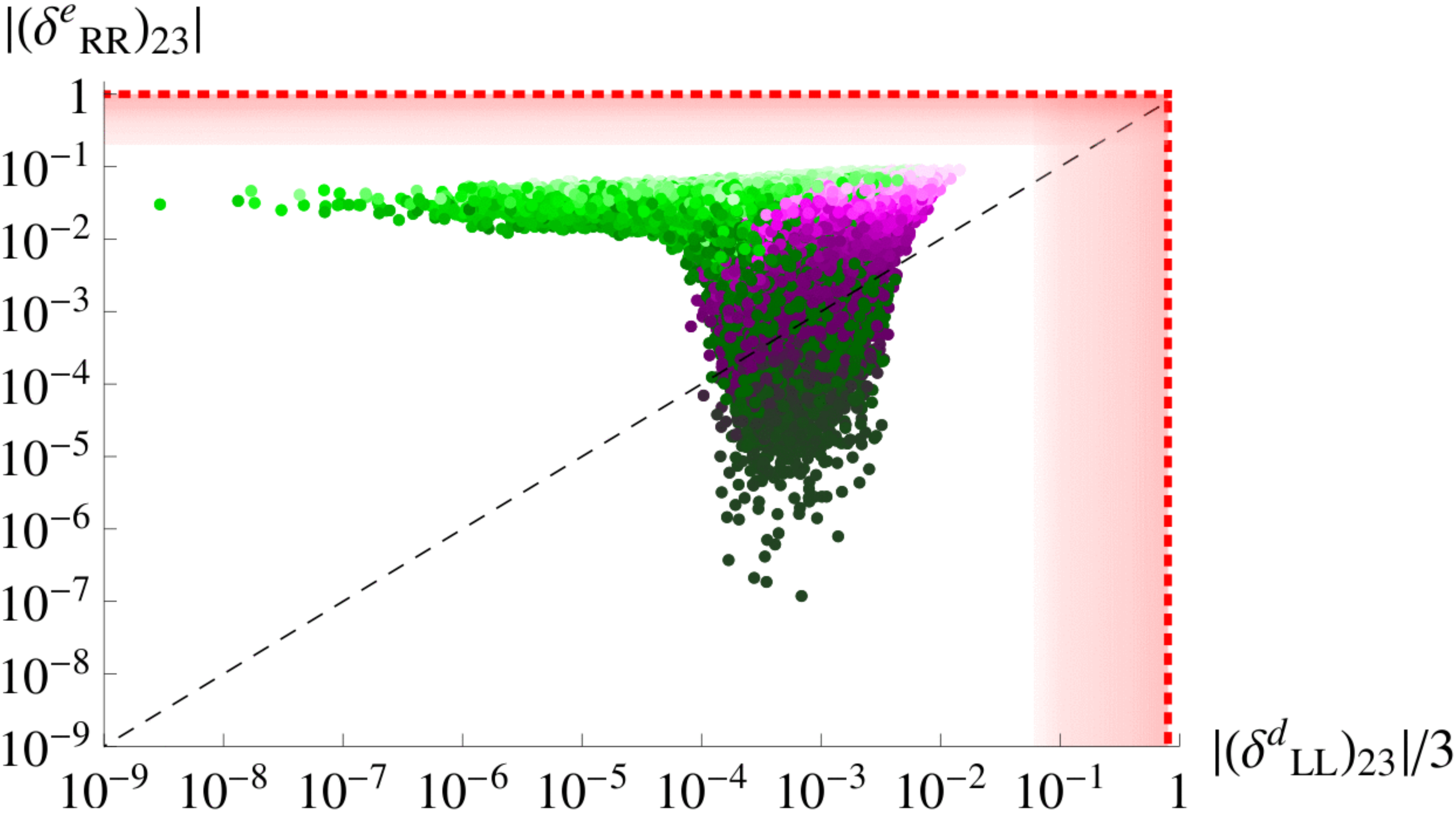}
\endminipage\hfill
\minipage{0.48 \textwidth}
  \includegraphics[width=\linewidth]{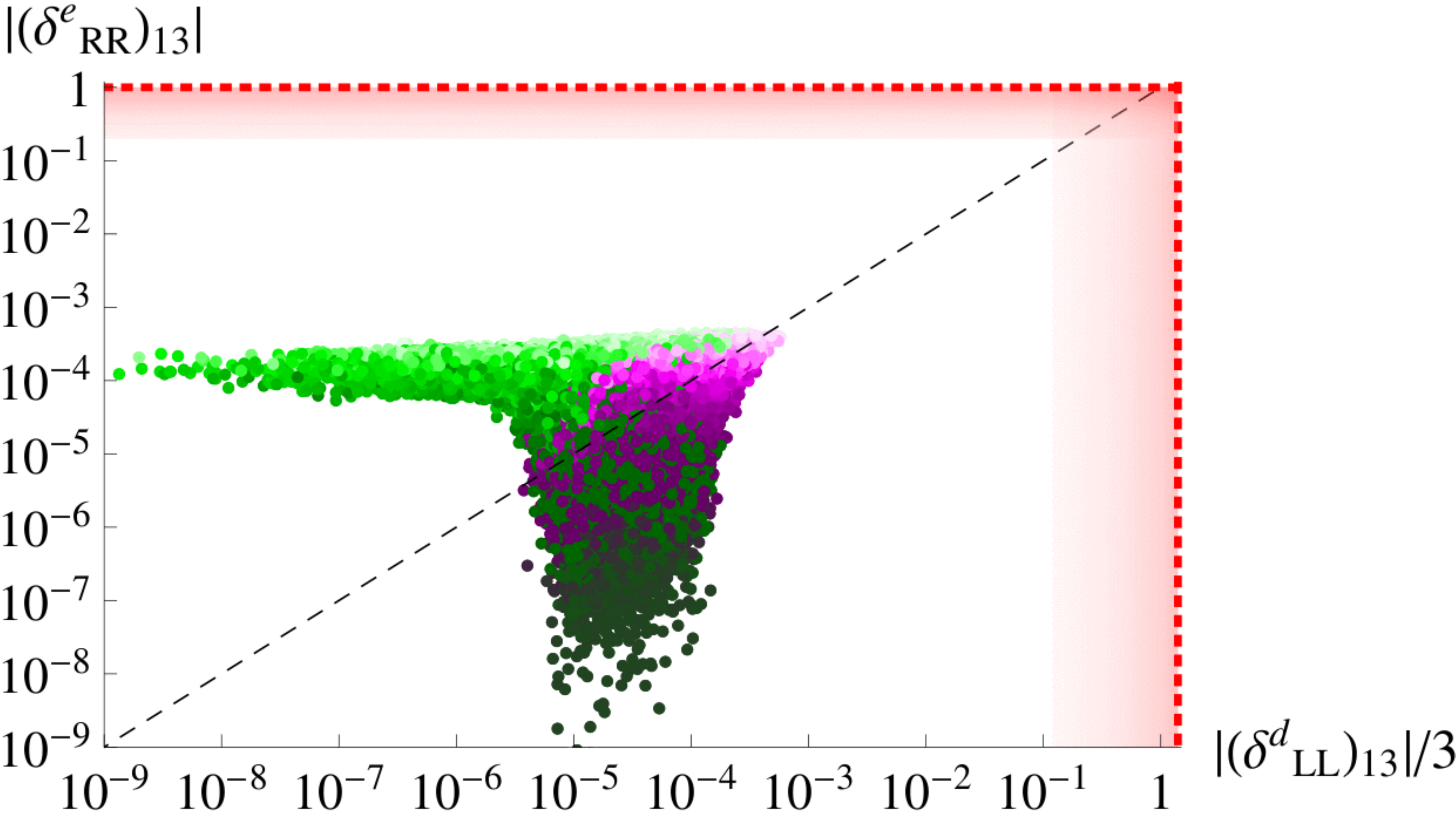}
\endminipage\hfill
\begin{center}
  \includegraphics[scale=0.48]{leg_b01b02.pdf}
\end{center}
\caption{The low energy lepton mass insertion parameters
  $(\delta^e_{AB})_{ij}$, $A,B=L,R$, plotted against the down-type 
  $\delta$s to which they are related via the $SU(5)$ framework.
  The dashed lines represent their GUT scale relations, while the
  red shaded areas denote experimental limits on the parameter space 
  according to the third column of Tables~\ref{down (12) limits}-\ref{electron
    limits}. Scanning over the input parameters within the ranges shown in
  Table~\ref{Ranges}, we observe that in particular $|(\delta^e_{LL})_{12}|$
   exceeds its limit for much of our parameter space. 
 Note that
 $|(\delta^e_{LL})_{12}|=|(\delta^e_{LL})_{23}|=|(\delta^e_{LL})_{13}|$ 
and $|(\delta^e_{RL})_{12}|=|(\delta^e_{LR})_{12}|=|(\delta^e_{RL})_{13}|$.}
\label{Fig:electron MIs}
\end{figure}

At the GUT scale, the mass insertion parameter $(\delta^e_{LL})_{12}\sim
\lambda^4$ is proportional to $\tilde{R}_{12}=B_3-K_3$. Its absolute value is
equal to $|(\delta^d_{RR})_{12}|$ due to the $SU(5)$ framework. However, the
parameter of the lepton sector, given in Eq.~\eqref{eLL12Low},
receives large RG corrections which encode seesaw effects. 
At the low energy scale, it is non-zero even for $B_3=K_3$, due to the term
proportional to the small parameter $\eta_N$ which is defined in
Eq.~\eqref{etas} and originates from the running between the GUT scale and the
scale of the right-handed neutrinos. 
In the second column of Table~\ref{electron limits}, we estimate this effect
by considering $B_3=K_3$.
We then expand to first order in $\eta_N$ and consider $R_l\approx R_l'$,
where $R_l$ and $R_l'$ are defined in Eqs.~(\ref{Rl},\ref{Rlp}). 
For $x\approx 1$, $R_l\approx 3 y_D^2+1$ and $y_D\approx 0.5$, we expect the
low energy $|(\delta^e_{LL})_{12}|$ to vary around $2\times 10^{-4}$. 
However, the non-trivial expression of $\tilde{E}_{12}$,
cf. Eqs.~(\ref{eLL12Low},\ref{E12}), creates a spread of about two orders of
magnitude around this value. 
As $|\tilde{R}_{12}|$ increases, the mass insertion parameter lies above the
limits given in Table~\ref{electron limits}. As can be seen from
Figure~\ref{Fig:electron MIs}, the non-observation of $\mu\to e\gamma$  places
stronger constraints on the down-type quark $\delta$s than the direct bounds
from the quark sector. Analogous to the down-type $RR$ parameters, the
absolute values of the (12), (23) and (13) lepton $LL$ parameters are
identical, see Eqs.~(\ref{eLL12Low},\ref{eLL12Low-13}).

\begin{figure}[H]
\minipage{0.48 \textwidth}
  \includegraphics[width=\linewidth]{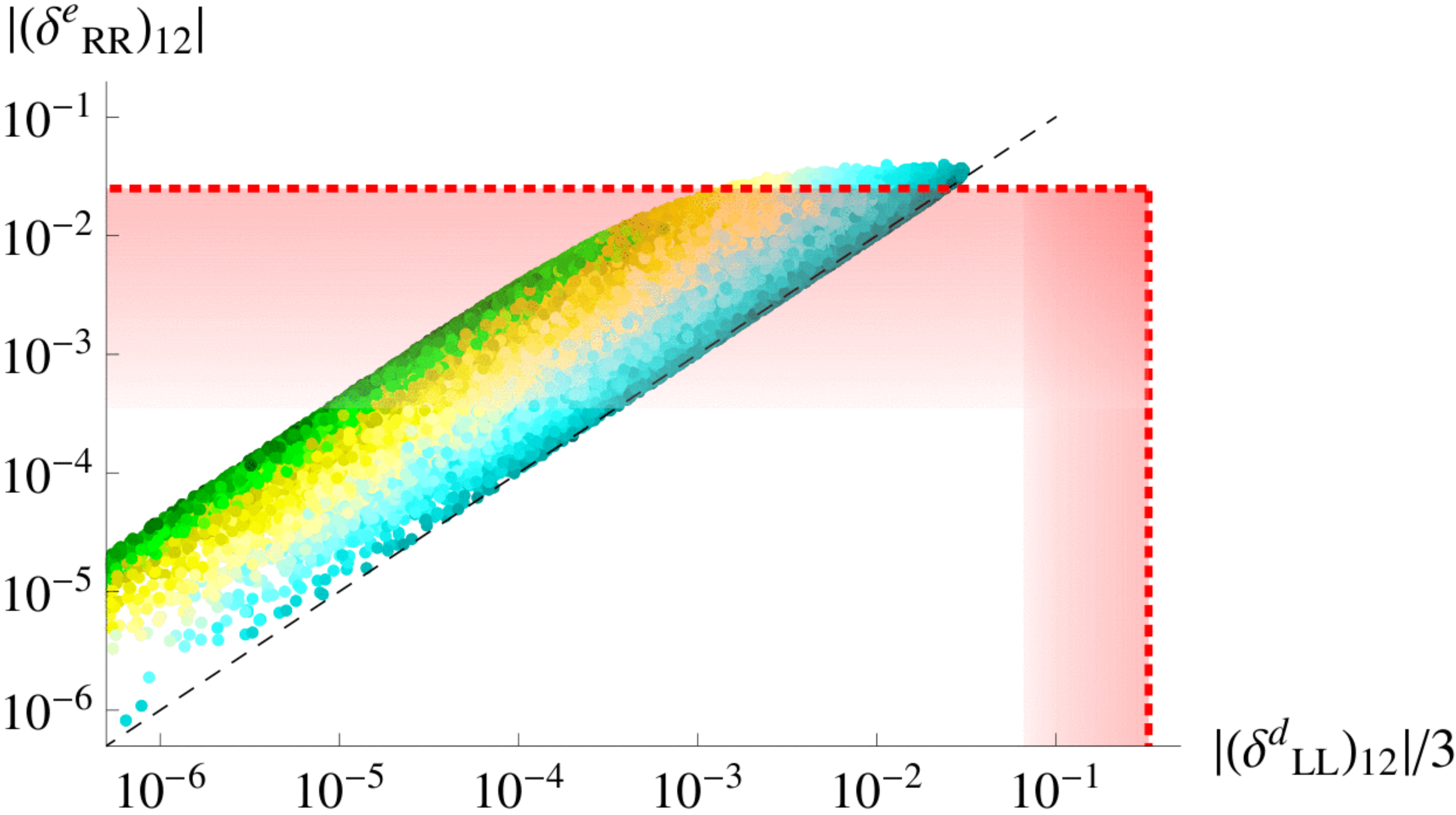}
\endminipage\hfill
\minipage{0.48 \textwidth}
  \includegraphics[width=\linewidth]{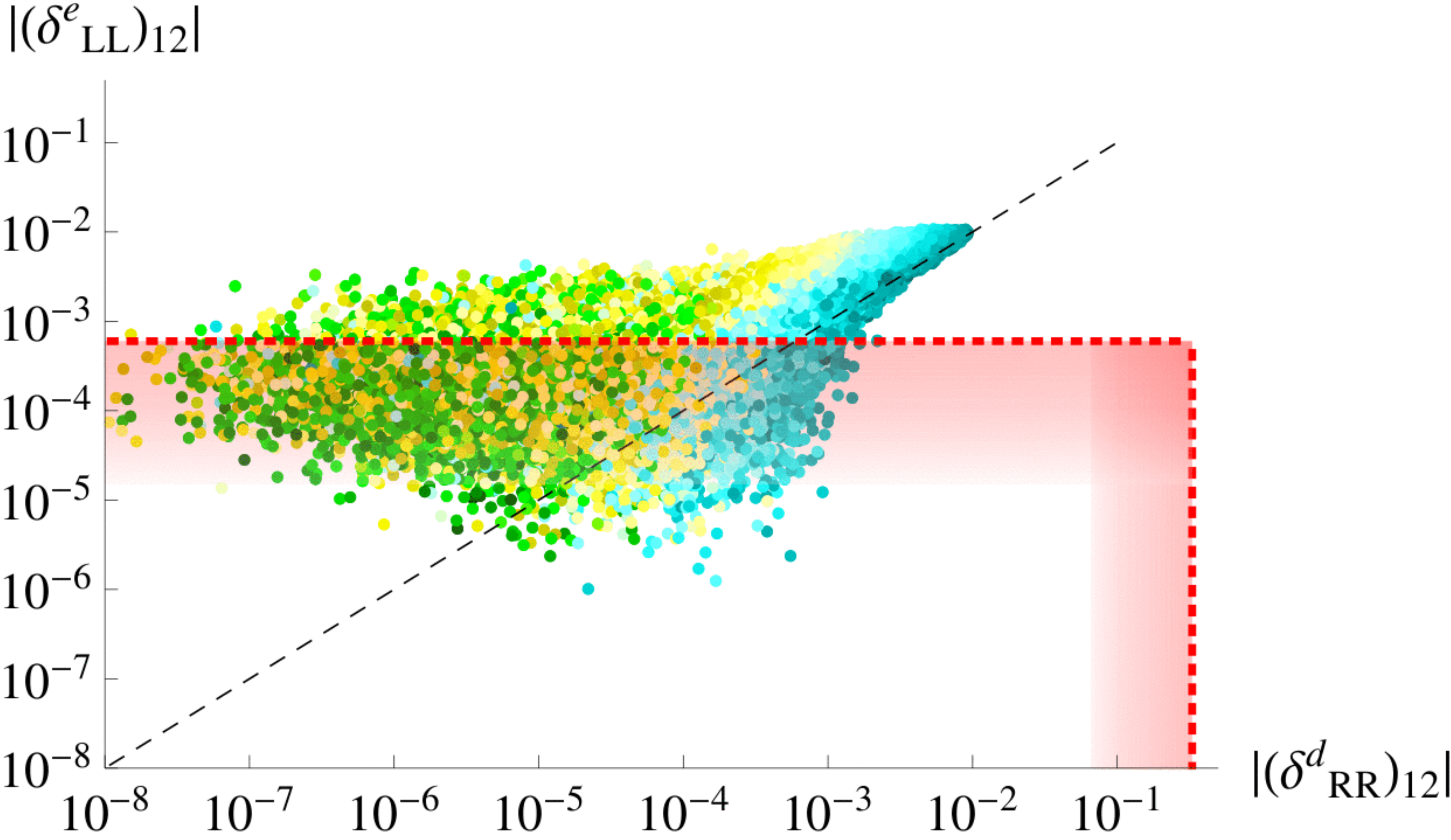}
\endminipage\hfill
\begin{center}
  \includegraphics[scale=0.48]{leg_x.pdf}
\end{center}
\minipage{0.48 \textwidth}
  \includegraphics[width=\linewidth]{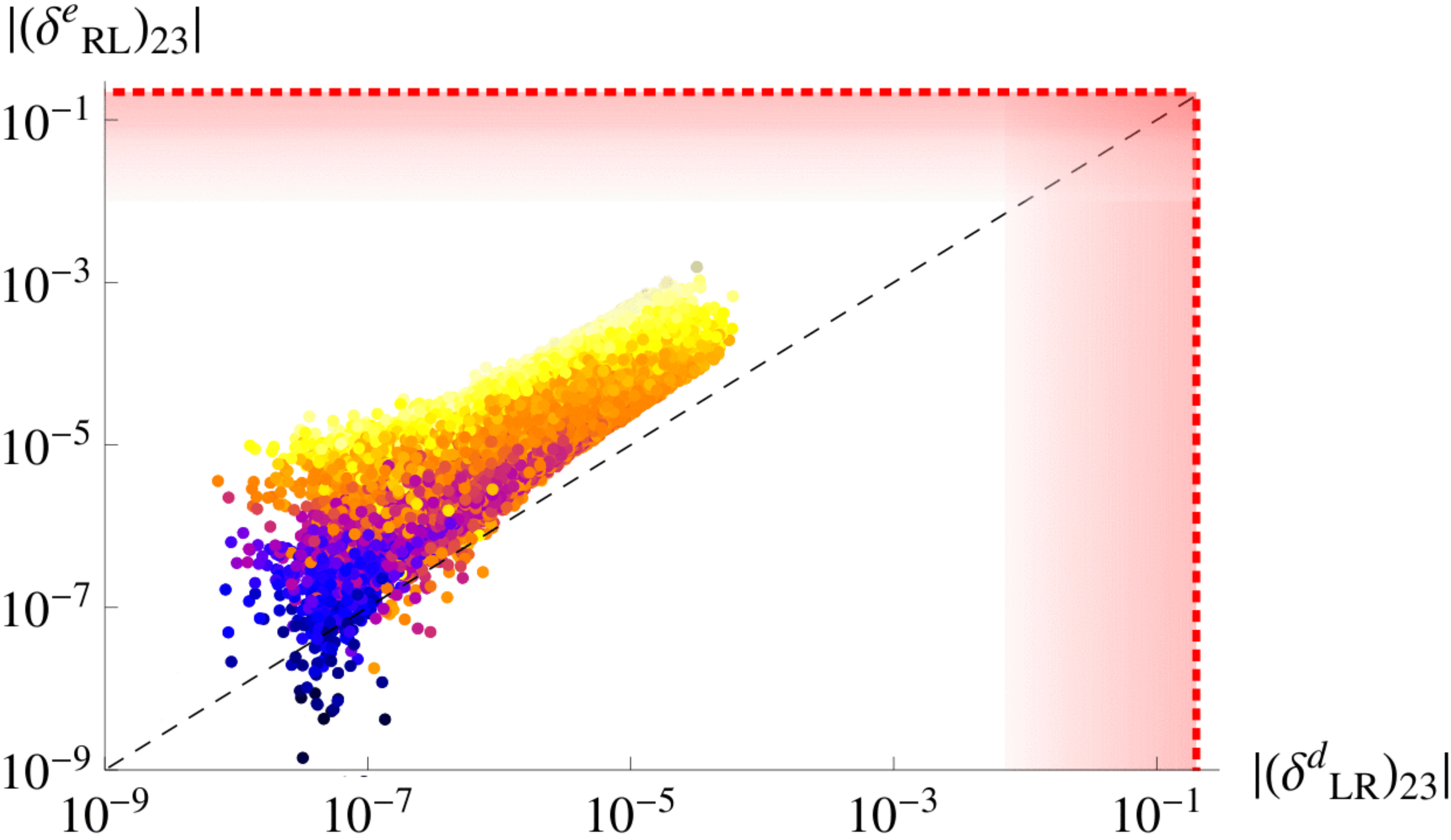}
\endminipage\hfill
\minipage{0.48 \textwidth}
  \includegraphics[width=\linewidth]{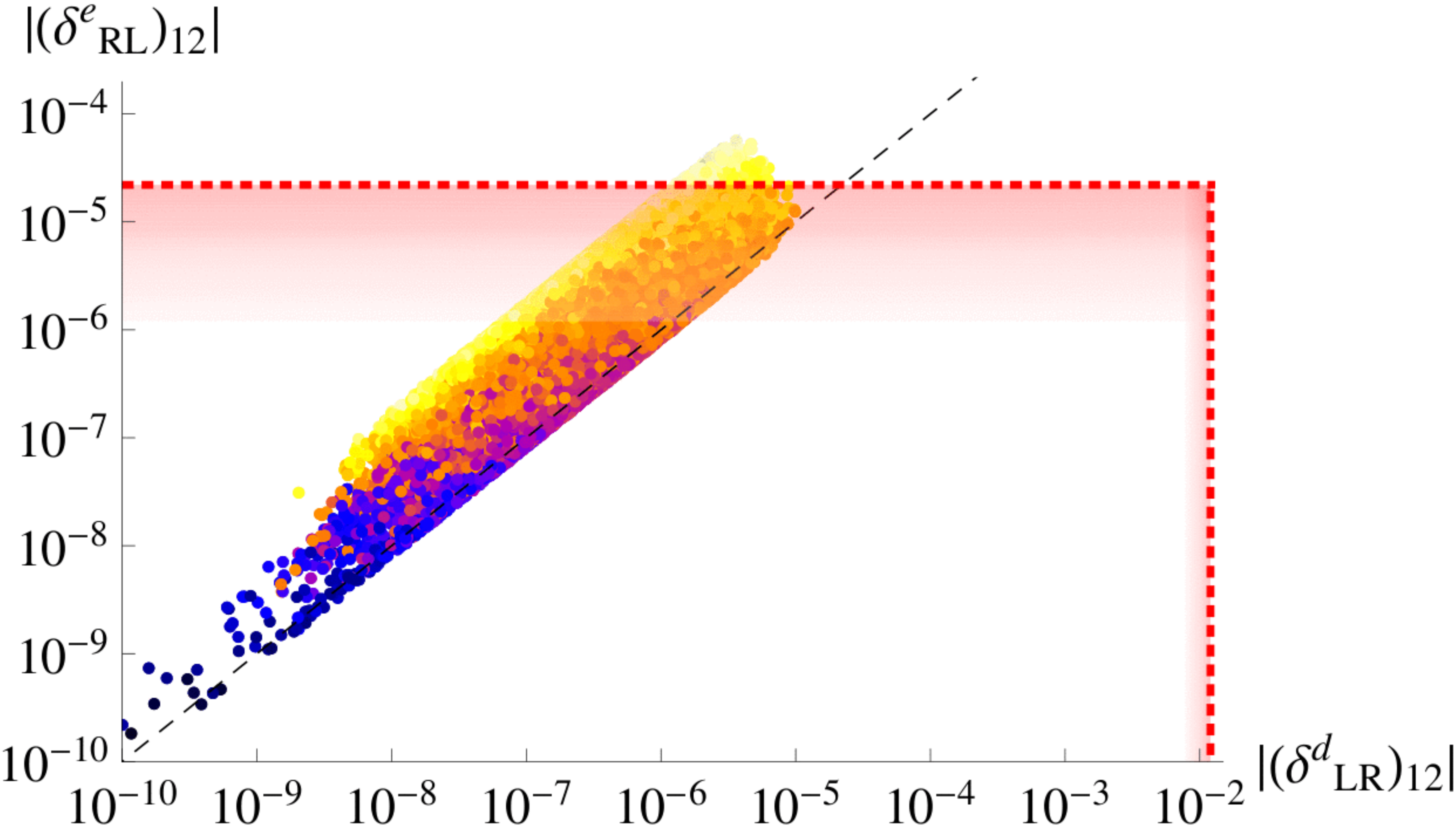}
\endminipage\hfill
\minipage{0.48 \textwidth}
  \includegraphics[width=\linewidth]{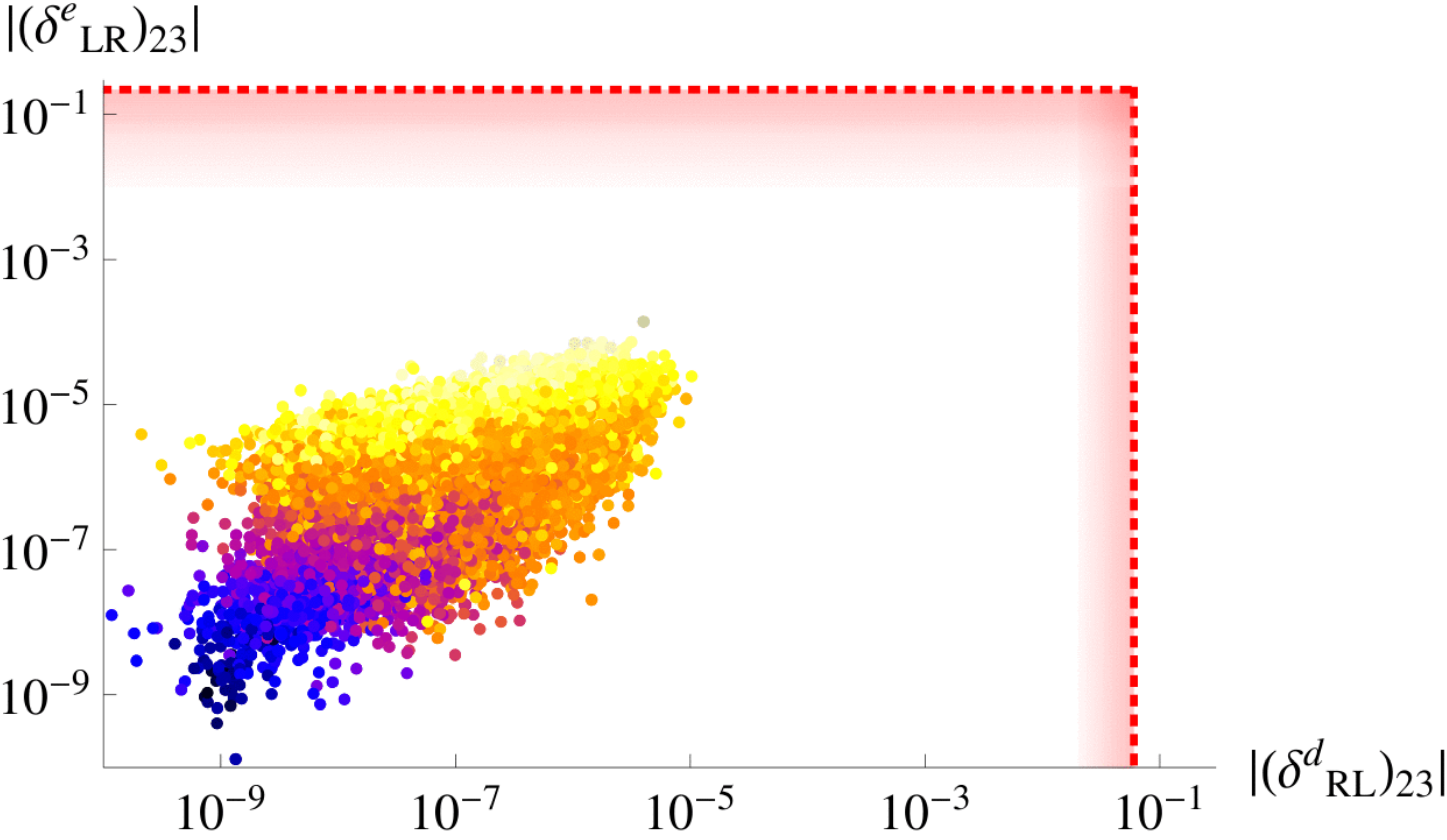}
\endminipage\hfill
\minipage{0.48 \textwidth}
  \includegraphics[width=\linewidth]{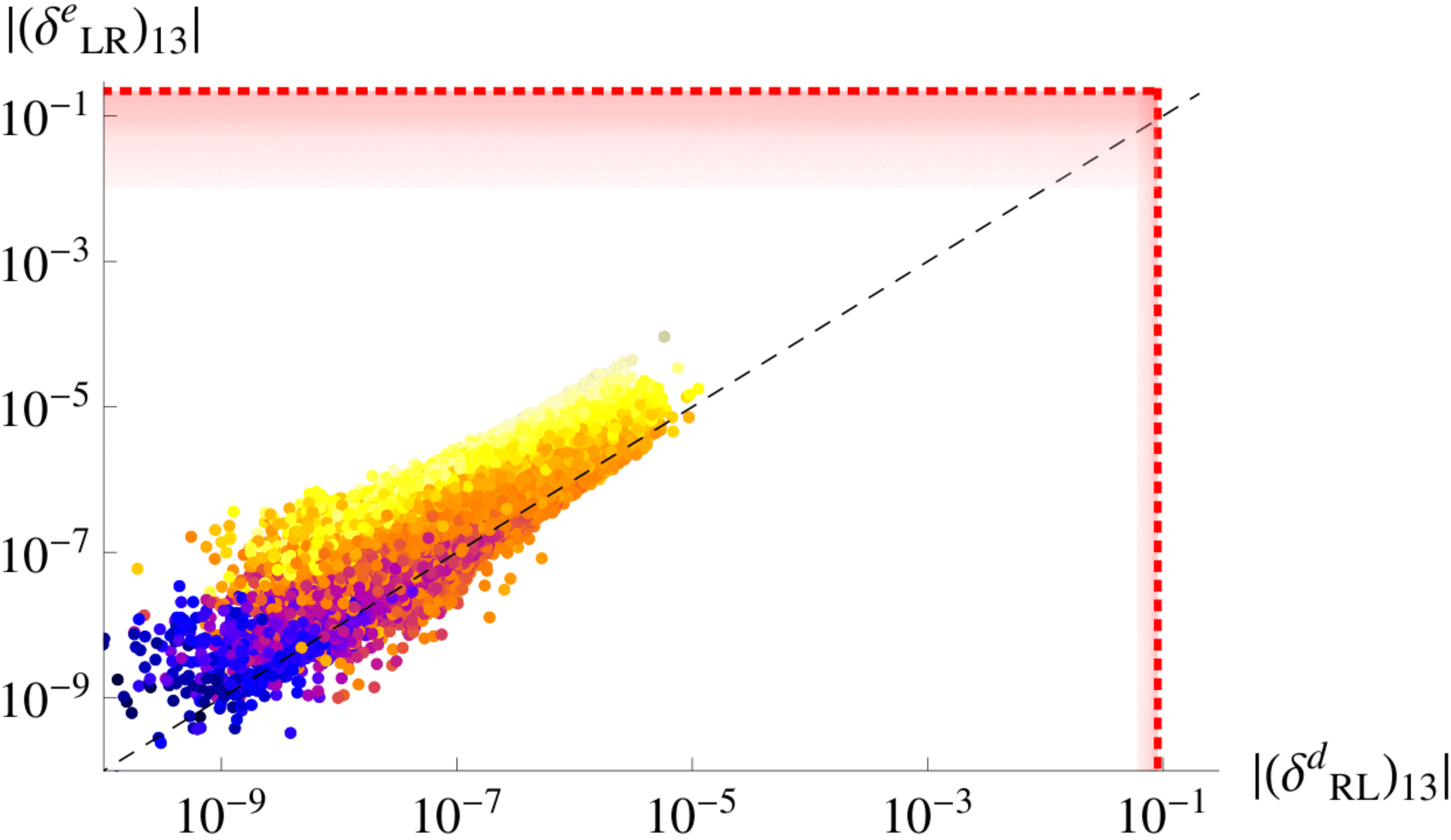}
\endminipage\hfill
\begin{center}
  \includegraphics[scale=0.44]{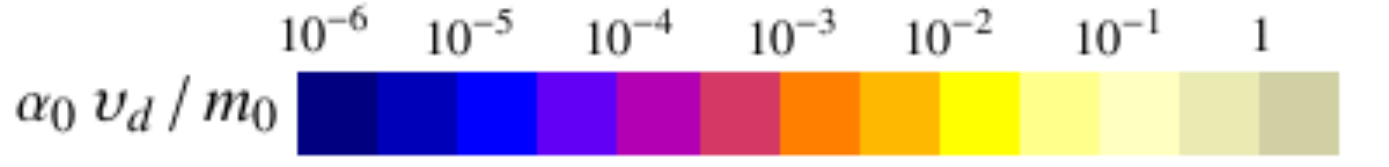}
\end{center}
\end{figure}

Similarly, at the GUT scale, the absolute values of the $RR$ parameters in the
lepton sector are equal to the $LL$ ones of the down-type sector times the
Georgi-Jarlskog factor of $1/3$. For the (12) $\delta$s, the RG running
effects are trivial, consisting only of a suppression through $x$, which is
milder in the lepton sector where the numerical prefactor of $x$ is $0.15$, as
compared to a factor of  $6.5$ in the quark one. 
For the (13) and (23) parameters, the non-trivial
running effects in the quark sector are obvious in Figure~\ref{Fig:electron
  MIs}, where we see that even though $|(\delta^d_{LL})_{23,13}|$ can get very
small for negative $b_{01}-b_{02}$, $|(\delta^e_{RR})_{23,13}|$ can only
receive such small values when $b_{01}\to b_{02}$, see
e.g. Eqs.~(\ref{dLL13Low},\ref{eRR13Low}).

Finally, the variation of the $LR$ parameters can be understood in an
analogous way to the one described in the quark
sector. $|(\delta^e_{LR})_{{ij}_{\text{GUT}}}|=|(\delta^d_{RL})_{{ij}_{\text{GUT}}}|$,
with the exception of the (23) parameters which are not equal due to a term
which involves a $H_{\bar{45}}$, thereby receiving an extra factor of 9 for
the leptons, see Eqs.~(\ref{dLR32Low},\ref{eLR23Low}) together with
Eq.~\eqref{adt}. As in the down-type sector,
$|(\delta^e_{RL})_{12}|=|(\delta^e_{LR})_{12}|=|(\delta^e_{RL})_{13}|$ and we 
only show the (12) parameter in Figure~\ref{Fig:electron MIs} which features the
strongest experimental constraint.




\section{\label{sec:pheno}Phenomenological implications}
\cleqn

In the preceding section, we found that parts of the parameter space spanned
by the (12)~mass insertion parameters of the down-type and charged lepton
sector are excluded due to experimental limits set by $\mu \to e\gamma$
and  Kaon mixing observables. 
The corresponding bounds are available in the literature and their derivation
is highly dependent on the assumed SUSY mass spectra. Possible interference
effects between contributions from multiple $\delta$ parameters to a given
observable can additionally have significant effects. These are usually
ignored when setting ``model independent'' limits on mass insertion parameters. 

In this section, we therefore investigate the phenomenological implications of the
deviations of our model from MFV. In particular, we focus on the predictions
 for $BR(\mu \to e\gamma)$ and $\epsilon_K$. We also scrutinise
whether the phase structure of our model can survive the strong limits set by
electric dipole moments.
Since the analysis in~\cite{Catania1}, which provides the limits on
$(\delta^d_{AB})_{23}$, assumes real parameters throughout, we also study
how our model contributes to the time-dependent CP asymmetry associated with
the decay $B_s\to J/\psi\phi$.  For completeness, we check that the limits set
by the decay $B_d\to J/\psi K_S$ and the mass differences $\Delta M_{B_{s,d}}$
are satisfied. 
Finally, we also consider the branching ratios of $b\to s\gamma$ and
$B_{s,d}\to\mu^+\mu^-$

Adopting the leading logarithmic approximation, the low energy 
gaugino masses~\cite{SPrimer}
\begin{eqnarray}
 M_i&=&\frac{g^2_i}{g^2_U}M_{1/2}\approx \frac{M_{1/2}}{1+2\,\eta
  \,g_U^2\,\beta_i}, \qquad  i=1,2,3,
\end{eqnarray}
with $\beta_1={33}/{5}$, $\beta_2=1$ and $\beta_3=-3$, are given by
\begin{eqnarray}
M_1\approx 0.43\,M_{1/2},\quad ~~M_2\approx 0.83\,M_{1/2},\quad ~~M_{3}\approx 2.53\,M_{1/2}.\label{gauginos}
\end{eqnarray}


\subsection {Electron EDM}\label{Electron-EDM}

\begin{figure}[t]
\minipage{0.48 \textwidth}
  \includegraphics[width=\linewidth]{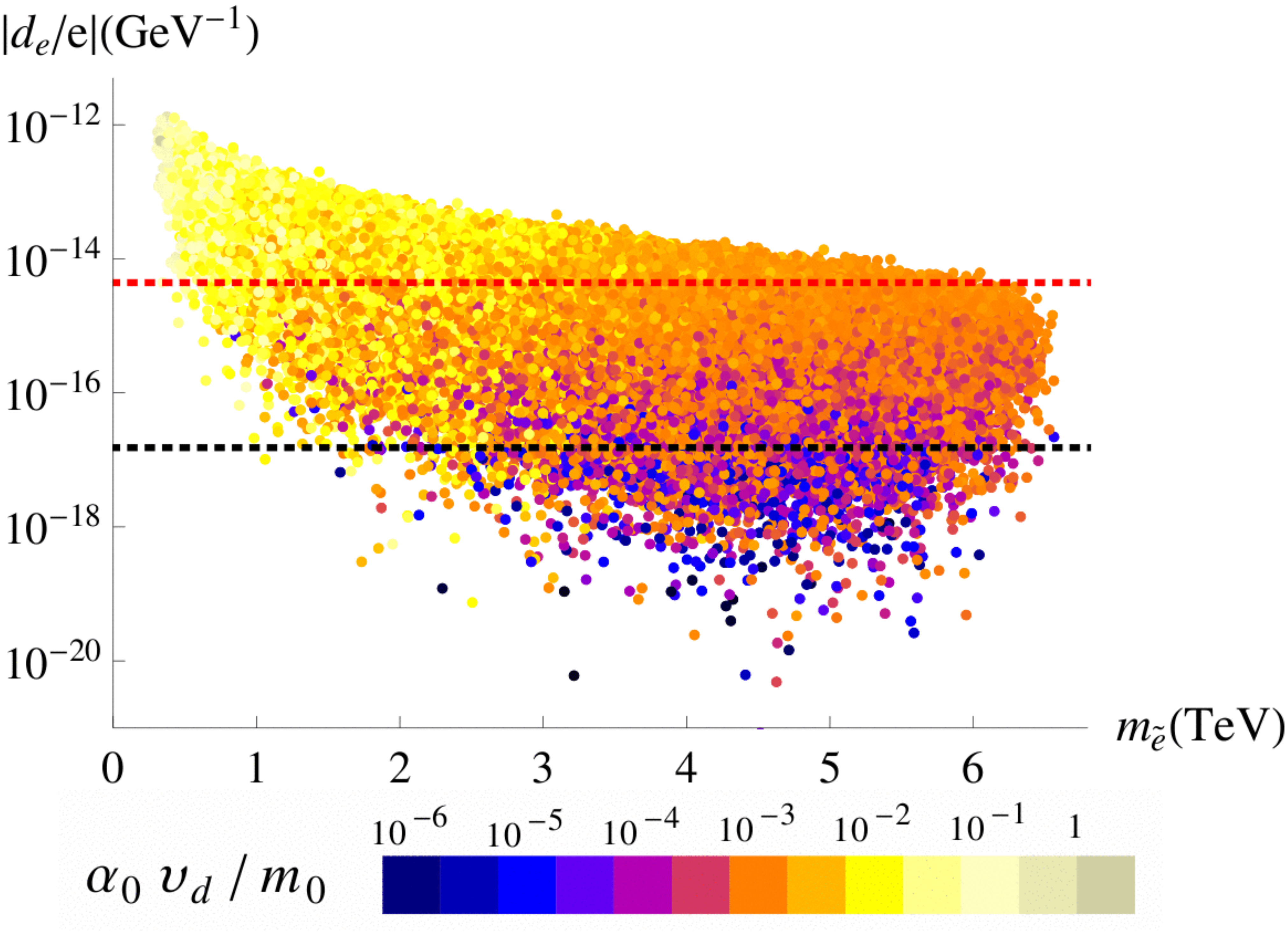}
\endminipage\hfill
\minipage{0.43 \textwidth}
  \includegraphics[width=\linewidth]{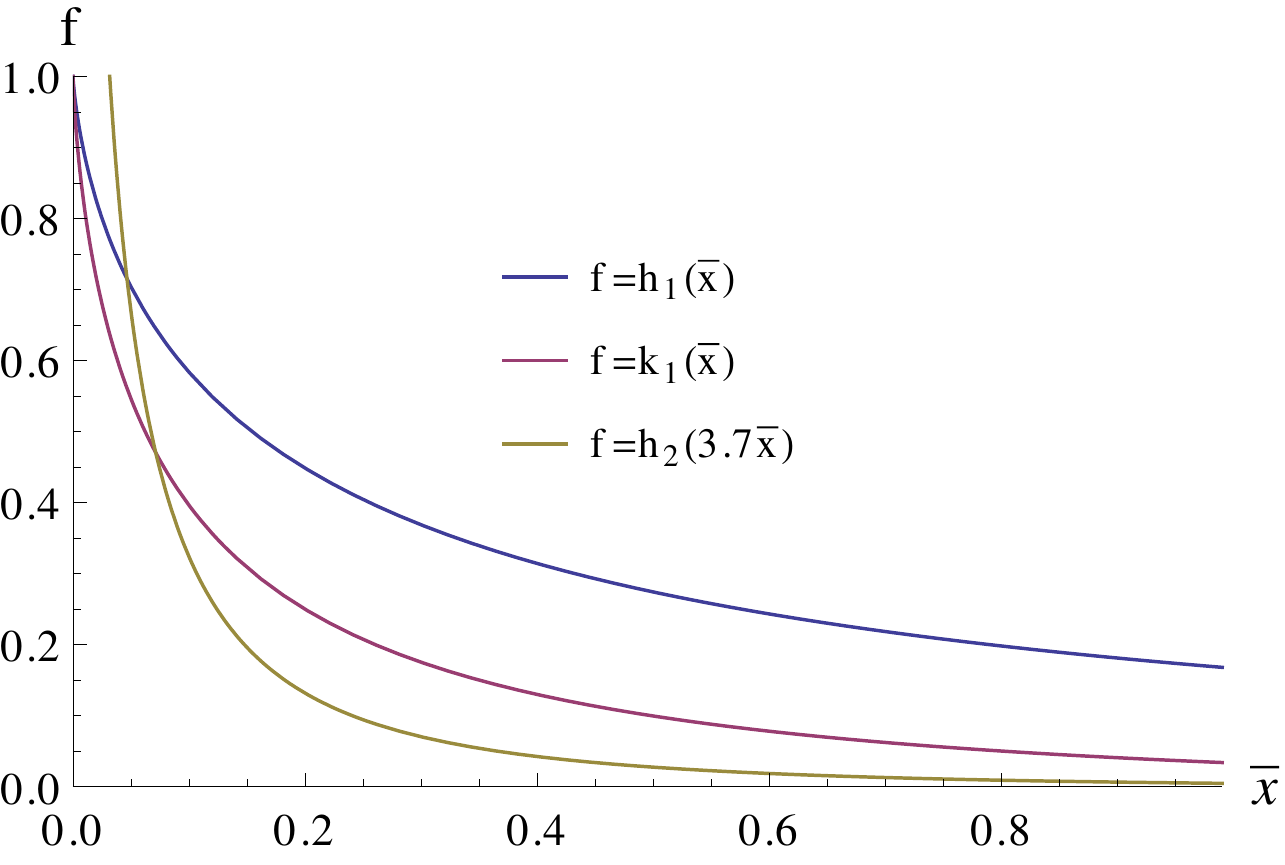}
\endminipage\hfill
\caption{Left panel: the prediction for the SUSY contribution to the electron EDM
  versus $m_{\tilde{e}}=\sqrt{m_{\tilde{e}_{LL}}m_{\tilde{e}_{RR}}}$. The red
  dotted line represents the current experimental limit of
  Eq.~\eqref{EDMlimit}, while the black dotted line corresponds to the
  expected future limit of $|d_e/e|\lesssim 3\times10^{-31}$\,cm~$\approx
  1.52\times10^{-17} \,\text{GeV}^{-1}$~\cite{Hisano15}. Right panel: the
  behaviour of the functions $h_1$, $k_1$ and (in anticipation of the
  discussion in Section~\ref{mutoegamma})  $h_2$.}
\label{Fig:eEDM}
\end{figure}

The current experimental limit for the electric dipole moment of the electron
stems from the ACME collaboration~\cite{delimit} and is given by
\begin{equation}
|d_e/e|~\lesssim ~8.7\times10^{-29}\,\text{cm}~\approx 4.41~\times10^{-15}\,
\text{GeV}^{-1} .\label{EDMlimit}
\end{equation}
This tiny value poses a strong constraint on the phases of any model. 
The supersymmetric contributions depend on the mass insertion parameters as
follows~\cite{Sleptonarium}\footnote{The corresponding expression
  in~\cite{AnatomyandPhenomenology} also includes triple mass insertions of type
  $(LR)(RR)(RR)$ and $(LL)(LL)(LR)$. In our model, these give suppressed
  contributions  to $d_e/e$  of order $\lambda^{11}$ and $\lambda^{13}$,
  respectively, which can be safely neglected.} 

\begin{eqnarray}
\nonumber \frac{d_e}{e}&=&\frac{\alpha}{8\pi\cos^2\theta_W}0.43\frac{\sqrt{x}}{~~m_0^3}\,
m_{\tilde{e}_{LL}}\mathrm{Im}\Big[-(\delta^e_{LR})_{11}C_B\,m_{\tilde{e}_{RR}}+\\
\nonumber &+&\Big{\{}(\delta^e_{LL})_{1 i}(\delta^e_{LR})_{i 1}C'_{B,L}+(\delta^e_{LR})_{1i}(\delta^e_{RR})_{i1}C'_{B,R}\Big{\}}m_{R_{ii}}-\\
 &-&\Big{\{}(\delta^e_{LL})_{1 i}(\delta^e_{LR})_{ij}(\delta^e_{RR})_{j1}+(\delta^e_{LR})_{1 j}(\delta^e_{RL})_{ji}(\delta^e_{LR})_{i1}\Big{\}}
C''_B\,m_{R_{jj}}\Big],\label{EDM}
\end{eqnarray}
where $m_{\tilde{e}_{LL}}$ and $m_{\tilde{e}_{RR}}$
are given in Eq.~\eqref{eLR}. Moreover $m_{R_{ii}}=m_{\tilde{e}_{RR}}$ for $i=1,2$ 
and $m_{R_{33}}=m_{\tilde{\tau}_{RR}}$ with the latter being defined in
Eq.~\eqref{eLR}.
The expression of Eq.~\eqref{EDM} is actually proportional to the bino mass
$M_1$, which we have approximated by Eq.~\eqref{gauginos} using
$x=(M_{1/2}/m_0)^2$. 
The dimensionless loop functions $C_i$, whose expressions can be found in
Appendix~\ref{Loop Functions} encode the contributions from the pure bino
($i=B$) and the bino-higgsino with left- ($i=B,L$) and right-handed ($i=B,R$)
slepton diagrams. 
For $x\ll 1$, all ratios of different $C_i$ functions are close to one. With
increasing $x$, $C_B$ takes slightly larger values than the rest of the
functions, reaching up to twice the value of $C'_{B,L(R)}$ and three times the
value of $C''_{B}$. 
This can be seen in the limit where the left- and right-type slepton masses
are not very different, such that the loop functions take the
form~\cite{Sleptonarium} 
\begin{eqnarray}
\nonumber C_B&\approx& \frac{m_0^4}{m_{\tilde{e}}^4}h_1(\bar{x}),
\qquad C''_{B}\approx \frac{m_0^4}{3m_{\tilde{e}}^4}\left(h_1(\bar{x})+2k_1(\bar{x})\right),\\
C'_{B,L}&\approx& C'_{B,R}\approx\frac{m_0^4}{2m_{\tilde{e}}^4}\left(h_1(\bar{x})+k_1(\bar{x})\right),
\end{eqnarray}
where we consider $m_{\tilde{e}}=\sqrt{m_{\tilde{e}_{LL}}m_{\tilde{e}_{RR}}}$
as the average slepton mass\footnote{$m_{\tilde{e}_{RR}}$ and
$m_{\tilde{\tau}_{RR}}$ only differ in the order one coefficients $b_{01}$ and
  $b_{02}$ which take values in the same range. Since the dominant term in
Eq.~\eqref{EDM} involves the first generation masses, we use
$m_{\tilde{e}}=\sqrt{m_{\tilde{e}_{LL}}m_{\tilde{e}_{RR}}}$ rather than
$m_{\tilde{e}}=\sqrt{m_{\tilde{e}_{LL}}\sqrt{m_{\tilde{e}_{RR}}m_{\tilde{\tau}_{RR}}}}$
as the average slepton mass.} 
and $\bar{x}=(M_1/m_{\tilde{e}})^2$. The function $h_1$  is given in
Appendix~\ref{Loop Functions} while $k_1$ denotes the derivative 
$k_1(\bar x)\equiv d(\bar xh_1(\bar x))/d\bar x$. Their behaviour is shown in
the right panel of Figure~\ref{Fig:eEDM}.

The dominant contribution to the electron EDM comes from the single chirality
flipping diagonal mass insertion $(\delta^e_{LR})_{11}\propto \lambda^6$, such
that we can make the approximation 
\begin{eqnarray}
|d_e/e|&\approx&\frac{\alpha}{8\pi\cos^2\theta_W}0.43\sqrt{x}\,\frac{|\alpha_0|\upsilon_d}{m_0^2}(1+R^y_e)\frac{1}{3}\,\left|\mathrm{Im}[\tilde{a}^d_{11}]\right|\,\lambda^6\,C_B,\label{EDMapprox}
\end{eqnarray}
where $R^y_e$ is an RG running factor defined in Eq.~\eqref{Rys} and
$\tilde{a}^d_{11}/3$, defined in Eq.~\eqref{adt}, is the (11) element of
$\tilde{A}^e_\text{GUT}/A_0$, with $\tilde{A}^e_\text{GUT}$ denoting the GUT
scale soft trilinear matrix in the SCKM basis. 
Its imaginary part is non-zero when allowing the phases of the soft trilinear
sector to be different from the phases of the corresponding Yukawa sector.
Then, for $|\alpha_0\upsilon_d/m_0|\approx 10^{-2}$, $m_0\approx 1$~TeV and $x\approx1$, we expect $|d_e/e|$ to vary around $10^{-13}\,\text{GeV}^{-1}$.

As can be seen in the left panel of Figure~\ref{Fig:eEDM}, which was produced
using the full expression in Eq.~\eqref{EDM}, the numerical choice for the
suppression factor $|\alpha_0\,\upsilon_d/m_0|$ corresponds to the yellow
points and brings our prediction for the EDM above its current experimental
limit, represented by the red dotted line. 

In the case where the phases of the soft trilinear and Yukawa sectors are
equal, $\tilde{a}^d_{11}$ and all factors in Eq.~\eqref{adt} become
real. In that case, the dominant imaginary part originates from the NLO
contribution\footnote{The SCKM rotation which
  renders the Yukawa sector diagonal and real does not do the same to the
  $A$-terms beyond leading order.}  to $(\delta^e_{LR})_{11}$
and is proportional to $\sin(4\theta^d_2+\theta^d_3)$. 
Setting $\theta^d_2=\pi/2$, as is preferred by the Jarlskog invariant
$J^q_{CP}$, given in Eq.~\eqref{JCPq}, we see that also the NLO contribution
  vanishes for $\theta^d_3=0$, such that $|d_e/e|$ would only arise at
  order $\lambda^8$.

Concerning the terms of Eq.~\eqref{EDM} with double mass insertions, they
enter at orders $(\delta^e_{LR})_{12}(\delta^e_{RR})_{21}\sim\lambda^8$, 
$(\delta^e_{LR})_{13}(\delta^e_{RR})_{31}\sim\lambda^{10}$ 
and $(\delta^e_{LL})_{12}(\delta^e_{LR})_{21}\sim
(\delta^e_{LL})_{13}(\delta^e_{LR})_{31}\sim \lambda^9$ in our model. 
In the situation described in the preceding paragraph, the first two terms are
real, while the contributions of the latter two cancel against each other. 
Finally, the contributions of the triple mass insertions are further
suppressed,  with the largest one,
$(\delta^e_{LL})_{13}(\delta^e_{LR})_{33}(\delta^e_{RR})_{31}\sim\lambda^{10}$,
being real in the case at hand, while all other triple insertions entail
contributions which lie below the experimental limit.


\subsection[$BR(\mu\to e\gamma)$]{$\boldsymbol{BR(\mu\to e\gamma)}$}\label{mutoegamma}

According to Figure~\ref{Fig:electron MIs}, a large part of our parameter
space in the (12) charged lepton sector appears to be excluded by the
experimental limit set by the non-observation of $\mu\to e\gamma$.
In this section, we  therefore study in detail the contributions to this
LFV process within our model. 
The current experimental limit for the branching ratio
\begin{equation}
BR(\mu\to e\,\gamma)~\lesssim ~5.7\times 10^{-13}\ ,\label{MEGlimit}
\end{equation}
is set by the MEG collaboration~\cite{MEG-limit}. The expression for the
corresponding SUSY contribution is given by~\cite{Sleptonarium}
\begin{eqnarray}
\nonumber &\,&BR(\mu\to e\gamma)~=~3.4\times 10^{-4}\times0.43^2\,M_W^4\,x\,\frac{\mu^2\,t_\beta^2}{m_0^6}\times\\
\nonumber &\times&\Bigg(\left|(\delta^e_{LL})_{12}\left(-(\delta^e_{LR})_{22}\frac{m_{\tilde{e}_{LL}}m_{\tilde{e}_{RR}}}{\mu\,t_\beta\,m_\mu}C'_{B,L}+\frac{1}{2}C'_L+C'_2\right)+(\delta^e_{LR})_{12}\frac{m_{\tilde{e}_{LL}}m_{\tilde{e}_{RR}}}{\mu\,t_\beta\,m_\mu}C_B\right|^2\\
&+&\left|(\delta^e_{RR})_{12}\left(-(\delta^e_{LR})^*_{22}\frac{m_{\tilde{e}_{LL}}m_{\tilde{e}_{RR}}}{\mu\,t_\beta\,m_\mu}C'_{B,R}-C'_R\right)+(\delta^e_{LR})^*_{21}\frac{m_{\tilde{e}_{LL}}m_{\tilde{e}_{RR}}}{\mu\,t_\beta\,m_\mu}C_B\right|^2\Bigg).\label{MEG}
\end{eqnarray}
It is proportional to the bino mass squared, that has been approximated by
Eq.~\eqref{gauginos} and expressed as $M_1^2=0.43^2x\,m_0^2$, where
$x=(M_{1/2}/m_0)^2$. The loop function $C'_2$ encodes the wino-higgsino
contribution and is defined in Appendix \ref{Loop Functions}, along with the
rest of the functions $C_i$.

In our model, $(\delta^e_{LL})_{12}\sim \lambda^4$, 
$(\delta^e_{RR})_{12}\sim \lambda^3$, 
$(\delta^e_{LR})_{12(21)}\sim\lambda^5$ and 
$(\delta^e_{LR})_{22}\sim \lambda^4$. 
To get an estimate of the dominant $\delta$s in Eq.~\eqref{MEG}, we first
compare the $SU(2)$ ($\propto C'_2$) and the $U(1)$ ($\propto C_{B,L}',C_L'$)
contributions to the $(\delta^e_{LL})_{12}$ term by studying the ratio
\begin{eqnarray}
 R&=&\left|C'_2\Big{/}\left(\left(1-\frac{A_0}{\mu\,t_\beta}\frac{\tilde{a}^d_{22}}{y_s}\right)C'_{B,L}+\frac{1}{2}C'_L\right)\right|,\label{SU2U1Ratio}
\end{eqnarray}
which, in the limit where $m_{\tilde{e}_{RR}}$ and $m_{\tilde{e}_{LL}}$ are not very different, can be written as
\begin{eqnarray}
R ~ \approx ~\bar{R}&=&2\frac{M_2}{M_1} \cot^2\theta_W\left|\frac{\frac{1}{\bar{y}-\bar{x}'}\left(h_2(\bar{x}')-h_2(\bar{y})\right)}{h_1(\bar{x})+k_1(\bar{x})+\frac{1}{\bar{y}-\bar{x}}\left(h_1(\bar{x})-h_1(\bar{y})\right)}\right|.\label{SU2U1Ratiobar}
\end{eqnarray}%
\begin{figure}[t]
\minipage{0.45 \textwidth}
  \includegraphics[width=\linewidth]{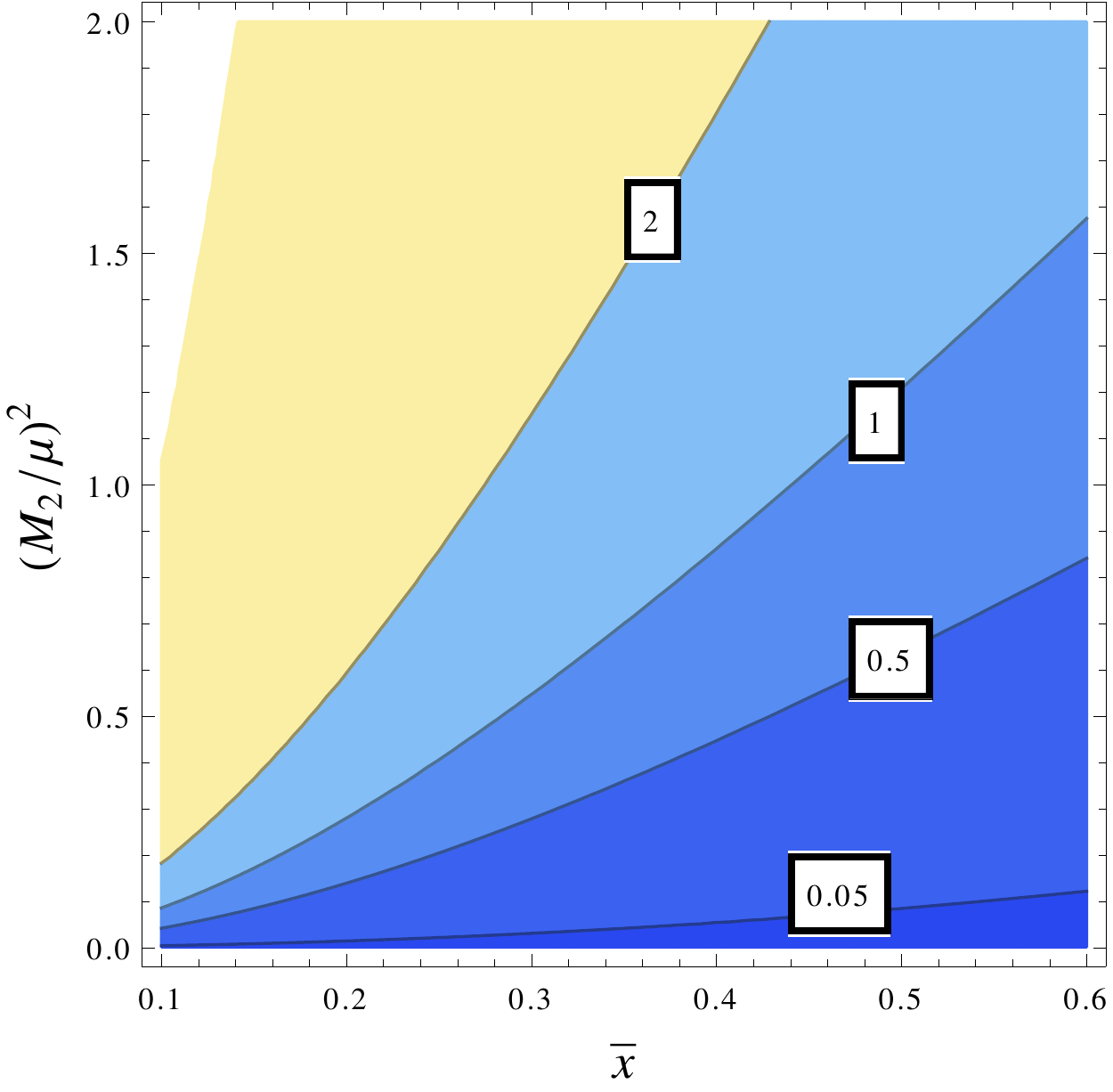}
\endminipage\hfill
\minipage{0.48 \textwidth}
  \includegraphics[width=\linewidth]{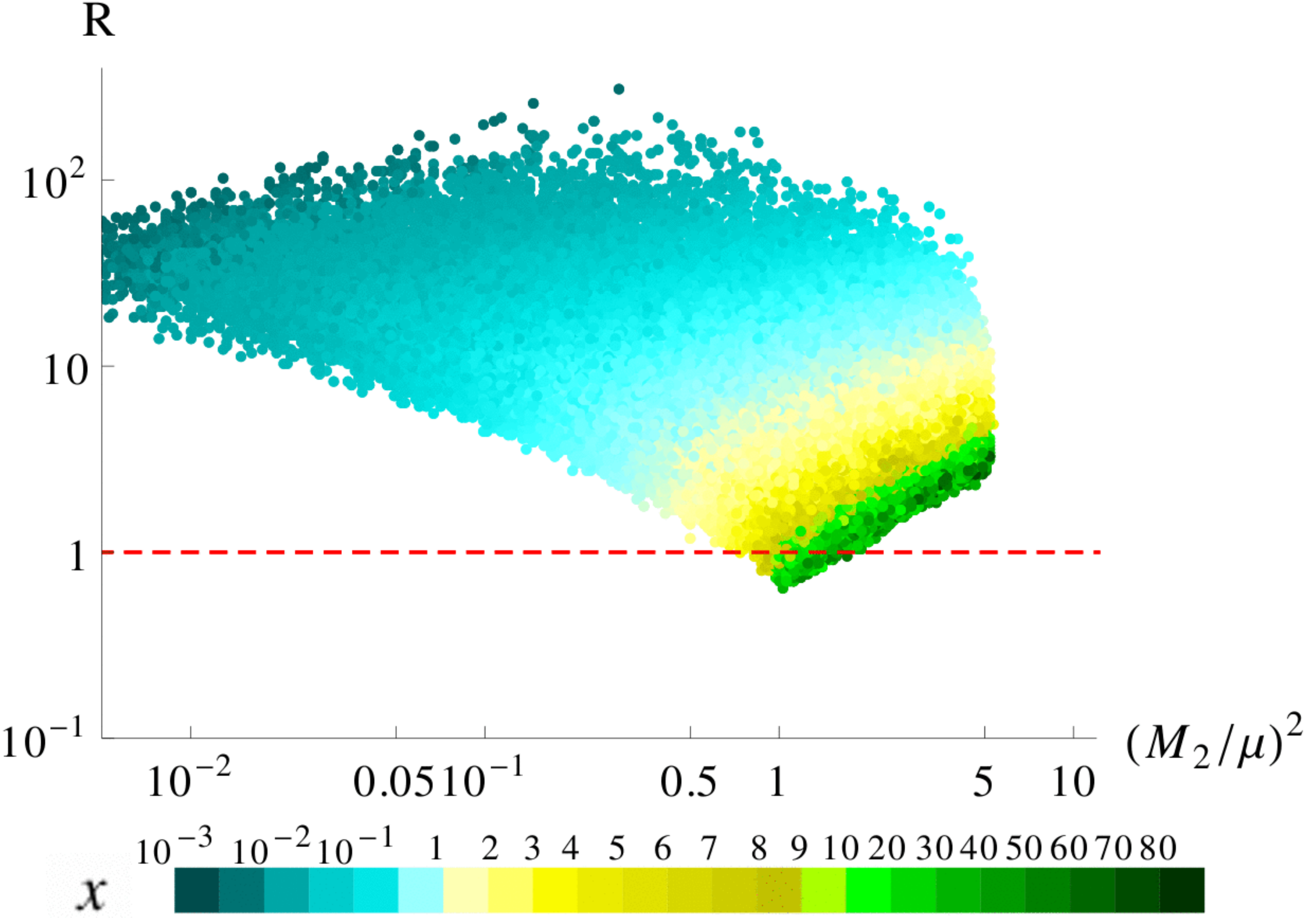}
\endminipage\hfill
\caption{Left panel: the contour lines for $\bar R$, the approximate ratio of
  the $SU(2)$ over the $U(1)$ contributions to the $(\delta^e_{LL})_{12}$ term
  in Eq.~\eqref{MEG}, as defined in Eq.~\eqref{SU2U1Ratiobar}. 
For the average slepton mass $m_{\tilde{e}}=\sqrt{ m_{\tilde{e}_{LL}}
  m_{\tilde{e}_{RR}}}$, $\bar{x}=(M_1/m_{\tilde{e}})^2\approx 0.43^2
x/(1+0.3\,x)$, with $x=(M_{1/2}/m_0)^2$. 
Right panel: the ratio $R$ (without approximation), as defined in
Eq.~\eqref{SU2U1Ratio} and produced in our scan. The dependence of
$(M_2/\mu)^2$ and $\bar{x}$ on $x$ is such that the $SU(2)$ contributions
dominate for most of the parameter space.} 
\label{Fig:MEG_contr}
\end{figure}%
The behaviour of the loop functions $h_1$ and $h_2$, which are defined in 
Appendix~\ref{Loop Functions}, as well as
$k_1(\bar x)\equiv d(\bar xh_1(\bar x))/d\bar x$ is shown in the right panel of
Figure~\ref{Fig:eEDM}, and $\bar{x}=(M_1/m_{\tilde{e}})^2$,
$\bar{x}'=(M_2/m_{\tilde{e}})^2$, 
$\bar{y}=(\mu/m_{\tilde{e}})^2$, 
with $m_{\tilde{e}}=\sqrt{ m_{\tilde{e}_{LL}} m_{\tilde{e}_{RR}}}$.
The contours in the left panel of Figure~\ref{Fig:MEG_contr} show the
dependence of $\bar{R}$, as defined in Eq.~\eqref{SU2U1Ratiobar}, on
$(M_2/\mu)^2$ and $\bar{x}$. 
We see that for $(M_2/\mu)^2\gtrsim1.5$, $\bar{R}$ is larger than one 
for all $\bar{x}\approx 0.43^2 x/(1+0.3 \,x) \lesssim 0.6$, while for
$(M_2/\mu)^2\sim\mathcal{O}(1)$ and smaller, the $U(1)$ contributions can
dominate if $\bar{x}$ does not decrease faster than $(M_2/\mu)^2$. 
The right panel in Figure~\ref{Fig:MEG_contr} is based on our scan and shows
that the correlation of $(M_2/\mu)^2$ and $\bar{x}$ through $x$ is such that
$R$, as defined in Eq.~\eqref{SU2U1Ratio}, stays larger than one in most of
our parameter space, making the $SU(2)$ contribution to the
$(\delta^e_{LL})_{12}$ term in Eq.~\eqref{MEG} the most important
one. 

Similarly, one can show that the $RR$ contribution to $\mu\to e\gamma$ in
Eq.~\eqref{MEG} is comparable to the $LL$ one only
when  $|(\delta^e_{RR})_{12}\lambda|/|(\delta^e_{LL})_{12}|\gtrsim1$, 
although $(\delta^e_{LL})_{12}$ is suppressed by an order of $\lambda$ with
respect to $(\delta^e_{RR})_{12}$. This happens because the 
$RR$ parameter has only two $U(1)$ contributions which come in with opposite
signs, allowing even for a complete cancellation.

Finally, we study the relative size of the $LL$ and $LR$ contributions by
considering the ratio 
\begin{eqnarray} R'=\Big{|}\frac{\mu\,t_\beta\,m_\mu(\delta^e_{LL})_{12}\,C_2'}{m_{\tilde{e}_{LL}}m_{\tilde{e}_{RR}}(\delta^e_{LR})_{12}C_B}\Big{|}=\lambda^3\kappa\Big{|}\frac{\mu\,t_\beta}{A_0}\frac{C_2'}{C_B}\Big{|},\label{MEG_LL_LR_contr}
\end{eqnarray}
where $\kappa=
\Big{|}3\,y_s(\tilde{R}_{12}-2\eta_N\,\tilde{E}_{12})/
(\tilde{a}^d_{12}(p^e_L)^2)\Big{|}$, with 
$\tilde{R}_{12}$, $\tilde{a}^d_{12}$, $p^e_L$, $\tilde{E}_{12}$ and
$\eta_N$ defined in
Eqs.~(\ref{Bt},\ref{adt},\ref{pes},\ref{E12},\ref{etas}), respectively.  
The absolute value of the right-hand side of Eq.~\eqref{MEG_LL_LR_contr}
exhibits a similar behaviour as the ratio $R$, defined in
Eq.~\eqref{SU2U1Ratio} and shown in the right panel of
Figure~\ref{Fig:MEG_contr}.   
Taking into account the $\lambda$-suppression ($\lambda^3\sim 10^{-2}$) and
the range of $\kappa$ which can vary within two orders of magnitude, 
we find that the $(\delta^e_{LR})_{12}$ contribution to the branching ratio
can be comparable to the $(\delta^e_{LL})_{12}$ one when $(M_2/\mu)^2\sim 1$.

Considering situations in which the $(\delta^e_{LR})_{12}$ contribution to
Eq.~\eqref{MEG} dominates, we obtain the approximate expression
\begin{eqnarray}
BR(\mu\to e\gamma)|_{_{(\delta^e_{LR})_{12}}}&\approx&\mathcal{O}\left(10^2\,\alpha_0^2\frac{m_0^4}{m_{\tilde{e}}^8}h^2_1(\bar{x})\right)\left(\frac{|\tilde{a}^d_{12}|}{3\,y_s}\right)^2.\label{MEGapproxLR}
\end{eqnarray}
In the case where $(\delta^e_{LL})_{12}$ is more important, e.g. when
$(M_2/\mu)^2\ll1$, cf. right panel of Figure~\ref{Fig:MEG_contr}, we obtain
\begin{eqnarray}
BR(\mu\to e\gamma)|_{_{(\delta^e_{LL})_{12}}}&\approx&\mathcal{O}\left(\frac{x\,t_\beta^2}{\mu^2}\frac{m_0^6}{~~m_{\tilde{e}_{LL}}^8}h^2_2(3.7\,x_L)\right)\Big{|}\tilde{R}_{12}-2\eta_N\,\tilde{E}_{12}\Big{|}^2.\label{MEGapproxLL}
\end{eqnarray}
For $x_L\equiv(M_1/m_{\tilde{e}_{LL}})^2\approx\bar{x}\approx0.1$,
$x\approx1$, $\alpha_0\approx 1$, $t_\beta\approx10$, $\mu\approx m_0\approx
1$~TeV and $m_{\tilde{e}_{LL}}\approx 750$~GeV, the approximations of
Eqs.~(\ref{MEGapproxLR},\ref{MEGapproxLL}) both produce a value of the order of 
$10^{-10}$ times the relevant order one coefficients squared. 
In order to gain an extra suppression of at least an order of magnitude, 
the latter are preferred to be smaller than one.

\begin{figure}[t]
\minipage{0.48 \textwidth}
  \includegraphics[width=\linewidth]{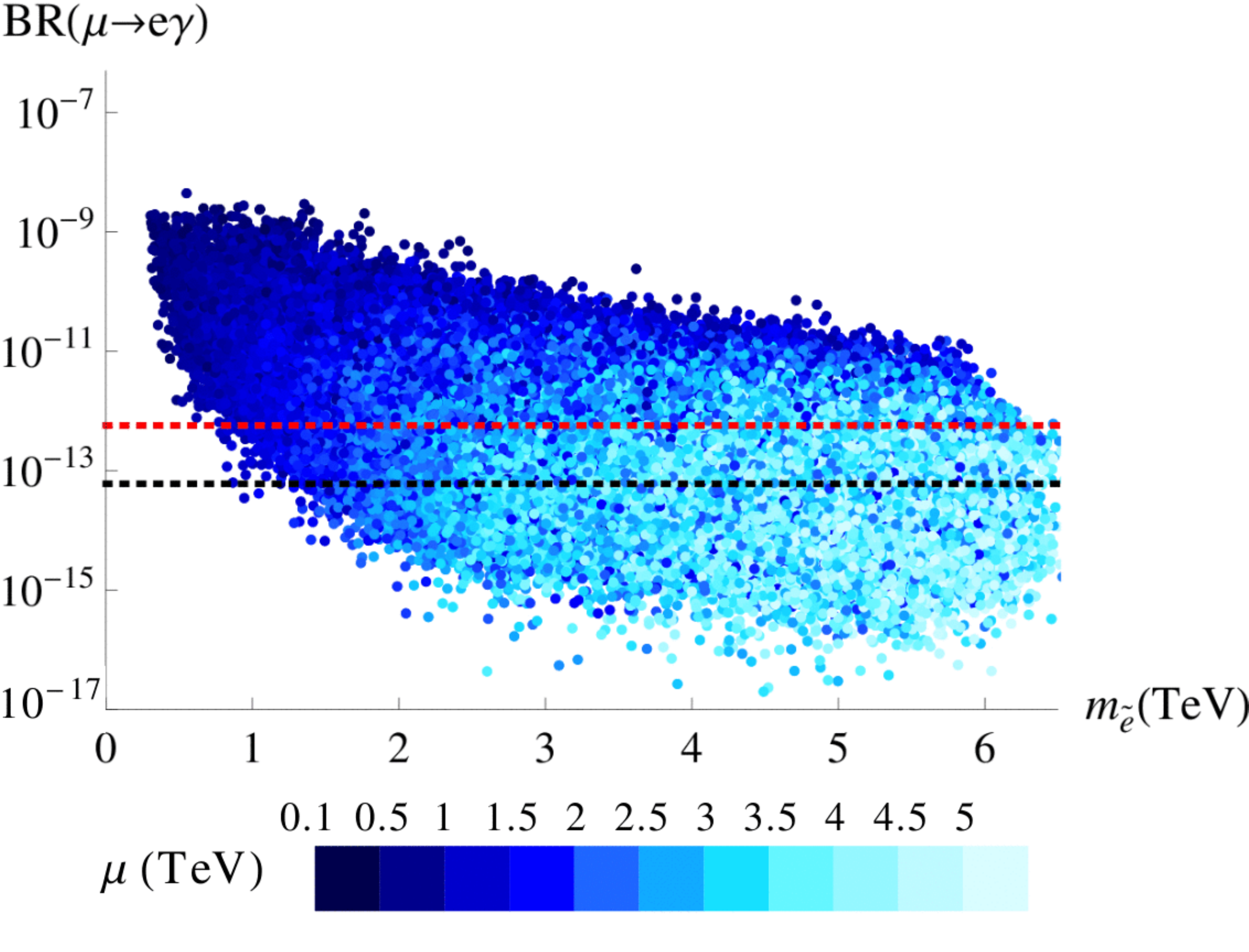}
\endminipage\hfill
\minipage{0.48 \textwidth}
  \includegraphics[width=\linewidth]{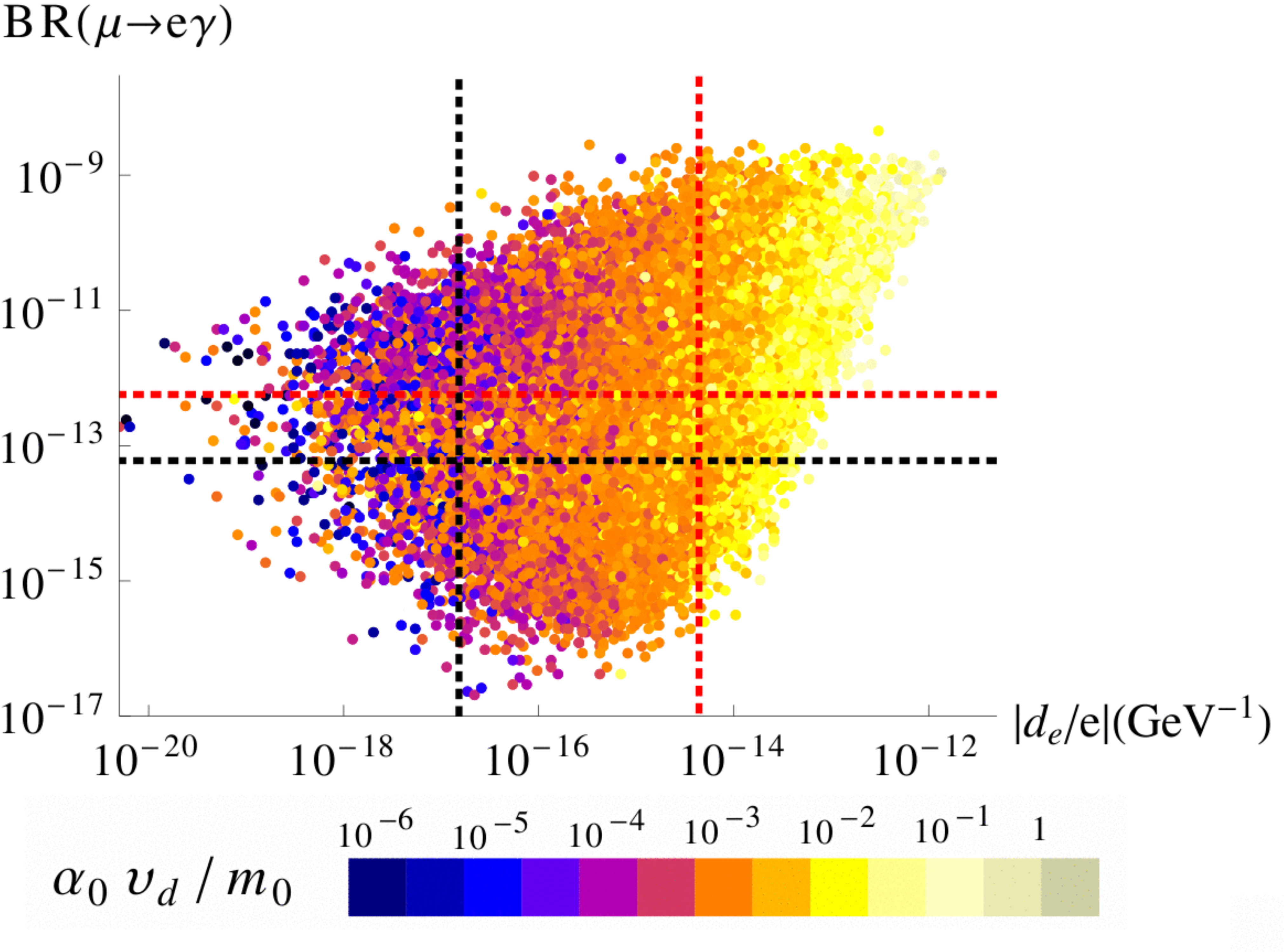}
\endminipage\hfill
\caption{The supersymmetric contribution to the branching ratio of $\mu\to
  e\,\gamma$ versus the average slepton mass 
$m_{\tilde{e}}=\sqrt{m_{\tilde{e}_{LL}}m_{\tilde{e}_{RR}}}$ (left panel) as
  well as $|d_e/e|$ (right panel). 
The red dotted lines represent the current experimental limits given in
Eqs.~(\ref{EDMlimit},\ref{MEGlimit}) while the black dotted lines show the
expected future limits, that is $BR(\mu\to e\,\gamma)\lesssim 6\times
10^{-14}$~\cite{MEG-limit-Future} and 
$|d_e/e|\lesssim 1.52\times10^{-17}~\text{GeV}^{-1}$~\cite{Hisano15}.}

\label{Fig:MEG-tot}
\end{figure}

The total supersymmetric contribution to the branching ratio of $\mu \to e
\gamma$ of Eq.~\eqref{MEG} as produced in our scan is shown in
Figure~\ref{Fig:MEG-tot}. There it is plotted against the average slepton mass
(left panel) as well as $|d_e/e|$ (right panel).
From the left panel we observe that our model requires rather heavy sleptons,
in the TeV range, in order to survive the current experimental limit in
Eq.~\eqref{MEGlimit}, which is denoted by the red dotted line. 
As can be seen in Eqs.~(\ref{MEG},\ref{MEGapproxLL}), there is also a strong $\mu$
dependence, with a preference for large values. 
The right panel of Figure~\ref{Fig:MEG-tot} shows that the $\mu\to e\gamma$
branching ratio is correlated with the electron EDM, mainly through the
slepton masses and the bino-slepton mass ratio. The combination of the current
limits on both observables highly restricts our parameter space. Reaching the
expected future limits, denoted by the black dotted lines, would nearly
exclude our model. 
\begin{figure}[t]
\minipage{0.48 \textwidth}
  \includegraphics[width=\linewidth]{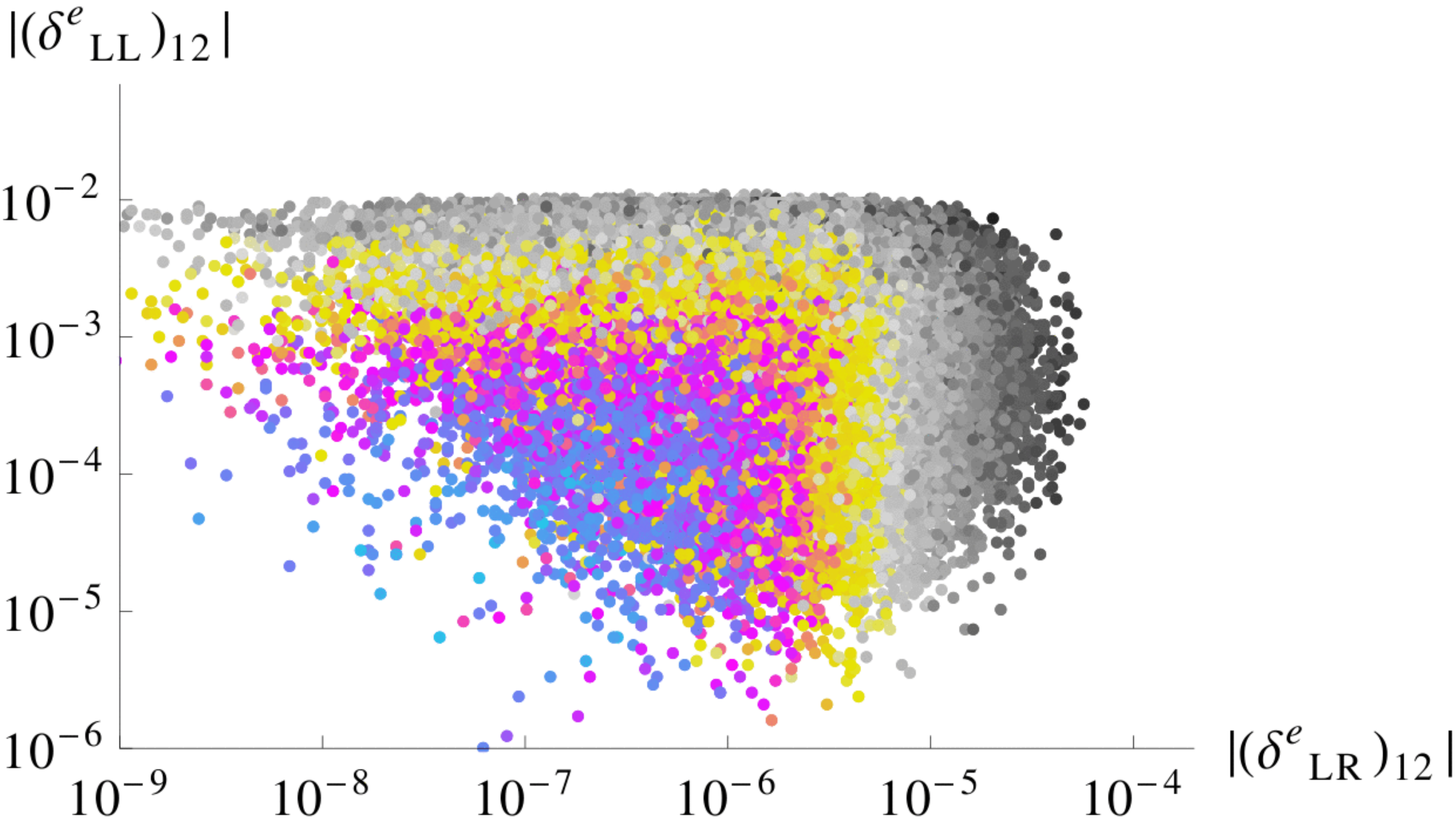}
\endminipage\hfill
\minipage{0.48 \textwidth}
  \includegraphics[width=\linewidth]{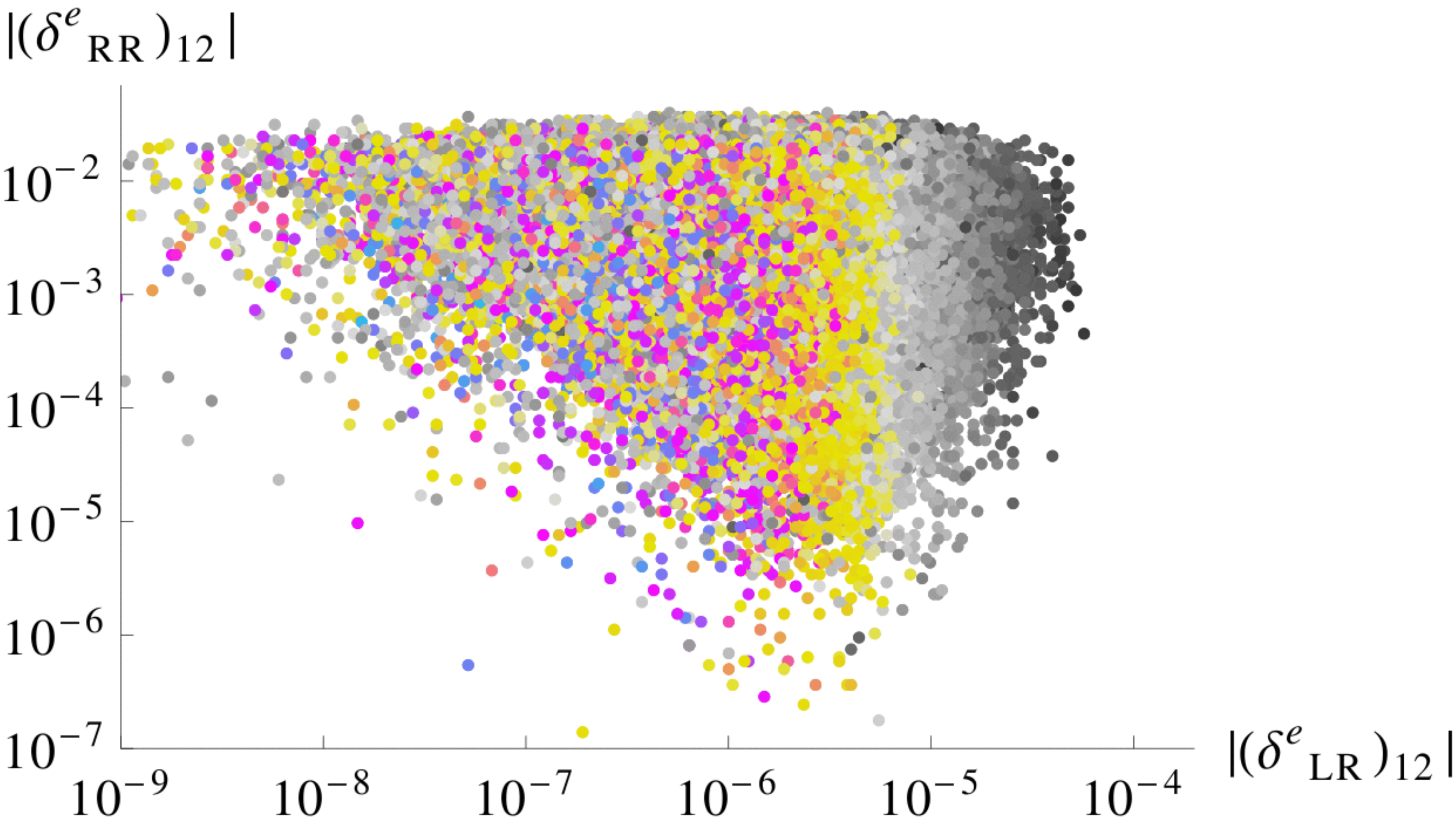}
\endminipage\hfill
\begin{center}
  \includegraphics[scale=0.49]{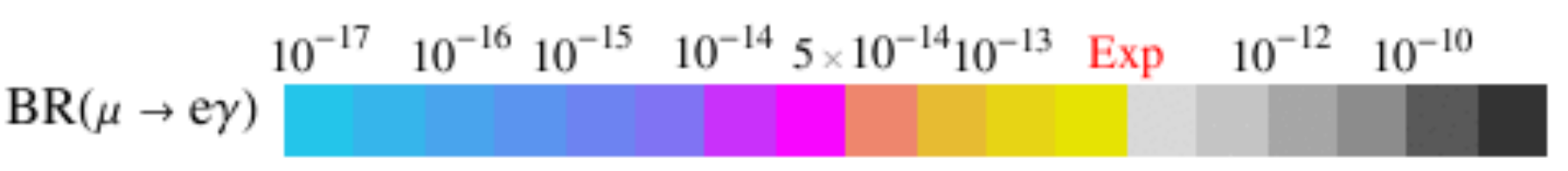}
\end{center}
\caption{The range of the (12) lepton mass insertion parameters as produced in
  our scan, together with the resulting prediction for the branching ratio of $\mu\to
  e\gamma$. The grey points do not satisfy the current experimental limit
  given in Eq.~(\ref{EDMlimit}).}
\label{Fig:MEG_deltas}
\end{figure}

In Figure~\ref{Fig:MEG_deltas} we show our predictions for $BR(\mu\to e\gamma)$ in
the plane of two (12) mass insertion parameters as produced in our
scan. Comparing this to the discussion of 
Section~\ref{Numerics:Charged-lepton sector} reveals that, with the present
MEG bound, 
$|(\delta^e_{LL})_{12}|\lesssim 5\times 10^{-3}$ and
$|(\delta^e_{LR})_{12}|\lesssim 5\times10^{-6}$ are not excluded as it was
suggested by the limits in Figure~\ref{Fig:electron MIs}. On the other hand,
$|(\delta^e_{RR})_{12}|$ can take its maximum values produced by the scan. 
The reason for these weaker bounds is twofold. Firstly, the analysis
in~\cite{Catania2}  sets the limits on the mass insertion parameters by
choosing $t_\beta$ as large as 60, whereas we only allow for maximum values of
25. Secondly, the derivation in~\cite{Catania2} requires that the discrepancy
of $(g-2)_\mu$ from its SM value is explained by SUSY contributions.


\newpage

\subsection{\label{sec:meson}Meson mixing}

Turning to $\Delta F=2$ transitions, we study the SUSY contributions to meson
mixing. The dispersive part of the mixing for a meson $P$  can be parametrised
as~\cite{LigetiDBs}
\begin{eqnarray}
M^P_{12}=M^{P,\,\text{SM}}_{12}+M^{P,\,\text{NP}}_{12}=M^{P,\,\text{SM}}_{12}\left(1+h_{P} e^{2 i \sigma_P}\right)\label{Mq12},
\end{eqnarray}
and the corresponding mass difference is given by
\begin{eqnarray}
\Delta M_P=2|M^P_{12}|.\label{MP}
\end{eqnarray}
We express the SM contribution as $M^{P,\,\text{SM}}_{12}=
|M^{P,\,\text{SM}}_{12}|\,e^{2i\phi_P^\text{SM}}$. The New Physics (NP) contribution, 
$M^{P,\,\text{NP}}_{12}=|M^{P,\,\text{NP}}_{12}|\,e^{2i\theta_P}$, is encoded
in the real parameters 
\begin{eqnarray}
h_{P}&=&\frac{|M^{P,\,\text{NP}}_{12}|}{|M^{P,\,\text{SM}}_{12}|},\qquad\label{hP_sigmaP}
\sigma_P=\theta_P-\phi_P^\text{SM}.
\end{eqnarray}
The contributions of the gluino-squark box diagram in terms of mass
insertion parameters read~\cite{Gabbiani:1996hi,AnatomyandPhenomenology} \\[-7mm]
\begin{eqnarray}
\nonumber M^{P,(\tilde{g})}_{12}&=&A_1^{P,(\tilde{g})}\Bigg(A_2^{P,(\tilde{g})}\left[(\delta^{d}_{LL})_{ji}^2+(\delta^{d}_{RR})_{ji}^2\right]+A_3^{P,(\tilde{g})}(\delta^{d}_{LL})_{ji}(\delta^{d}_{RR})_{ji}\\[-3mm]
&+&A_4^{P,(\tilde{g})}\left[(\delta^{d}_{LR})_{ji}^2+(\delta^{d}_{RL})_{ji}^2\right]+A_5^{P,(\tilde{g})}(\delta^{d}_{LR})_{ji}(\delta^{d}_{RL})_{ji}\Bigg),\label{MPSUSY}
\end{eqnarray}
where 
\begin{eqnarray}
\label{APG}
A_1^{P,(\tilde{g})}&=&-\frac{\alpha_s^2}{216\,m^2_{\tilde{q}}}\frac{1}{3}M_{P}f^2_P,~~\qquad
A_2^{P,(\tilde{g})}=24\,yf_6(y)+66\tilde{f}_6(y),\\
\nonumber A_3^{P,(\tilde{g})}&=&\Bigg(384\left(\frac{M_P}{m_j+m_i}\right)^2+72\Bigg)yf_6(y)+\Bigg(-24\left(\frac{M_P}{m_j+m_i}\right)^2+36\Bigg)\tilde{f}_6(y),\\\nonumber
A_4^{P,(\tilde{g})}&=&-132\left(\frac{M_P}{m_j+m_i}\right)^2yf_6(y),~~~A_5^{P,(\tilde{g})}=\Bigg(-144\left(\frac{M_P}{m_j+m_i}\right)^2-84\Bigg)\tilde{f}_6(y).
\end{eqnarray}
$M_P$ denotes the mass of the meson under consideration and $f_P$ is the
associated decay constant.
$m_i$ and $m_j$ are the masses of the meson's constituent quarks while
$m_{\tilde{q}}$ is an average squark mass which we define as 
\begin{eqnarray}
m_{\tilde{q}}&=&\Bigg\{\begin{array}{ccc}
\sqrt{m_{\tilde{d}_{LL}}m_{\tilde{d}_{RR}}},  &P=K , \\[2mm]
\sqrt{\sqrt{m_{\tilde{d}_{LL}}m_{\tilde{b}_{LL}}}m_{\tilde{d}_{RR}}},  &~~P=B_{s,d}, \label{msqav}
\end{array}
\end{eqnarray}
with $m_{\tilde{d}_{LL}}$, $m_{\tilde{b}_{LL}}$ and $m_{\tilde{d}_{RR}}$
defined in Eq.~\eqref{dLR}. 
The loop functions $f_6(y)$ and $\tilde{f}_6(y)$, where
$y=(m_{\tilde{g}}/m_{\tilde{q}})^2$, are given in Appendix~\ref{Loop
  Functions} and the gluino mass has been approximated by
Eq.~\eqref{gauginos}.


\subsubsection[$B_q-\bar{B}_q$]{$\boldsymbol{B_q-\bar{B}_q}$ mixing}\label{Bmixing}

The SM contribution to $B_q$, $q=s,d$ meson mixing given by~\cite{Buras13}
\begin{eqnarray}
M^{{B_q},\text{SM}}_{12}&=&\frac{G_F^2 M_{B_q}}{12
  \pi^2}M_W^2(V_{tb}V_{tq}^*)^2\eta_B S_0(x_t) f^2_{B_q}\hat{B}_{B_q},
\end{eqnarray}
with
\begin{eqnarray}
V_{ts}&=&-|V_{ts}|e^{i\beta_s},\qquad
V_{td}=|V_{td}|e^{-i\beta},\label{SMphases1}\\
\phi_{B_s}^\text{SM}&=&-\beta_s,\qquad
\phi_{B_d}^\text{SM}=\beta .\label{SMphases2}
\end{eqnarray}
Here $\eta_B$ is a QCD factor, $\hat{B}_{B_q}$ a perturbative parameter related
to hadronic matrix elements and $S_0(x_t\equiv \bar{m}_t^2(\bar m_t)/M_W^2)$
is the Inami-Lim loop function~\cite{Inami:1980fz}. 
The calculation of the pure SM contribution to the $B_s$ mass difference
gives~\cite{DeltaMBsSM}
\begin{eqnarray}
\Delta M^\text{(SM)}_{B_s}=125.2^{+13.8}_{-12.7}\times 10^{-13}\text{ GeV},\label{DMBsSM}
\end{eqnarray}
with the largest uncertainty stemming from the non-perturbative factor
$f_{B_s}\sqrt{\hat{B}_{B_s}}$, for which the value $275 \pm
13$~MeV~\cite{FBsBBs} has been used.\footnote{We note that the 2014 average of
the FLAG collaboration~\cite{FLAG} corresponds to a lower central value but
with a larger error: 
$f_{B_s}\sqrt{\hat{B}_{B_s}}\,\Big{|}_\text{FLAG}=266\pm 18\text{ MeV}$.}
The SM prediction for $\Delta M_{B_d}$ can be deduced from the ratio \cite{DeltaMBsSM}
\begin{eqnarray}
\frac{\Delta M^\text{(SM)}_{B_d}}{\Delta M^\text{(SM)}_{B_s}}=
0.02835\pm 0.00187,\label{DMBratioSM}
\end{eqnarray}
which is less sensitive to theoretical uncertainties. 
On the other hand, the associated experimental averages as of summer 2014,
provided by the HFAG group, read~\cite{HFAG}
\begin{eqnarray}
\Delta M^\text{(exp)}_{B_s}&=&(116.9\pm0.1)\times 10^{-13} \text{ GeV },\label{DMBsexp}\\
\Delta M^\text{(exp)}_{B_d}&=&(3.357\pm0.020)\times 10^{-13} \text{ GeV },\label{DMBdexp}\\
\frac{\Delta M^\text{(exp)}_{B_d}}{\Delta M^\text{(exp)}_{B_s}}&=&0.02879\pm0.0002.\label{DMBratioexp}
\end{eqnarray}
Comparing Eq.~(\ref{DMBsSM}) with Eq.~(\ref{DMBsexp}) leads to a negative
central value for the experimentally allowed NP contribution to $\Delta
M_{B_s}$, with a similar result being obtained for $\Delta M_{B_d}$. The main
source for the errors are the uncertainties of the SM
calculation.\footnote{For a recent discussion on theoretical uncertainties and
  comparison with experimental results, see~\cite{BsSystemReview}.} 
In view of Eqs.~(\ref{DMBsSM}-\ref{DMBratioexp}),  and in anticipation of
reduced theoretical uncertainties, we conclude that the largest NP
effects that could still be allowed should be consistent with 
\begin{eqnarray}
|\Delta M^\text{(NP)}_{B_s}|&\leq&2\times 10^{-12} \text{ GeV },\qquad
|\Delta M^\text{(NP)}_{B_d}|\leq1\times 10^{-13} \text{ GeV }.\label{DMBsdNP}
\end{eqnarray}

Using Eqs.~(\ref{MP},\ref{MPSUSY}), we can estimate the effects of the
gluino-squark box diagrams. 
Taking into account the $\lambda$-suppression of each $\delta$ parameter
entering Eq.~\eqref{MPSUSY}, we can write $\Delta M_{B_{s,d}}^{(\tilde{g})}$
in the schematic form 
\begin{eqnarray}
\nonumber\Delta M_{B_{s}}^{(\tilde{g})}&\propto& \lambda^4\left(A_2^{B_s,(\tilde{g})}+A_3^{B_s,(\tilde{g})}\lambda^2+A_4^{B_s,(\tilde{g})}\lambda^4+A_5^{B_s,(\tilde{g})}\lambda^6\right),\\
\Delta M_{B_{d}}^{(\tilde{g})}&\propto& \lambda^8\left(A_2^{B_d,(\tilde{g})}+A_3^{B_d,(\tilde{g})}+A_4^{B_d,(\tilde{g})}\lambda^2+A_5^{B_d,(\tilde{g})}\lambda^3\right).
\label{MBsdGapprox}
\end{eqnarray}
\begin{figure}[t]
\minipage{0.49 \textwidth}
  \includegraphics[width=\linewidth]{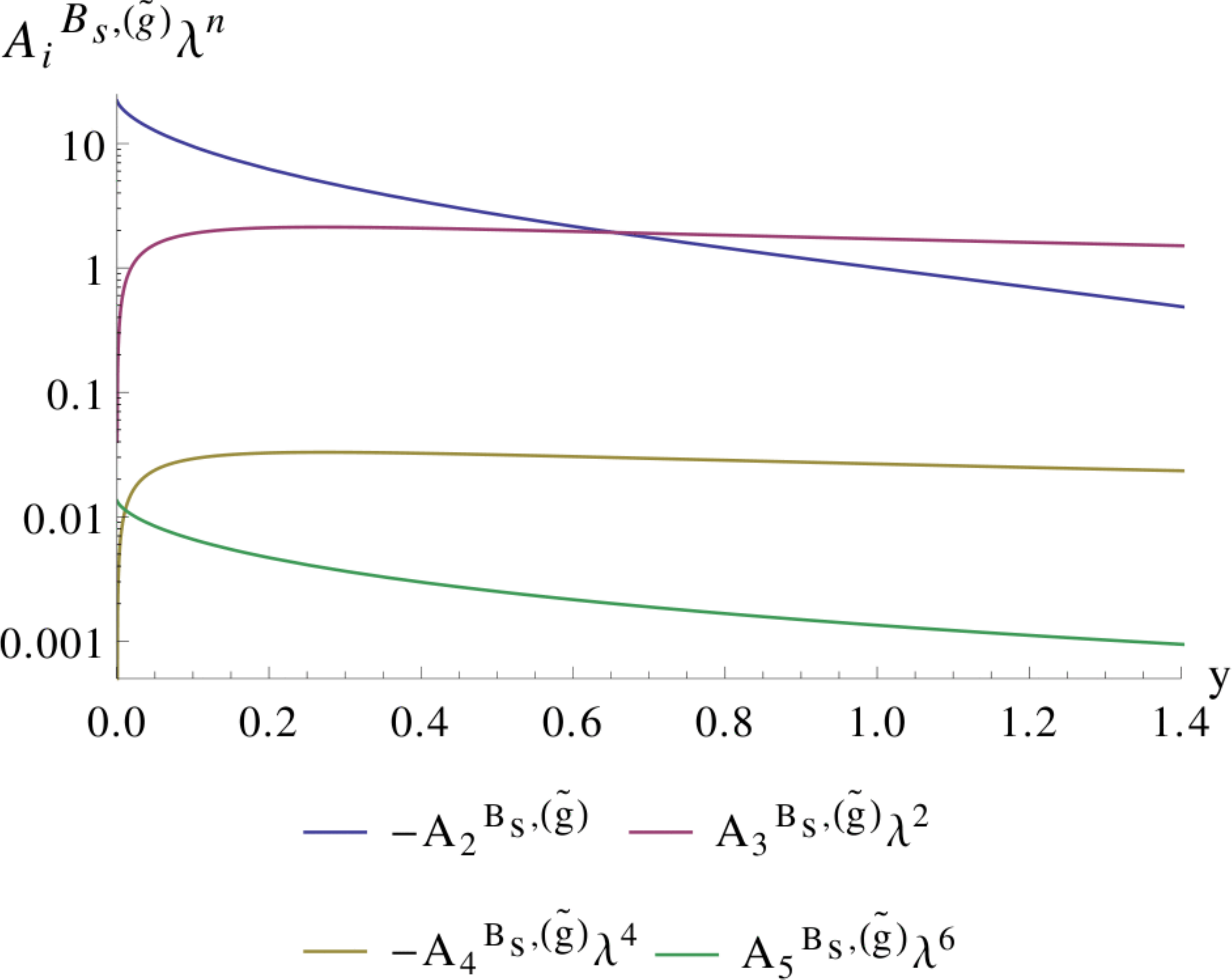}
\endminipage\hfill
\minipage{0.48 \textwidth}
  \includegraphics[width=\linewidth]{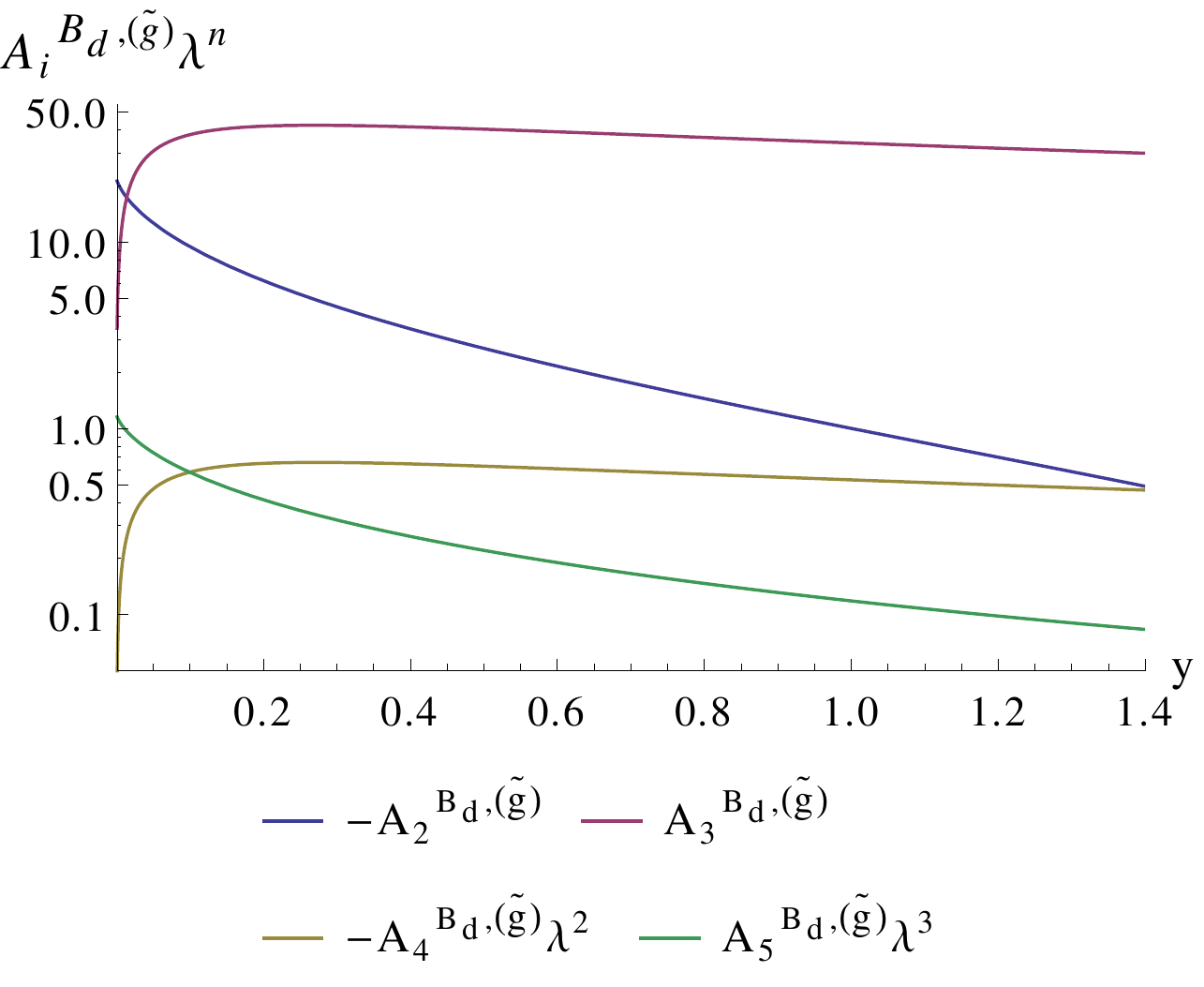}
\endminipage\hfill
\caption{The dependence of the individual contributions in Eq.~\eqref{MBsdGapprox}
  on $y=(m_{\tilde{g}}/m_{\tilde{q}})^2$. The average squark mass 
$m_{\tilde q}$ is defined in Eq.~\eqref{msqav} while the functions
  $A^{B_{s,d},(\tilde{g})}_i$ can be found in Eq.~\eqref{APG}.}
    \label{Fig:Bmixing2}
\end{figure}
Figure~\ref{Fig:Bmixing2} shows the individual contributions as a function of
$y=(m_{\tilde{g}}/m_{\tilde{q}})^2$. The largest contributions originate from
the terms proportional to $A^{B_{s,d},(\tilde{g})}_2$ and
$A^{B_{s,d},(\tilde{g})}_3$, i.e. the terms associated with the $\delta^d_{LL}$ and
$\delta^d_{RR}$, cf. Eq.~\eqref{MPSUSY}. The contributions from the
$LR$-type mass insertion parameters, proportional to
$A^{B_{s,d},(\tilde{g})}_{4,5}$,  are negligible. 
The maximum effect of the gluino-squark box diagrams is obtained when
$x=(M_{1/2}/m_0)^2$ and $y$ are smaller than one, with the
$(\delta^d_{LL(RR)})^2_{i3}$ and $(\delta^d_{LL})_{i3}(\delta^d_{RR})_{i3}$
terms interfering constructively.  
For relatively light $m_{\tilde{q}}$ around 2~TeV, 
$|A^{B_{s,d},(\tilde{g})}_1|_{\text{max}}\sim \mathcal{O}(10^{-12})$~GeV. 
Assuming furthermore $|(\delta^d_{LL})_{13}|\approx 10^{-3}$, 
$|(\delta^d_{LL})_{23}|\approx 2\times 10^{-2}$ and 
$|(\delta^d_{RR})_{13}|=|(\delta^d_{RR})_{23}|\approx10^{-2}$ 
(cf. Figure~\ref{Fig:down MIs}) as well as $y\approx 0.3$, we can use
Eqs.~(\ref{MP},\ref{MPSUSY}) together with Figure~\ref{Fig:Bmixing2} to
estimate the maximum gluino effects as
$|\Delta M^{(\tilde{g})}_{{B_{s}}}|_{\text{max}}\sim$ 
$\mathcal{O}(10^{-14})$~GeV
and 
$|\Delta M^{(\tilde{g})}_{{B_{d}}}|_{\text{max}}\sim$ 
$\mathcal{O}(10^{-15})$~GeV. This is about two orders of magnitude
smaller than the corresponding SM and experimental values.

\begin{figure}[t]
\minipage{0.32\textwidth}
  \includegraphics[width=\linewidth]{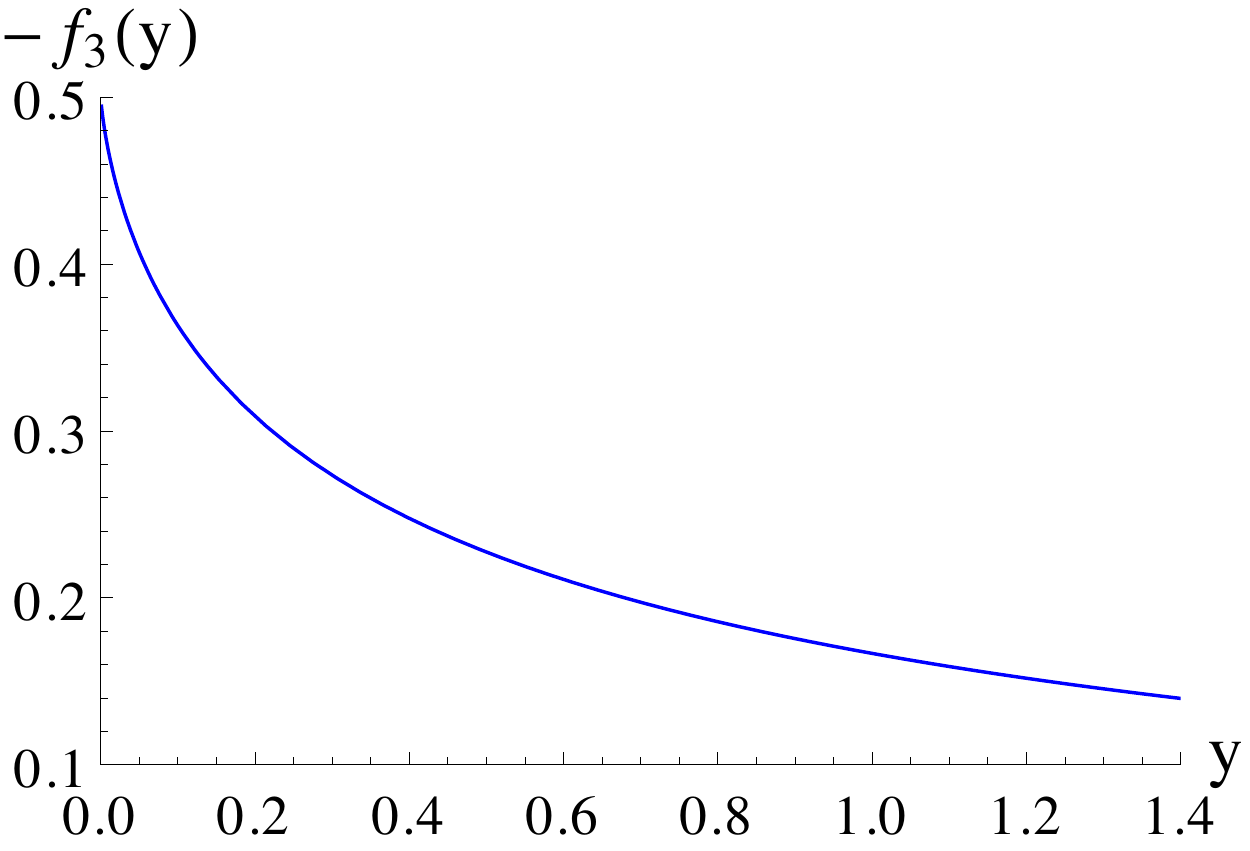}
\endminipage\hfill
\minipage{0.32\textwidth}
\includegraphics[width=\linewidth]{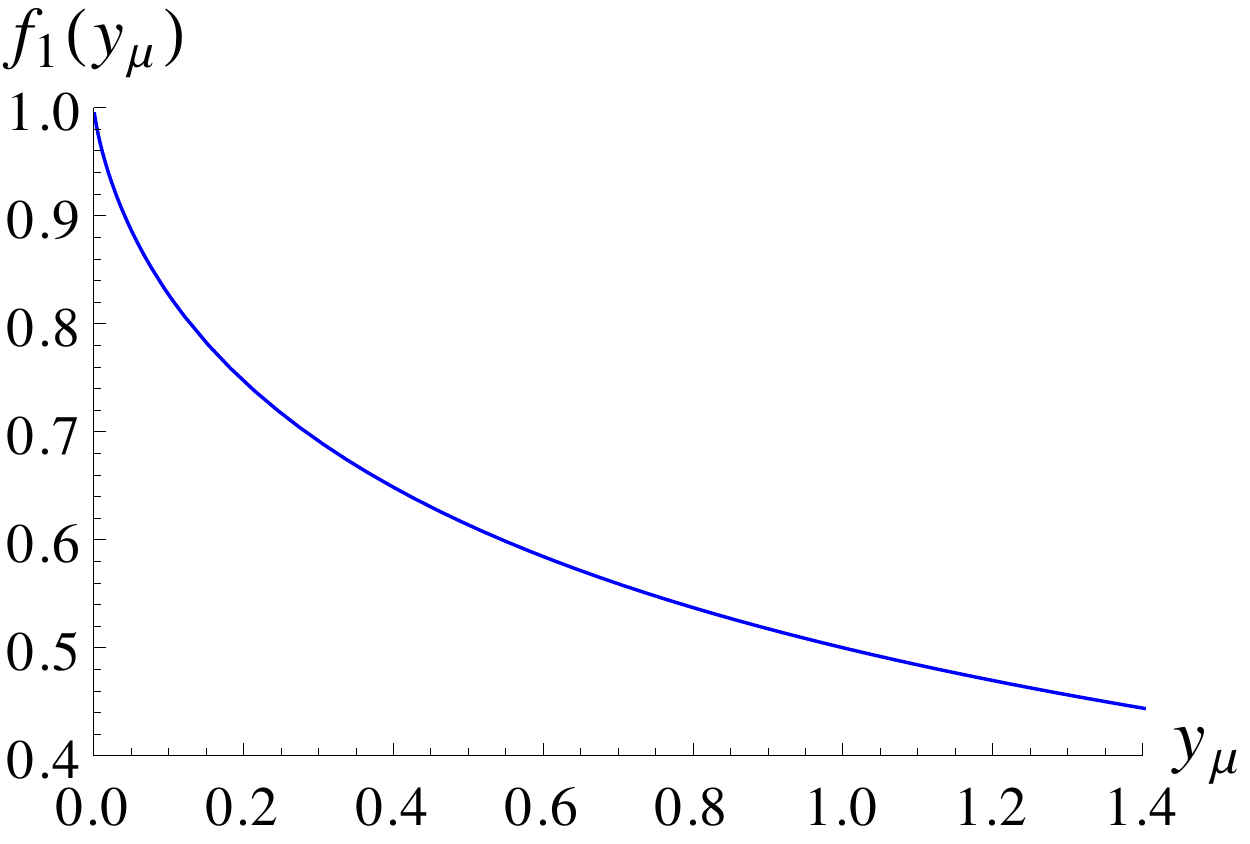}
\endminipage\hfill
\minipage{0.32\textwidth}
\includegraphics[width=\linewidth]{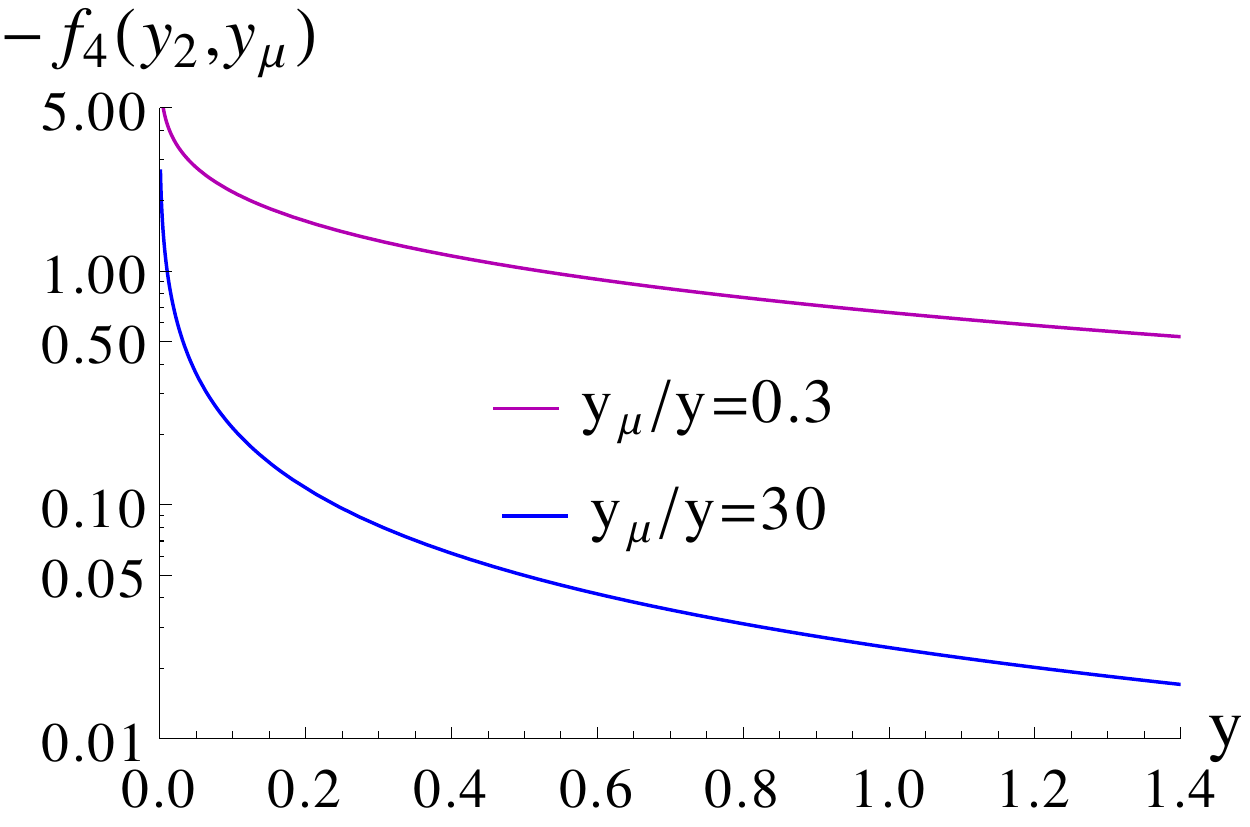}
\endminipage\hfill
\minipage{0.33\textwidth}
  \includegraphics[width=\linewidth]{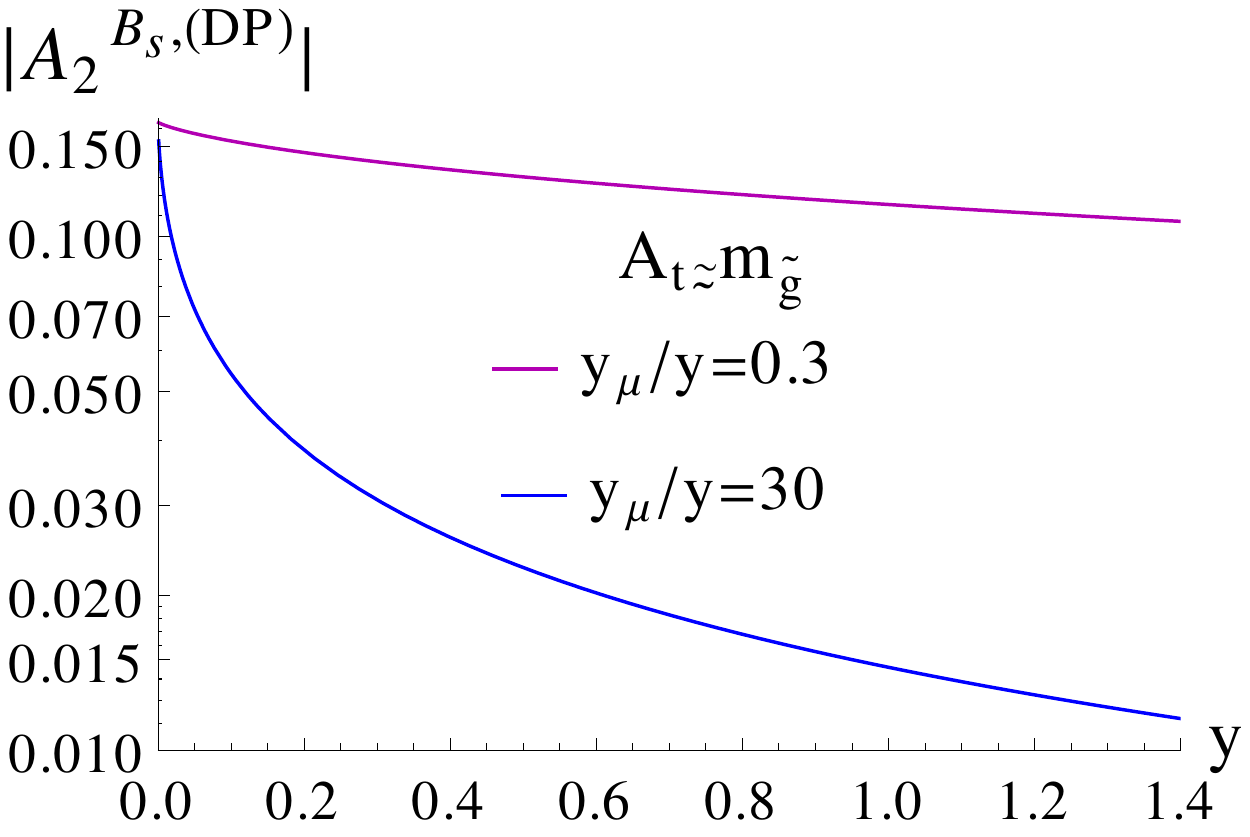}
\endminipage\hfill
\minipage{0.33\textwidth}
\includegraphics[width=\linewidth]{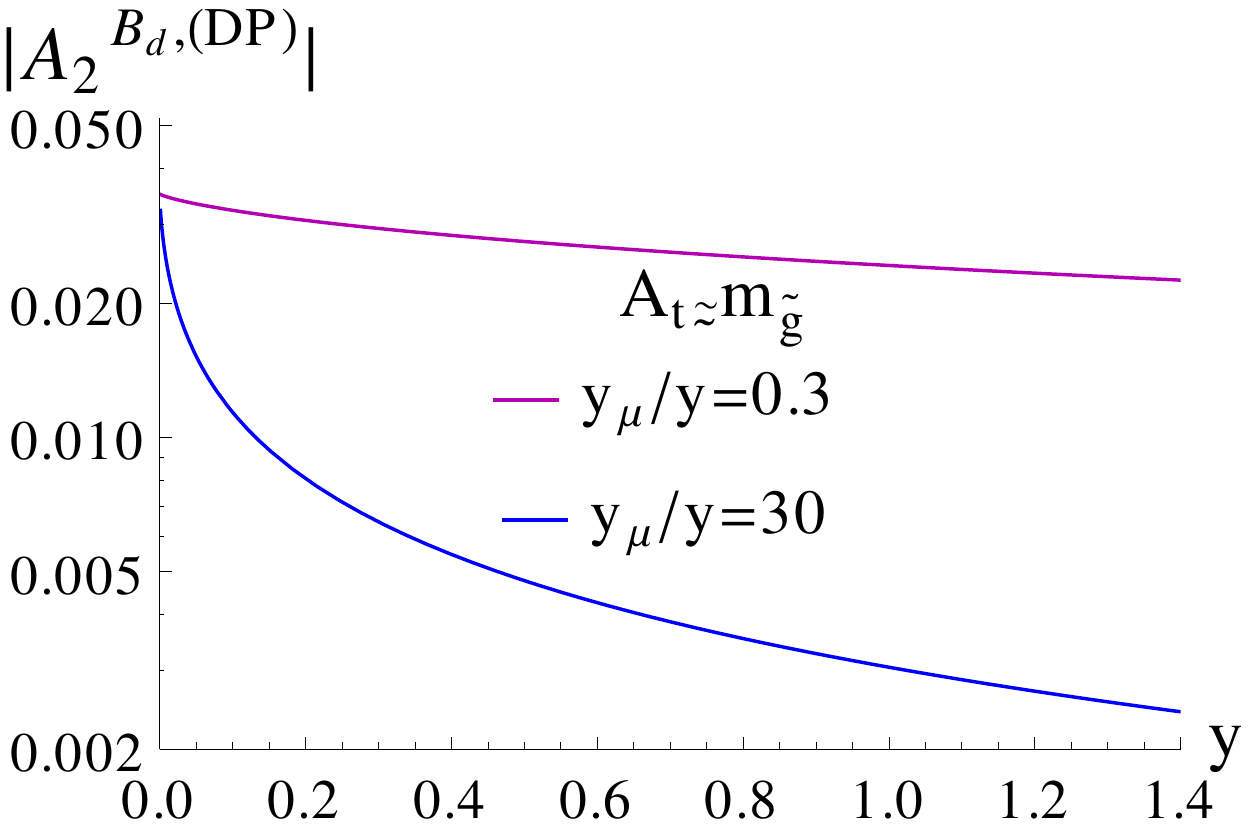}
\endminipage\hfill
\minipage{0.33\textwidth}
\includegraphics[width=\linewidth]{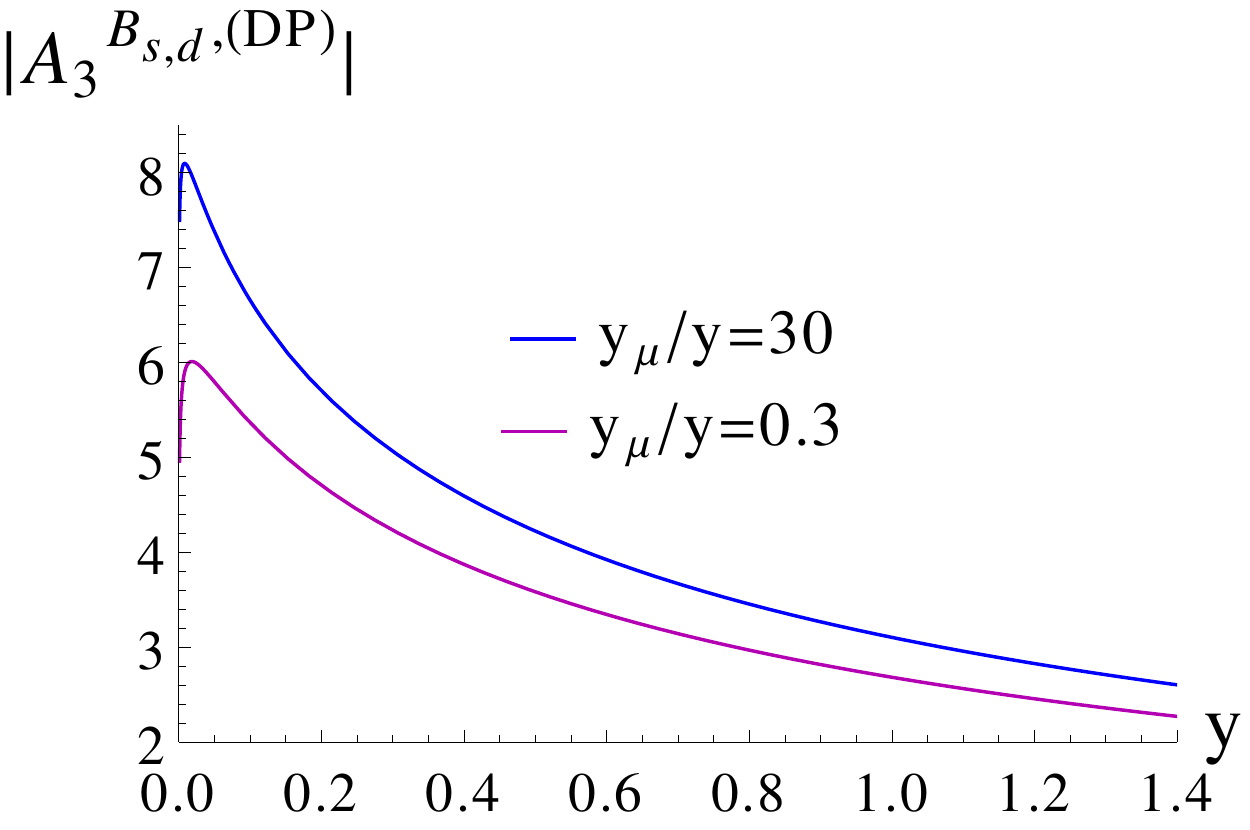}
\endminipage\hfill
\caption{The dependence of the loop functions as well as
  $|A^{B_q,\text{(DP)}}_{2,3}|$ appearing in Eq.~\eqref{BqDP} on
  $y=(m_{\tilde{g}}/m_{\tilde{q}})^2$, $y_\mu=(\mu/m_{\tilde{q}})^2$ and
  $y_2=(M_2/m_{\tilde{q}})^2\approx 0.11 \, y$.
The blue lines correspond to $y_\mu/y=30$ and
  the magenta ones to $y_\mu/y=0.3$. In the plots for $|A^{B_q,\text{(DP)}}_{2}|$, we
  have assumed that $A_t\approx m_{\tilde{q}}$.}
    \label{Fig:DPfunctions}
\end{figure}

For relatively large values of $t_\beta$ and a light CP-odd Higgs mass $M_A$, 
the contributions of the double penguin (DP) diagrams, which scale as
$t_\beta^4\,\mu^2/M_A^2$, become important. Considering diagrams with ($i$) two
gluino,  ($ii$) one gluino and one Higgsino and  ($iii$) one gluino and one
Wino loops, the associated part of $M^{B_q}_{12}$ can be approximated
by~\cite{AnatomyandPhenomenology}
\begin{eqnarray}
 M^{B_q,\text{(DP)}}_{12}&=&A^{B_q,\text{(DP)}}_1(\delta^d_{RR})_{3i}\,t_\beta^4\frac{\mu^2}{M_A^2}\Bigg{\{}A^{B_q,\text{(DP)}}_2
+(\delta^d_{LL})_{3i}A^{B_q,\text{(DP)}}_3\Bigg{\}},\label{BqDP}
\end{eqnarray}
where $i=1(2)$  for $q=d(s)$ and
\begin{eqnarray}
\nonumber A^{B_q,\text{(DP)}}_1&=&\frac{\alpha_s\,\alpha_2^2}{16\pi}\frac{M_{B_q}f^2_{B_q}}{m^2_{\tilde{q}}}\left(\frac{M_{B_q}}{m_b+m_q}\right)^2\frac{2m_b^2}{3M_W^2}\,y\,f_3(y),\\
\nonumber A^{B_q,\text{(DP)}}_2&=&\frac{A_t}{m_{\tilde{g}}}\,\frac{m_t^2}{M_W^2}\,V_{tb}V_{tq}^*\,f_1(y_\mu),\\
A^{B_q,\text{(DP)}}_3&=&2\left(\frac{M_2}{m_{\tilde{g}}}\,f_4(y_2,y_\mu)-\frac{8}{3}\frac{\alpha_s}{\alpha_2}f_3(y)\right).\label{ADP}
\end{eqnarray}
$y_\mu=(\mu/m_{\tilde{q}})^2$ 
and $y_2=(M_2/m_{\tilde{q}})^2$ where the latter is related to $y=(m_{\tilde
  g}/m_{\tilde q})^2$ via the approximations of Eq.~\eqref{gauginos}.  
The loop functions $f_3(y)$, $f_1(y_\mu)$, $f_4(y_2,y_\mu)$ are 
given in Appendix~\ref{Loop Functions}. 
Their behaviour is sketched in Figure~\ref{Fig:DPfunctions}, along with that of
$|A^{B_q,\text{(DP)}}_{2,3}|$.  
For $|A_t|>500$ GeV, the dominant contribution to Eq.~\eqref{BqDP} comes from 
$A^{B_d,\text{(DP)}}_2$ in the $B_d$ sector, 
even for our maximum values of $|(\delta^d_{LL})_{13}|$, 
while for $B_s$, where $|(\delta^d_{LL})_{23}|$ assumes larger values 
(cf. Figure~\ref{Fig:down MIs}), the two terms in the curly brackets are comparable. 
For light average squark masses $m_{\tilde{q}}$ around 2~TeV,
$A^{B_q,\text{(DP)}}_1$ can reach values up to $\mathcal{O}(10^{-16})$~GeV, 
while $|(\delta^d_{RR})_{i3}|_{\text{max}}\approx 10^{-2}$ 
(cf. Figure \ref{Fig:down MIs}). 
Then, for $A_t\gtrsim m_{\tilde{g}}$ and $\mu\ll m_{\tilde{q}}$, 
$|A^{B_{s(d)},\text{(DP)}}_2|\approx\mathcal{O}(10^{-1(-2)})$, such that 
$|\Delta M^{B_{s(d)},\text{(DP)}}_{12}|\approx
2\times10^{-19(-20)}\times t_\beta^4\,\mu^2/M_A^2$~GeV, 
barring contributions from the $A^{B_q,\text{(DP)}}_3$ term. 
When $t_\beta$ takes its maximum value of 25 and $\mu\sim M_A$, the double
penguin contributions to $\Delta M_{B_q}$ increase to about an order of
magnitude above the gluino-box contributions, which is however still
significantly below the SM and experimental values.

\begin{figure}[t]
\minipage{0.5 \textwidth}
  \includegraphics[width=\linewidth]{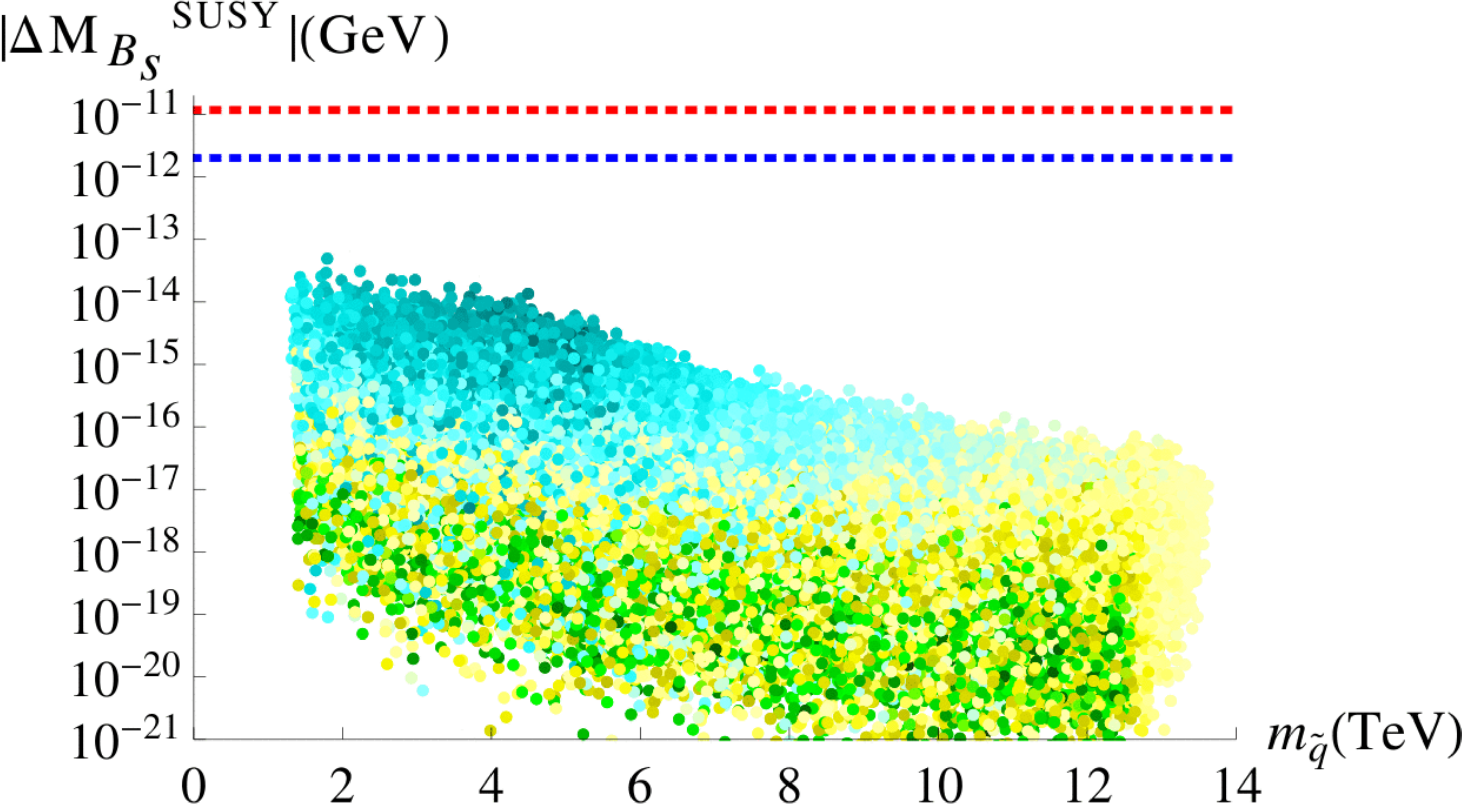}
\endminipage\hfill
\minipage{0.505 \textwidth}
  \includegraphics[width=\linewidth]{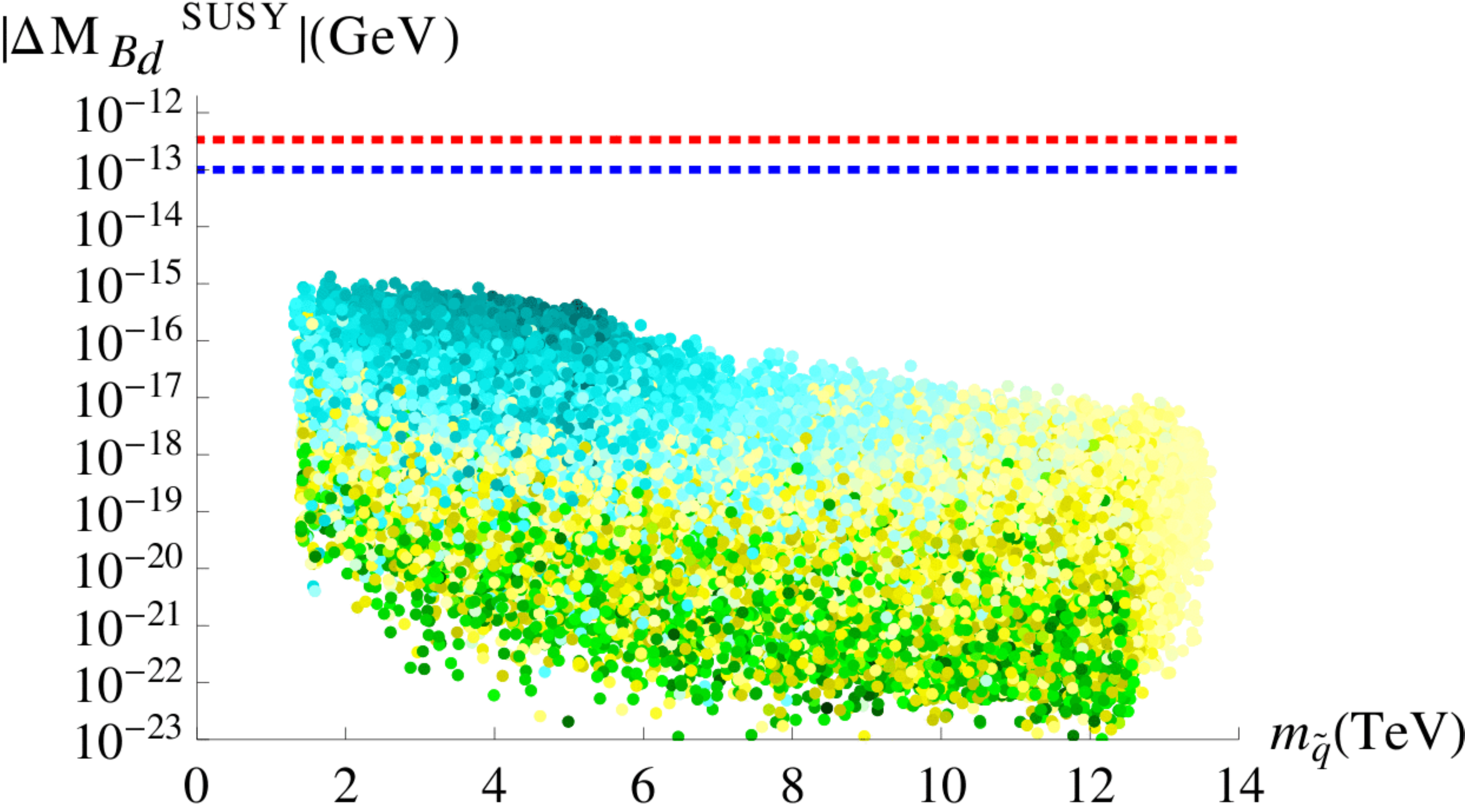}
\endminipage\hfill
\begin{center}
  \includegraphics[scale=0.491]{leg_x.pdf}
\end{center}
\caption{The absolute value of the gluino and double penguin contributions to 
$\Delta M_{B_{s(d)}}$ versus the average squark mass as defined in
Eq.~\eqref{msqav}. The colour coding corresponds to different values of
$x=(M_{1/2}/m_0)^2$.  The red dotted lines denote the experimental central
values of Eqs.~(\ref{DMBsexp},\ref{DMBdexp}), while the blue dotted lines
indicate the maximum allowed NP contributions according to
Eq.~\eqref{DMBsdNP}.} 
\label{Fig:DMB}
\end{figure}

Figure~\ref{Fig:DMB} shows the predicted SUSY contributions to the $B_q$ meson
mixings as produced in our scan. They are plotted against the average squark
mass defined in Eq.~\eqref{msqav} and lie below both the experimental
measurements (red dotted lines) and the NP limits (blue dotted lines) by at
least an order of magnitude. This result is in agreement with the findings in
Section~\ref{Numerics:Down-quark sector}, where we have compared our
predictions for the mass insertion parameters with existing limits in the
literature.

The effects of the complex down-type mass insertion parameters 
of the (23) and (13) sectors can be studied through the time dependent
CP asymmetries associated with the decays
$B_s\to J/\psi\,\phi$ and $B_d\to J/\psi\,K_S$. Focusing on the
mixing-induced CP asymmetries, we have~\cite{Aushev:2010bq}
\begin{eqnarray}
S_f&=&\frac{2\,\mathrm{Im}(\lambda_f)}{1+|\lambda_f|^2},
\end{eqnarray}
with
\begin{eqnarray}
\lambda_f&=&\frac{q}{p}\frac{\bar{\mathcal{A}}(\bar{B}_q\to
  f)}{\mathcal{A}(B_q\to f)},\qquad
\frac{q}{p}=\sqrt{\frac{M^{B_q*}_{12}-\frac{i}{2}\Gamma^{B_q*}_{12}}{M^{B_q}_{12}-\frac{i}{2}\Gamma^{B_q}_{12}}},
\end{eqnarray}
where $f$ denotes the final state of the decay and $\mathcal A$ is the
corresponding amplitude. 
As the absorptive part~$\Gamma^{B_q}_{12}$ of the $B_q$ meson mixing 
is much smaller than the dispersive one~$M^{B_q}_{12}$, i.e. 
$\Gamma^{B_q}_{12}\ll M^{B_q}_{12}$, we can approximate  
$q/p\approx \sqrt{M^{B_q*}_{12}/M^{B_q}_{12}}$. 
Then, the $\lambda_f$ factors associated with the decays
$B_s\to J/\psi\,\phi$ and $B_d\to J/\psi\,K_S$
take the form
\begin{eqnarray}
\nonumber \lambda_{J/\psi\phi}&=&e^{-i\phi_s},\quad \:\:~~~~\phi_s=-2\beta_s+\text{arg}\left(1+h_{B_s}e^{2i\sigma^{}_{B_s}}\right),\\
\lambda_{J/\psi K_S}&=&-e^{-i\phi_d},\quad ~~~\phi_d=2\beta+\text{arg}\left(1+h_{B_d}e^{2i\sigma^{}_{B_d}}\right),
\end{eqnarray}
where the parameters $h_{B_q}$ and $\sigma_{B_q}$ are defined in
Eq.~\eqref{hP_sigmaP}, while the SM phases $\beta_s$ and~$\beta$ 
can be found in Eqs.~(\ref{SMphases1},\ref{SMphases2}). 
The mixing-induced time dependent asymmetries can then be simply written as
\begin{eqnarray}
S_{J/\psi\phi}&=&-\sin(\phi_s),\quad ~~~~S_{J/\psi K_S}=\sin(\phi_d).
\end{eqnarray}
The current measurements are~\cite{HFAG}\footnote{LHCb recently published
their first measurements of 
$S_{J/\psi K_S}=0.746\pm0.030$~\cite{LHCb_SKs} in the limit of a vanishing direct CP
  asymmetry, i.e. 
$\frac{1-|\bar{\mathcal{A}}(\bar{B}_q\to J/\psi K_S)/
\mathcal{A}(B_q\to J/\psi K_S)|^2}{1+|\bar{\mathcal{A}}(\bar{B}_q\to J/\psi K_S)/
\mathcal{A}(B_q\to J/\psi K_S)|^2}=0$, 
thereby improving consistency with the SM expectation.}
\begin{eqnarray}
S_{J/\psi\phi}&=&0.015\pm 0.035,\quad ~~~~S_{J/\psi K_S}=0.682\pm 0.019, \label{Sexp}
\end{eqnarray}
while the SM expectations read~\cite{CKMFitter15}
\begin{eqnarray}
S_{J/\psi\phi}^\text{SM}=\sin(2\beta_s)=0.0365^{+0.0012}_{-0.0013},\quad~~~S_{J/\psi K_S}^\text{SM}=\sin(2\beta)=0.771^{+0.017}_{-0.041}.\label{SSM}
\end{eqnarray}
$S_{J/\psi\phi}^\text{SM}$ comes with a relatively small error, whereas
$S_{J/\psi K_S}^\text{SM}$ depends strongly on the value of $|V_{ub}|$, which
differs significantly when extracted via inclusive or exclusive decays, 
see e.g.~\cite{Buras13}, with the above data preferring the lower exclusive result. 
The value of $S_{J/\psi K_S}^\text{SM}$ quoted in Eq.~\eqref{SSM} has been
derived by averaging over inclusive and exclusive semileptonic determinations
of the relevant CKM elements and using the value of the CP-violating parameter
$\epsilon_K$, see Eq.~\eqref{eq:defepsK}, amongst the input parameters but not
the measurement of $\sin(2\beta)$ itself. 

Comparing Eq.~\eqref{Sexp} and Eq.~\eqref{SSM}, we observe that the NP
contributions to $S_{J/\psi\phi}$ and $S_{J/\psi K_S}$ can be as large as
$\sim100\%$ and $\sim10\%$ of the respective SM values. 
 In order to reach $10\%$ deviations, $h_{B_s}$ and $h_{B_d}$ should be larger
 than $\sim4\times 10^{-3}$ and $\sim 0.14$ respectively, corresponding to 
$|\Delta M^\text{(NP)}_{B_{s,d}}|\gtrsim 5\times 10^{-14}$. 
Here we have assumed  NP phases which maximise the effect.
In view of Figure~\ref{Fig:DMB}, we would expect a non-negligible contribution
to $S_{J/\psi\phi}$ in a small part of the parameter space. 
However, at leading order, $(\delta^d_{LL})_{23}$ and $(\delta^d_{RR})_{23}$
are real, cf. Eqs.~(\ref{dLL23Low},\ref{dRR23Low}). They only receive
non-trivial phase
factors at order~$\lambda^5$, suppressing the imaginary part of 
$\Delta M^\text{SUSY}_{B_{s}}$ by one power of $\lambda \approx 10^{-1}$ 
with respect to the real part. As a result, any deviation from
$S_{J/\psi\phi}^\text{SM}$ is only of the order of~$1\%$.
In the $B_d$ sector, $(\delta^d_{LL})_{13}$ and
$(\delta^d_{RR})_{13}$ are already complex at leading order in $\lambda$, 
cf. Eqs.~(\ref{dLL13Low},\ref{dRR13Low}). 
But as can be seen from Figure~\ref{Fig:DMB}, 
$|\Delta M^\text{SUSY}_{B_{d}}|_{\text{max}}\approx 10^{-15}$ is too small to
be relevant.
Even for $|\Delta M^\text{SUSY}_{B_{d}}|\approx 10^{-14}$, the maximum
deviation from $S_{J/\psi K_S}^\text{SM}$ would be $\sim 3\%$ at most.

In conclusion, our model would not be able to explain any persistent deviations
from SM expectations in observables related to $B$ meson mixing.


\subsubsection[$K-\bar{K}$]{$\boldsymbol{K-\bar{K}}$ mixing}

The SM contribution to the Kaon mixing reads~\cite{Buras13}
\begin{eqnarray}
\nonumber M^{K,\text{SM}}_{12}&=&\frac{G_F^2 M_{K}}{12 \pi^2}M_W^2\Big((V_{cs}V_{cd}^*)^2\eta_{cc}S_0(x_c)+(V_{ts}V_{td}^*)^2\eta_{tt}S_0(x_t)+\\
&+&2V_{cs}V_{cd}^*V_{ts}V_{td}^*\eta_{ct}S_0(x_c,x_t)\Big) f^2_{K}\hat{B}_K,
\end{eqnarray}
where $\eta_i$ are QCD factors, 
$\hat{B}_K$ denotes a perturbative parameter and 
$S_0(x_i\equiv \bar{m}_i^2 (\bar{m}_i)/M_W ^2)$ are the Inami-Lim
loop functions~\cite{Inami:1980fz}. 
From this, the SM value for the Kaon mass difference is numerically given
by~\cite{DeltaMKSM}
\begin{eqnarray}
\Delta M_K^\text{(SM)}=3.30(34)\times 10^{-15}\,\text{ GeV},
\end{eqnarray}
while the experimental measurement yields~\cite{DeltaMKexp} 
\begin{eqnarray}
\Delta M_K^\text{(exp)}=3.484(6)\times 10^{-15}\,\text{ GeV}.\label{DMKexp}
\end{eqnarray}
We therefore impose the constraint that the maximum allowed NP contribution
should be limited by
\begin{eqnarray}
\Delta M_K^\text{(NP)}\leq 5\times 
10^{-16}\,\text{ GeV}.\label{DMKNP}
\end{eqnarray}

For Kaon mixing, the relevant mass insertion parameters are those of the (12)
sector. Taking into account their $\lambda$-suppression, we can write the
gluino-box contribution to the mixing amplitude, given in Eq.~\eqref{MPSUSY},
in the schematic form
\begin{eqnarray}
\Delta M_{K}^{(\tilde{g})}&\propto&
\lambda^6\left(A_2^{K,(\tilde{g})}+A_3^{K,(\tilde{g})}\lambda+A_4^{K,(\tilde{g})}\lambda^4+A_5^{K,(\tilde{g})}\lambda^4\right).
\label{MKGapprox}
\end{eqnarray}
\begin{figure}[t]
 \centering
    \includegraphics[scale=0.6]{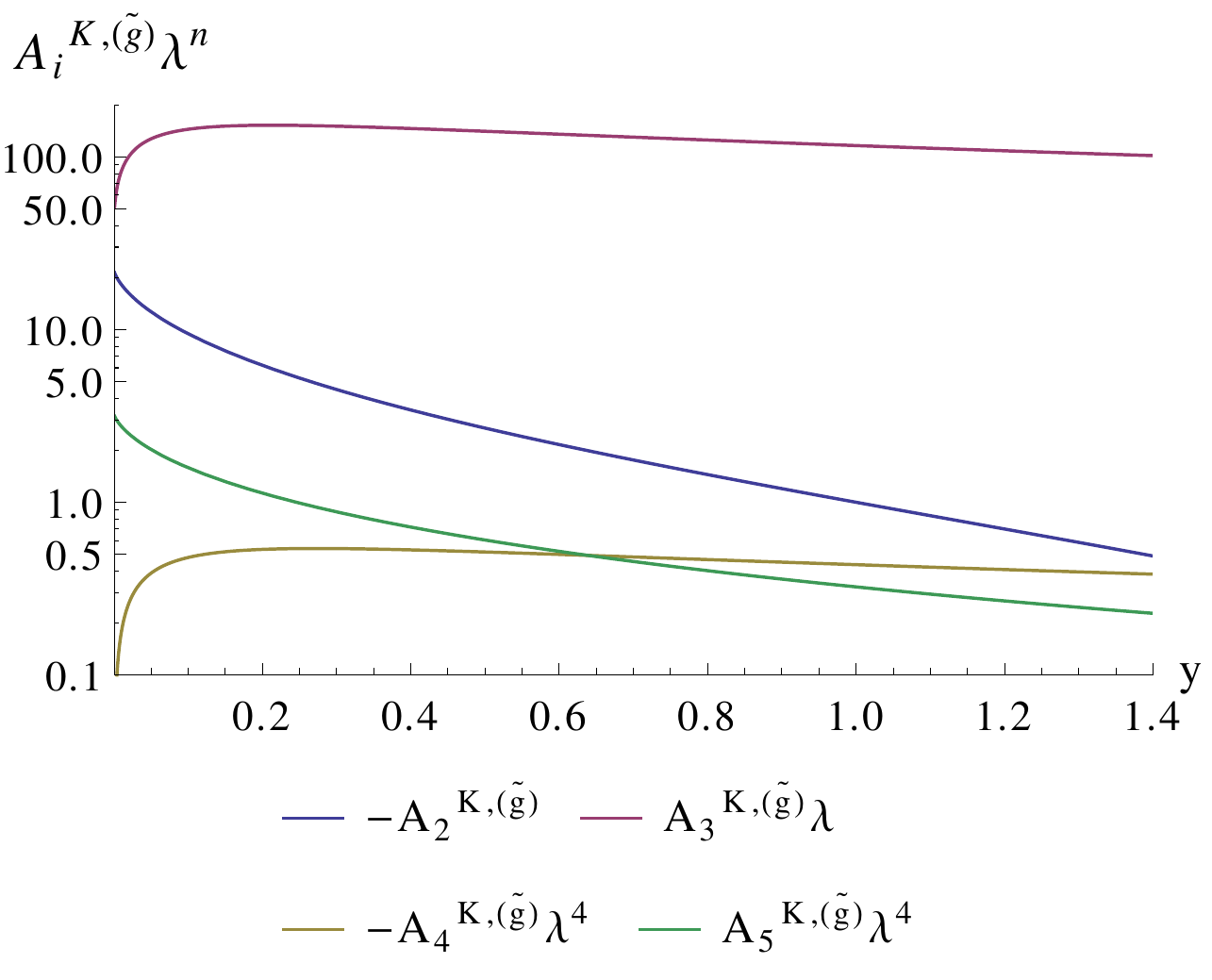}
\caption{The dependence of the individual contributions in Eq.~\eqref{MKGapprox}
  on $y=(m_{\tilde{g}}/m_{\tilde{q}})^2$. The average squark mass 
$m_{\tilde{q}}$ is defined in Eq.~\eqref{msqav} while the functions
  $A^{K,(\tilde{g})}_i$ can be found in Eq.~\eqref{APG}.}
    \label{Fig:Kmixing2}
\end{figure}%
Figure~\ref{Fig:Kmixing2} depicts the individual contributions as a
function of $y=(m_{\tilde{g}}/m_{\tilde{q}})^2$. It shows that the dominant
contribution originates from the term proportional to 
$A^{K,(\tilde{g})}_3$, i.e. the term proportional to
$(\delta^d_{LL})_{21}(\delta^d_{RR})_{21}$, see Eq.~\eqref{MPSUSY}.  
The effects of the $LR$-type $\delta$s, proportional to
$A^{K,(\tilde{g})}_{4,5}$, are negligible. 
Using Eqs.~(\ref{MP},\ref{MPSUSY}) together with Figure~\ref{Fig:Kmixing2}, 
we can estimate the maximum gluino contributions to $|\Delta M_K|$. 
Assuming $y\approx 0.3$, 
$A^{K,(\tilde{g})}_{1}\approx 10^{-13}$~GeV and 
$(\delta^d_{LL})_{21}\approx5\times 10^{-2}$, 
$(\delta^d_{RR})_{21}\approx 7\times 10^{-3}$ (cf. Figure~\ref{Fig:down  MIs}), 
we expect that 
$|\Delta M_K^{(\tilde{g})}|_{\text{max}}\approx 5\times10^{-14}$~GeV,
which is about one order of magnitude larger than the experimental
result of Eq.~\eqref{DMKexp}.

\begin{figure}[t!]
\minipage{0.482 \textwidth}
  \includegraphics[width=\linewidth]{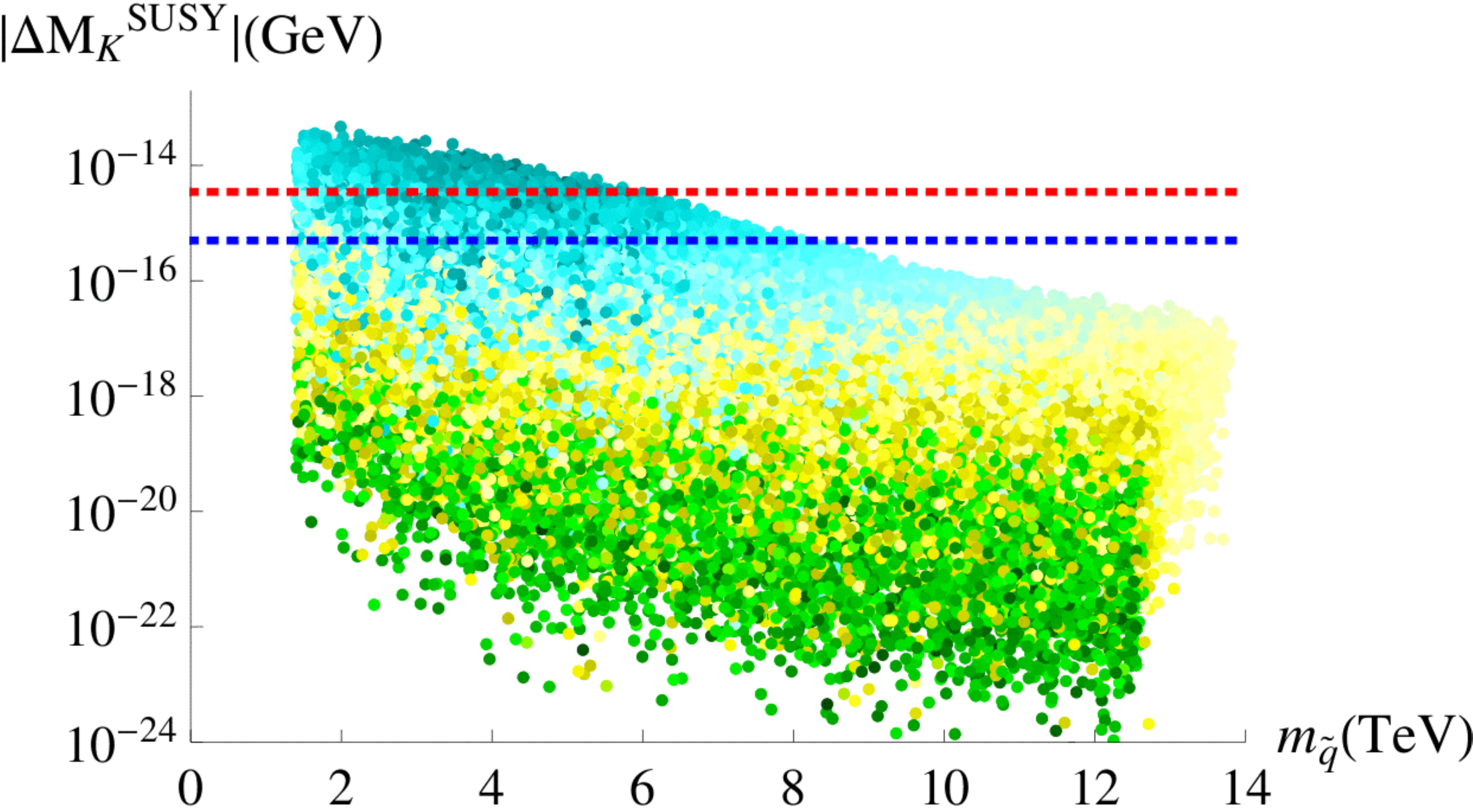}
\endminipage\hfill
\minipage{0.479 \textwidth}
  \includegraphics[width=\linewidth]{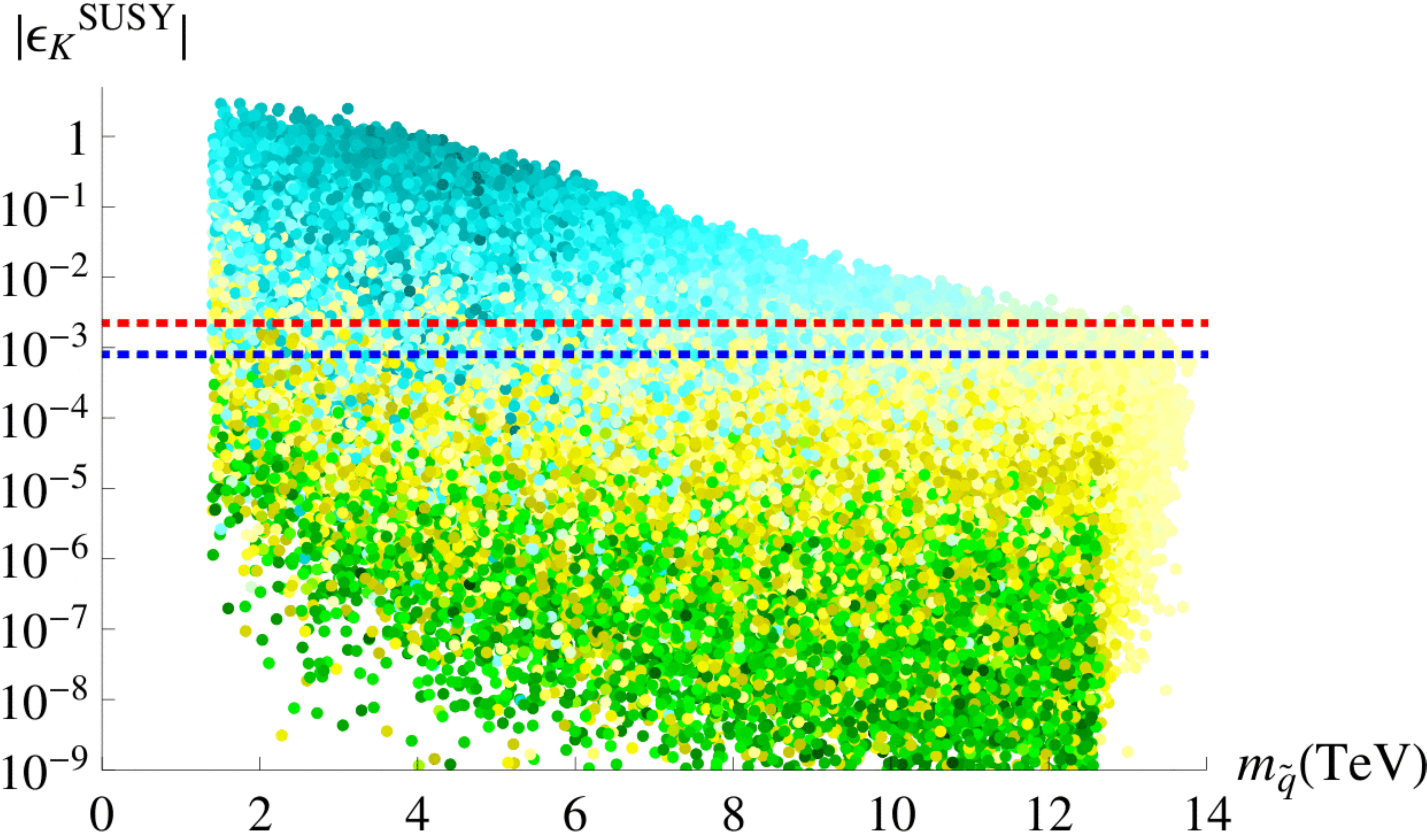}
\endminipage\hfill
\begin{center}
  \includegraphics[scale=0.48]{leg_x.pdf}
\end{center}
\minipage{0.482 \textwidth}
  \includegraphics[width=\linewidth]{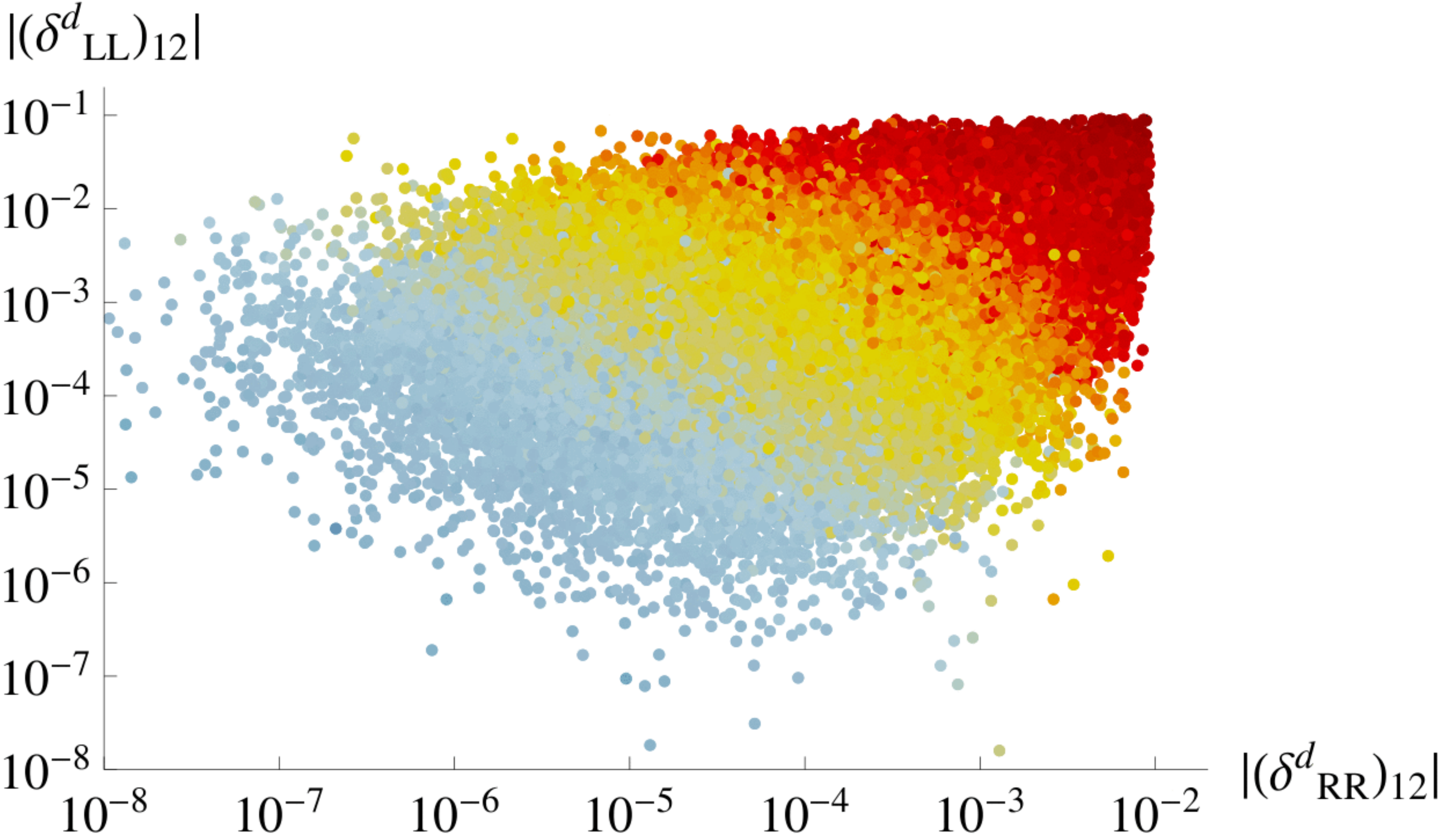}
\endminipage\hfill
\minipage{0.479 \textwidth}
  \includegraphics[width=\linewidth]{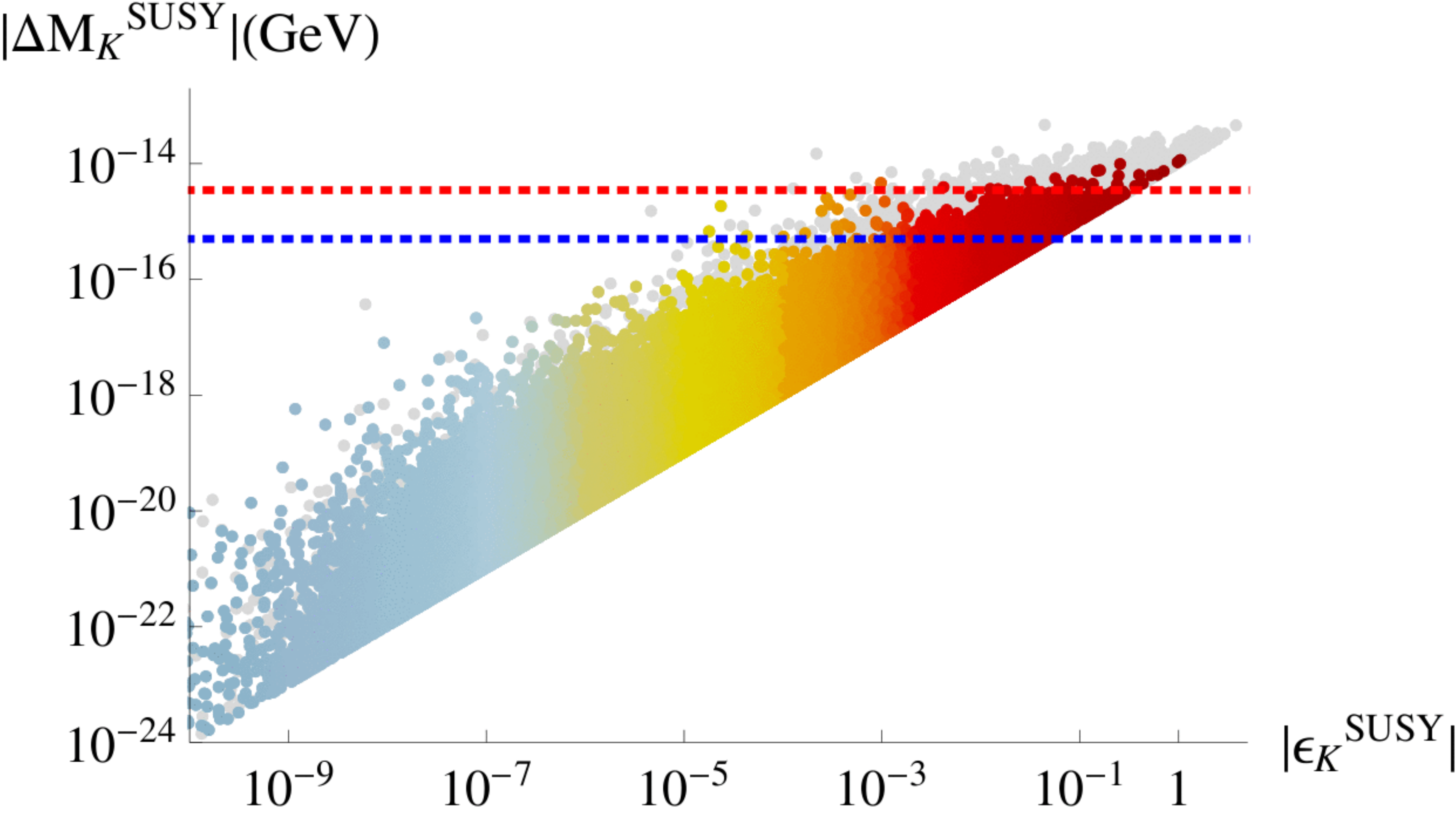}
\endminipage\hfill
\begin{center}
  \includegraphics[scale=0.49]{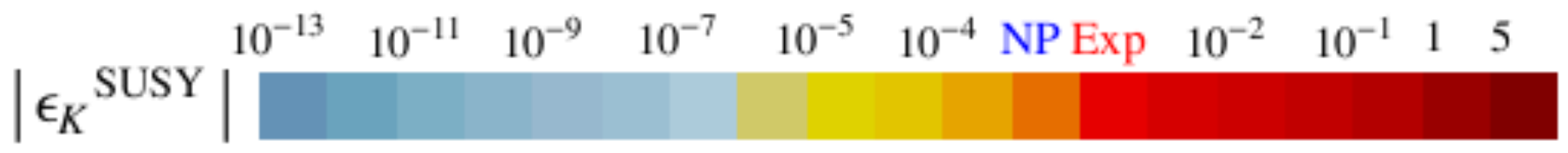}
\end{center}
\caption{Upper panels: the absolute value of SUSY contributions to $\Delta
  M_K$ (left) and $\epsilon_K$ (right) plotted against the average squark mass
  defined in Eq.~\eqref{msqav}, with the different colours corresponding to
  different values of $x=(M_{1/2}/m_0)^2$. 
Lower panels: the most important mass insertion parameters, relevant for 
$K$ mixing (left) with different colours representing the produced value of
$|\epsilon_K^{\text{SUSY}}|$; $|\Delta M_K^{\text{SUSY}}|$ versus
$|\epsilon_K^{\text{SUSY}}|$ (right), with the grey shaded points being
excluded by $BR(\mu\to e\gamma)$. The red dotted lines indicate the
experimentally observed values, while the blue dotted lines show the limits on NP
contributions.} 
\label{Fig:Kstuff}
\end{figure}

The double penguin (DP) contributions to $\Delta M_K$ arise at the level of
four mass insertions, by effectively generating the $(s\to d)$ transitions
through $(s\to b)$ followed by $(b\to d)$. The relevant part of
the mixing amplitude takes the form~\cite{AnatomyandPhenomenology}
\begin{eqnarray}
 M^{K,\,\text{(DP)}}_{12}&=&\frac{\alpha_s^2\,\alpha_2}{16\pi}M_{K}f^2_{K}\left(\frac{M_K}{m_s+m_d}\right)^2\frac{32m_b^2}{9M_W^2}\frac{t_{\beta}^2\,\mu^2}{M_A^2\,m^2_{\tilde{q}}}\,y\,(f_5(y))^2\times\\
&\times& (\delta^d_{LL})_{23}(\delta^d_{LL})_{31}(\delta^d_{RR})_{23}(\delta^d_{RR})_{31},\label{KDP}
\end{eqnarray}
with the loop function $f_5(y)$ given in Appendix~\ref{Loop Functions}. 
We find that this contribution is completely negligible, as it is proportional
to $\lambda^{14}$. 
The upper left panel of Figure~\ref{Fig:Kstuff} shows the combined 
gluino and DP SUSY contribution to $\Delta M_K$, as produced in our scan.
It can exceed the NP limit quoted in Eq.~\eqref{DMKNP} (blue dotted line)
for small values of $x$, even shooting above the experimental
value of Eq.~\eqref{DMKexp} (red dotted line) for $x\ll 1$.

We now turn to the CP-violating parameter $\epsilon_K$, defined as~\cite{Buras13}
\begin{eqnarray}
\epsilon_K=\frac{\kappa_\epsilon e^{i \varphi_{\epsilon}}}{\sqrt{2}\Delta M_K^{\text{exp.}}}\left(\mathrm{Im}(M_{12}^{K,\,\text{SM}})+\mathrm{Im}(M_{12}^{K,\,\text{SUSY}})\right),\label{eq:defepsK}
\end{eqnarray}
where the superweak phase\footnote{$\Delta\Gamma$ denotes the difference of the
  widths.} $\varphi_{\epsilon}=\text{arctan}(2\Delta 
M_{K}/\Delta\Gamma) = (43.52\pm 0.05)^\circ$~\cite{DeltaMKexp}, and the factor 
$\kappa_\epsilon=0.94\pm0.02$~\cite{kappaepsilon}
takes into account that $\varphi_{\epsilon} \neq \pi/4$  and includes long
distance contributions. 
The experimentally measured value of $\epsilon_K$ is~\cite{DeltaMKexp}
\begin{eqnarray}
\epsilon_K^\text{(exp)}&=&(2.228\pm0.011)\times 10^{-3}\times e^{i \varphi_{\epsilon}},\label{epsilonKexp}
\end{eqnarray}
while the SM prediction depends highly on the value of
$V_{cb}$~\cite{Buras13}. According to~\cite{epsilonK} and for the input set 
from the angle-only fit~\cite{AOF}, where the Wolfenstein parameters do not show
an unwanted correlation with $\epsilon_K$ and $\hat{B}_K$, one finds
\begin{eqnarray}
\nonumber
|\epsilon_K^\text{(SM)}|&=&2.17(24)\times 10^{-3}\text{ (inclusive $V_{cb}$)},\\
|\epsilon_K^\text{(SM)}|&=&1.58(18)\times 10^{-3}\text{ (exclusive $V_{cb}$)}.
\end{eqnarray}
We therefore demand that
\begin{eqnarray}
|\epsilon_K^\text{(NP)}|&\leq&0.8\times 10^{-3}.\label{epsilonKNP}
\end{eqnarray}

The upper right panel of Figure~\ref{Fig:Kstuff} shows the absolute value of
our predicted SUSY contribution to $\epsilon_K$, plotted against the average
squark mass. 
We find that it can exceed the limit of Eq.~\eqref{epsilonKNP} by more than
three orders of magnitude when $x<1$.  
In view of Figure~\ref{Fig:down MIs}, we would not have expected such a big effect.
However, the limits on the mass insertion parameters used in
Section~\ref{Numerics:Down-quark sector}, only take into account one non-zero
mass insertion at a time.
As we have seen in this section, the dominant contribution to the Kaon mixing
amplitude stems from the multiple $\delta$ term 
$A^{K,(\tilde{g})}_3  (\delta^d_{LL})_{21}(\delta^d_{RR})_{21}$ 
(cf. Figure~\ref{Fig:Kmixing2}). The non-zero phase of the $RR$ parameter is
the source of our prediction of a large $|\epsilon_K^{\text{SUSY}}|$. 

The lower left panel of Figure~\ref{Fig:Kstuff} shows
$|\epsilon_K^{\text{SUSY}}|$ 
in the $|(\delta^d_{LL})_{12}|-|(\delta^d_{RR})_{12}|$ plane. It indicates
that for $|(\delta^d_{LL})_{12}|\sim 5\times10^{-2}$, i.e. towards the largest possible
value according to Figure~\ref{Fig:down MIs}, 
$|(\delta^d_{RR})_{12}|\lesssim 10^{-5}$ is required.
When $|(\delta^d_{RR})_{12}|$ takes its maximum value of 
$\sim 10^{-2}$, $|(\delta^d_{LL})_{12}|$ should stay below $\sim 10^{-4}$.

Finally, from the lower right panel of Figure~\ref{Fig:Kstuff} we observe that 
$\epsilon_K$ places stronger bounds on the mass insertion parameters than 
$\Delta M_K$. Due to the $SU(5)$ framework of our model there is a correlation
between the $\delta$ parameters relevant in Kaon mixing and the ones 
that enter the branching ratio of $(\mu\to e\gamma)$. 
Denoting the points excluded by $BR(\mu\to e\gamma)$ with a grey shade 
reveals that there still remains a small area of parameter space which is excluded
by $\epsilon_K$.


\subsection[$BR(b\to s\gamma)$]{\label{sec:bsgamma}$\boldsymbol{BR(b\to s\gamma)}$}

We now consider the gluino contribution to the branching ratio of $b\to
s\gamma$. In terms of the relevant mass insertion parameters it is given
by~\cite{Gabbiani:1996hi} 
\begin{equation}
BR(b\to s\gamma)=\frac{\alpha_s^2\,\alpha}{81\pi^2m^4_{\tilde{q}}}m_b^3\tau_B\Big(|m_b\,M_3(y)(\delta^d_{LL})_{23}+m_{\tilde{g}}\,M_1(y)(\delta^d_{LR})_{23}|^2+L\leftrightarrow R \Big),~~
\end{equation}
where the loop functions $M_1(y),~M_3(y)$ are defined in 
Appendix~\ref{Loop Functions}, $\tau_B$ denotes the mean life of the $B$ meson
and $y=(m_{\tilde{g}}/m_{\tilde{q}})^2$. 
This observable does not constrain our parameter space.
Even for squark masses as low as 100~GeV and $y=1$, the $LL$ and $RR$
mass insertion parameters would only need to be smaller than 0.4 to be
consistent with the current experimental value of~\cite{HFAG}
\begin{eqnarray}
BR(B\to X_s\gamma)=(3.43\pm0.21\pm0.07)\times 10^{-4},
\end{eqnarray}
which is in good agreement with the SM prediction~\cite{Misiak:2006zs}.
Similarly, the chirality flipping mass insertion parameters would need to be
smaller than $3\times 10^{-3}$.
In our scan we find, cf. Figure~\ref{Fig:down MIs},
$({\delta^d_{LL}})_{23} \lesssim 10^{-2}$, 
$({\delta^d_{RR}})_{23} \lesssim 10^{-2}$, 
$({\delta^d_{LR}})_{23} \lesssim 10^{-5}$ and 
$({\delta^d_{RL}})_{23} \lesssim 10^{-6}$. Taking into account the squark mass
dependence and the fact that our scan excludes such light squarks, we 
have found that our model predicts a contribution to $BR(b\to s\gamma)$ which
is at least three orders of magnitude below the experimental measurement.



\subsection[$BR(B_{s,d}\to \mu^+\mu^-)$]{\label{sec:Bsdtomumu}$\boldsymbol{BR(B_{s,d}\to \mu^+\mu^-)}$}

The most recent SM predictions for the branching ratios of
$B_{s,d}\to\mu^+\mu^-$ are given by~\cite{BmumuSM}
\begin{eqnarray}
\nonumber BR(B_s\to\mu^+\mu^-)^\text{(SM)}&=&(3.65\pm 0.23)\times 10^{-9},\\
BR(B_d\to\mu^+\mu^-)^\text{(SM)}&=&(1.06\pm0.09)\times 10^{-10},
\end{eqnarray}
while the averages of the CMS and LHCb collaborations read~\cite{BmumuExp}
\begin{eqnarray}
\nonumber BR(B_s\to\mu^+\mu^-)^\text{(exp.)}&=&2.8^{+0.7}_{-0.6}\times 10^{-9},\\
BR(B_d\to\mu^+\mu^-)^\text{(exp.)}&=&3.9^{+1.6}_{-1.4}\times 10^{-10}.
\end{eqnarray}
The $B_d$ sector therefore still allows for rather large relative deviations from the
SM expectations. In the case of $B_s$ the experimental measurement yields a
value which is slightly lower than the SM prediction.\footnote{The calculations 
in~\cite{BmumuSM} have been performed using the inclusive value of 
$|V_{cb}|$. Working with the exclusive one would result in a lower central 
value of $BR(B_s\to\mu^+\mu^-)^\text{(SM)}=3.1\times 10^{-9}$ 
which fully agrees with the data~\cite{BurasZeptouniverse}.}
We therefore quote the allowed room for contributions from new physics as
\begin{eqnarray}
\nonumber BR(B_s\to\mu^+\mu^-)^\text{(NP)}&\leq&1.68\times10^{-9},\\
BR(B_d\to\mu^+\mu^-)^\text{(NP)}&\leq&4.53\times 10^{-10}.\label{BmumuNP}
\end{eqnarray}

The chargino and gluino contributions to the branching ratio of 
$B_{s,d}\to \mu^+\mu^-$ can be expressed as~\cite{AnatomyandPhenomenology}
\begin{eqnarray}\label{Bmumu}
BR(B_{q}\to \mu^+\mu^-)&=&\frac{\tau_{B_{q}}\,f_{B_q}^2\,M_{B_q}^3}{32\pi}\sqrt{1-4\frac{m_\mu^2}{M^2_{B_q}}}\times\\
\nonumber &\times&\Bigg\{
\left|\mathcal{A}^{B_q}_1\Bigg{[}\mathcal{A}^{B_q}_2-\frac{\alpha_s}{\alpha_2}f_3(y)\left((\delta^d_{LL})_{i3}-(\delta^d_{RR})_{i3}\right)\Bigg{]}\right|^2
\left(1-4\frac{m_\mu^2}{M^2_{B_q}}\right)\\
\nonumber &+&\left|2\frac{m_\mu}{M_{B_q}}C_{10}^\text{SM}+\mathcal{A}^{B_q}_1\Bigg{[}\mathcal{A}^{B_q}_2-\frac{\alpha_s}{\alpha_2}f_3(y)\left((\delta^d_{LL})_{i3}+(\delta^d_{RR})_{i3}\right)\Bigg{]}\right|^2\Bigg\},
\end{eqnarray}
where
\begin{eqnarray}
\nonumber \mathcal{A}^{B_q}_1&=&\alpha_2^2\,t_\beta^3\frac{M_{B_q}\,m_\mu}{4M_W^2}\frac{m_{\tilde{g}}\,\mu}{M_A^2\,m^2_{\tilde{q}}},~~~~~
\mathcal{A}^{B_q}_2=\frac{m_t^2}{M_W^2}\frac{A_t}{m_{\tilde{g}}}V_{tb}V_{tq}^*f_1(y_\mu)+\frac{M_2}{m_{\tilde{g}}}(\delta^u_{LL})_{i3}\,f_4(y_2,y_\mu),\\
C_{10}^\text{SM}&=&\frac{\alpha_2}{4\pi}\frac{4G_F}{\sqrt{2}}V_{tb}V_{tq}^*Y_0(x_t),~~~~~Y_0(x)=\frac{x}{8}\left(\frac{x-4}{x-1}+\frac{3x}{(x-1)^2}\ln(x)\right),
\end{eqnarray}
with $x_t={m_t^2}/{M_W^2}$ and $i=1(2)$ for $q=d(s)$ . 
The loop functions $f_1(y_\mu)$, $f_3(y)$ and
$f_4(y_2,y_\mu)$ are the ones which appear in the double penguin contributions
to $B_q$ mixing in Section~\ref{Bmixing}. 
With $C_{10}^\text{SM}=0$ and $A_t\gtrsim 100$~GeV, the dominant contribution
to Eq.~\eqref{Bmumu} originates from the flavour blind term of 
$\mathcal{A}_2^{B_q}$, such that we can make the approximation 
\begin{eqnarray}
BR(B_{s(d)}\to \mu^+\mu^-)&\approx&\mathcal{O}\Bigg(\frac{6\times
  10^{-6}(1\times 10^{-7})\text{GeV}^4}{m_{\tilde{q}}^4}t_{\beta}^6\,\frac{A_t^2\,\mu^2}{M_A^4}f^2_1(y_\mu)\Bigg)\label{Bmumuapprox}.
\end{eqnarray}
Then, for $|A_t\,\mu|/M_A^2\approx\mathcal{O}(1)$, 
$m_{\tilde{q}}\approx 2$~TeV, 
$t_\beta\approx25$ and  $f_1(y_\mu)$ receiving its maximum value of order one
(cf. Figure~\ref{Fig:DPfunctions}), we expect $BR(B_{s(d)}\to
\mu^+\mu^-)\approx\mathcal{O}(10^{-10(-12)})$.
\begin{figure}[t]
\minipage{0.48 \textwidth}
  \includegraphics[width=\linewidth]{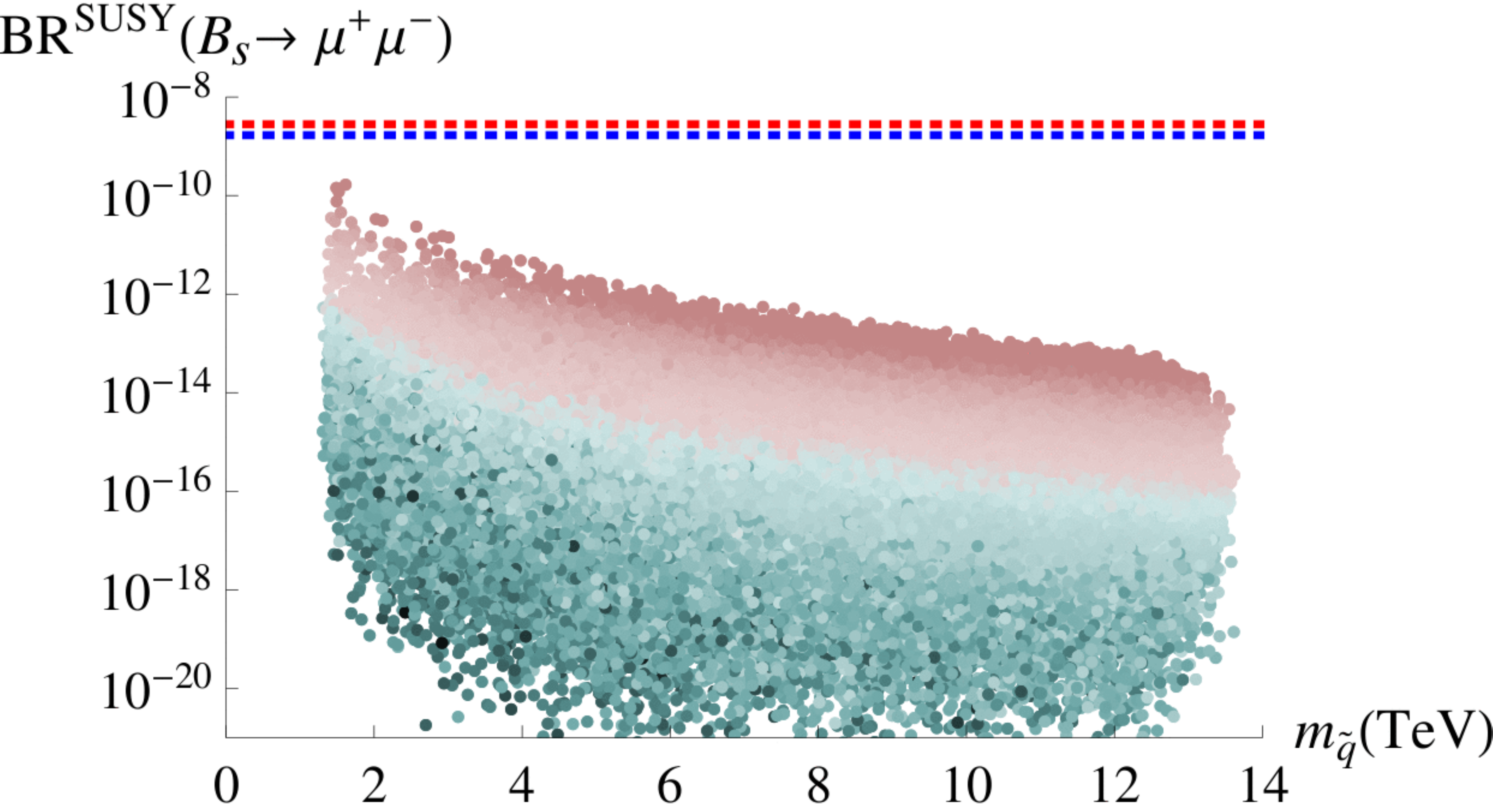}
\endminipage\hfill
\minipage{0.485 \textwidth}
  \includegraphics[width=\linewidth]{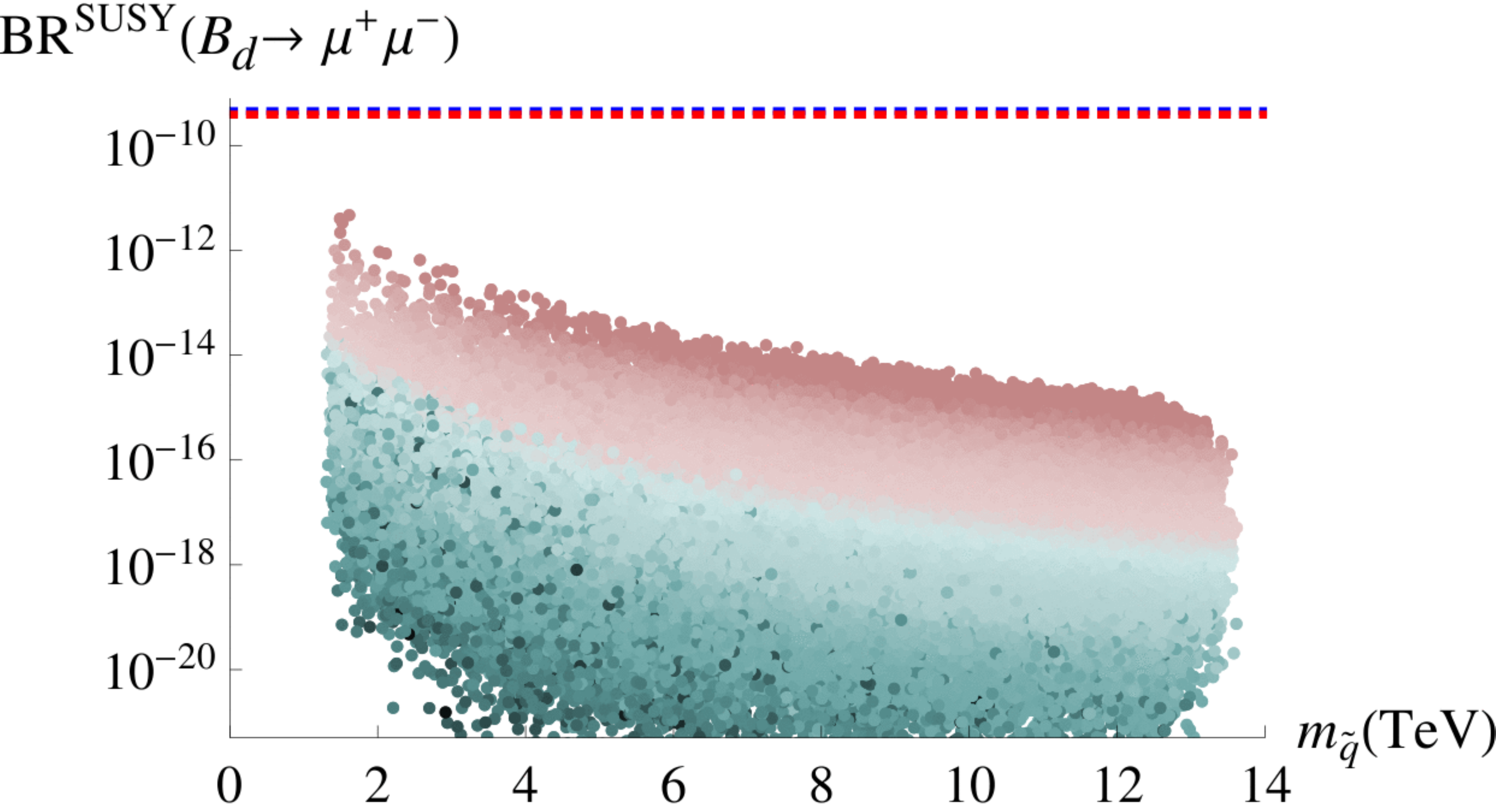}
\endminipage\hfill
\begin{center}
  \includegraphics[scale=0.23]{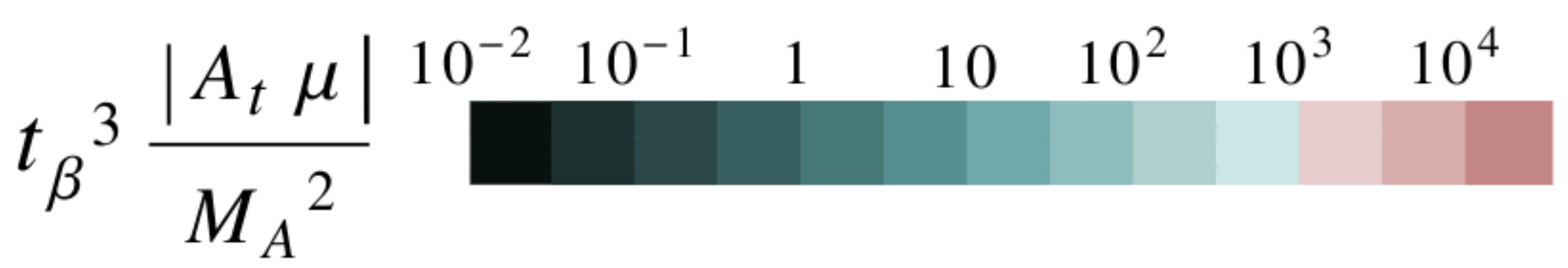}
\end{center}
\caption{The SUSY contributions to the branching ratios of
  $B_{q}\to\mu^+\mu^-$ versus the average squark mass $m_{\tilde{q}}$, defined
  in Eq.~\eqref{msqav}. The red dotted lines denote the experimental
  measurements, while the blue dotted lines indicate the maximum NP contributions.}
\label{Fig:Bmumu}
\end{figure}

In Figure~\ref{Fig:Bmumu}, we plot our predicted SUSY contributions to the
branching ratios of $B_{q}\to\mu^+\mu^-$ against the average squark 
mass $m_{\tilde{q}}$, defined in Eq.~\eqref{msqav}. 
The red dotted lines denote the experimental measurements, while the blue ones
correspond to the limits for the NP contributions as given in 
Eq.~\eqref{BmumuNP}. In both sectors, $B_s$ and $B_d$, our maximum predictions
fall about an order of magnitude  below these limits.\footnote{As 
discussed in~\cite{DeBruyn:2012wj} and also in~\cite{Buras:2012ru}, the theory
prediction in Eq.~\eqref{Bmumu} should take into account the large
width difference between the mass eigenstates of the $B_s$ system. This
correction enhances the corresponding branching ratio by about 10\%. Given the
smallness of the new physics contribution in our model, it does, however, not
change our results significantly.}



\subsection[Neutron and $^{199}$Hg EDMs]{\label{sec:hadrEdM}Neutron and $\boldsymbol{^{199}}$Hg EDMs}

CP-violating effects in the quark sector can manifest themselves through the
quark EDMs as well as the quark Chromo Electric Dipole Moments (CEDMs). 
The gluino contributions read~\cite{Hisano04,Hisano09,AnatomyandPhenomenology}
\begin{eqnarray}
\left\{\frac{d_{q_i}}{e},d_{q_i}^C\right\}&=&\frac{\alpha_s}{4\pi}\frac{m_{\tilde{g}}}{m^2_{\tilde{q}}}\mathrm{Im}
\left[(\delta^q_{LL})_{ik}(\delta^q_{LR})_{kj}(\delta^q_{RR})_{ji}\right]\left\{Q_{q}\mathcal{F}_{q}(y),\mathcal{F}^C_{q}(y)\right\},\label{QEDMs}
\end{eqnarray}
with
\begin{eqnarray}
\mathcal{F}_{q}(y)&=&-\frac{8}{3}N_1(y),\qquad
\mathcal{F}_{q}^C(y)=\left(\frac{1}{3}N_1(y)+3N_{2}(y)\right),
\end{eqnarray}
where $Q_q$ denotes the electric charge of quark $q$ and the loop functions
$N_1(y)$, $N_{2}(y)$, with $y=(m_{\tilde{g}}/m_{\tilde{q}})^2$, are given in
Appendix~\ref{Loop Functions}. 
As the first generation squarks dominate Eq.~\eqref{QEDMs}, we use the average
squark masses 
\begin{eqnarray}
m_{\tilde{u}}=\sqrt{m_{\tilde{u}_{LL}}m_{\tilde{u}_{RR}}},\qquad
m_{\tilde{d}}=\sqrt{m_{\tilde{d}_{LL}}m_{\tilde{d}_{RR}}},\label{msqavdn}
\end{eqnarray}
with $m_{\tilde{q}_{LL(RR)}}$ given in Eqs.~(\ref{uLR},\ref{dLR}).

Similar to the case of the electron EDM, we consider the most general scenario
where the phases of the soft trilinear sector are different from the
corresponding Yukawa ones. Then the dominant contributions of Eq.~\eqref{QEDMs} 
arise from the single mass insertions with $i=j=k=1$,\\[-7mm]
\begin{eqnarray}
\mathrm{Im}\left[(\delta^u_{LR})_{11}\right]\propto \mathrm{Im}\left[\tilde{a}^u_{11}\right]\lambda^8,\qquad
\mathrm{Im}\left[(\delta^d_{LR})_{11}\right]\propto \mathrm{Im}\left[\tilde{a}^d_{11}\right]\lambda^6,\label{ImLR}
\end{eqnarray}
where $\tilde{a}^f_{ij}$ is defined in Eq.~\eqref{adt}. The double and triple
mass insertions start contributing at orders $\lambda^{12}$ and $\lambda^8$ for the up and down quark (C)EDMs, respectively.

If, however, the phases of the soft trilinear and Yukawa sectors are aligned,
$\tilde{a}^f_{ij}$ is real. In the case of the up quark sector, one should
then check\footnote{We have truncated our expansion at the order of $\lambda^8$.} 
whether the NLO corrections to $\mathrm{Im}\left[(\delta^u_{LR})_{11}\right]$ 
also vanish, before assuming that the term
$\mathrm{Im}\left[(\delta^u_{LL})_{13}(\delta^u_{LR})_{33}
(\delta^u_{RR})_{31}\right]\propto \sin(4\theta^d_2-\theta^d_3)\lambda^{12}$ 
dominates. 
The situation in the down sector is such that the NLO correction to 
$(\delta^d_{LR})_{11}$ gives a non-vanishing contribution to the
(C)EDMs. Explicitly, we find 
$\mathrm{Im}\left[(\delta^d_{LR})_{11}\right]_\text{NLO}\propto 
\sin(4\theta^d_2+\theta^d_3)\lambda^7$, while the smallest contribution from
multiple mass insertions is 
$\mathrm{Im}\left[(\delta^d_{LL})_{{12}_{\text{NLO}}}(\delta^d_{LR})_{21}\right]
\propto\sin(\theta^d_2)\lambda^9$.

In order to compare the gluino contributions of our model according to
Eq.~\eqref{QEDMs} with the experimental limits, we take into account the RG
running from the SUSY scale down to the hadronic scale, using the LO results
of~\cite{dnRunning}, for 
$\alpha_s(\mu_S\approx 1 \text{TeV})\approx 0.089$ and 
$\alpha_s(\mu_H\approx 1 \text{GeV})\approx 0.358$~\cite{RunningStrong}. Then,
\begin{eqnarray}
\nonumber d^C_{q_i}(\mu_H)&\approx& 0.87 \,d^C_{q_i}(\mu_S),\\
\frac{d_{q_i}}{e}(\mu_H)&\approx& 0.38\, \frac{d_{q_i}}{e}(\mu_S)-0.39\,Q_{q}\,d^C_{q_i}(\mu_S),
\end{eqnarray}
with $d^{(C)}_{q_i}(\mu_S)$ as given in Eq.~\eqref{QEDMs}.

With these preparations, we can study the predictions for the neutron and the
$^{199}$Hg EDMs. 
Adopting the QCD sum rules approach, the neutron EDM at the renormalisation
scale $\mu=1$~GeV is given in terms of the QCD $\bar{\theta}$-term and the
quark (C)EDMs by~\cite{Hisano15} 
\begin{equation}
\frac{d_n}{e}=8.2\times 10^{-17}\, \text{cm}\,\bar{\theta}-0.12\,\frac{d_u}{e}+0.78\,\frac{d_d}{e}+\left(-0.3\,d^{C}_u+0.3\,d^{C}_d-0.014\,d^{C}_s\right),\label{dn}
\end{equation}
while the current experimental limit is~\cite{dnLimit}
\begin{eqnarray}
|d_n/e|\leq 2.9\times 10^{-26}\text{cm}\approx 1.47\times 10^{-12}\text{ GeV}^{-1}.\label{dn_limit}
\end{eqnarray}
The quark (C)EDMs can also be probed through measurements of the EDMs of
atomic systems, where $^{199}$Hg  provides the best upper limit amongst the
diamagnetic systems~\cite{dHgLimit} 
\begin{eqnarray}
|d_{\text{Hg}}/e|\leq 3.1\times 10^{-29}\text{cm}\approx 1.57\times 10^{-15}\text{ GeV}^{-1}.\label{dHdLimit}
\end{eqnarray}
However, large theoretical uncertainties in the atomic and in particular the
nuclear calculations prevent the extraction of bounds on $d^{(C)}_{q_i}$. 
Eq.~\eqref{dHdLimit} limits the nuclear Schiff moment as~\cite{HgCalc15}
\begin{eqnarray}
S_{\text{Hg}}\leq 1.45\times 10^{-12}|e|\,\text{fm}^3,\label{Slimit}
\end{eqnarray}
which, assuming  it is dominated by pion-nucleon interactions, can be
expressed as~\cite{EDMReview} 
\begin{eqnarray}
S_{\text{Hg}}=13.5\left(0.01\,\bar{g}^{(0)}_{\pi NN}+(\pm)0.02\,\bar{g}^{(1)}_{\pi NN}+0.02\, \bar{g}^{(2)}_{\pi NN}\right).\label{Sav}
\end{eqnarray}
In this equation, the $\bar{g}^{(i)}_{\pi NN}$ denote the pion-nucleon
couplings. Their coefficients in Eq.~\eqref{Sav} are the best fit values taken
from the review article~\cite{EDMReview}, which assesses the strengths and
weaknesses of  different, sometimes contradictory, nuclear calculations
provided in the literature. Combining 
Eqs.~(\ref{Slimit},\ref{Sav}) with the relation
\begin{eqnarray}
\bar{g}^{(1)}_{\pi NN}=2\times10^{-12}\left(d^C_u-d^C_d\right),
\end{eqnarray}
which was derived in~\cite{Pospelov}, it can be inferred that~\cite{HgCalc15}
\begin{eqnarray}\label{Hd_limit}
|(d^C_u-d^C_d)/e|\leq 2.8\times 10^{-26}\text{cm}\approx 1.42\times 10^{-12} \text{ GeV}^{-1}.
\end{eqnarray}
However, this bound only applies if the coefficient of 
$\bar{g}^{(1)}_{\pi NN}$ in Eq.~\eqref{Sav} takes its best fit value. In
principle, it could also be zero, in which case no bound on
$|(d^C_u-d^C_d)/e|$ could be extracted.

\begin{figure}[t]
\minipage{0.482 \textwidth}
  \includegraphics[width=\linewidth]{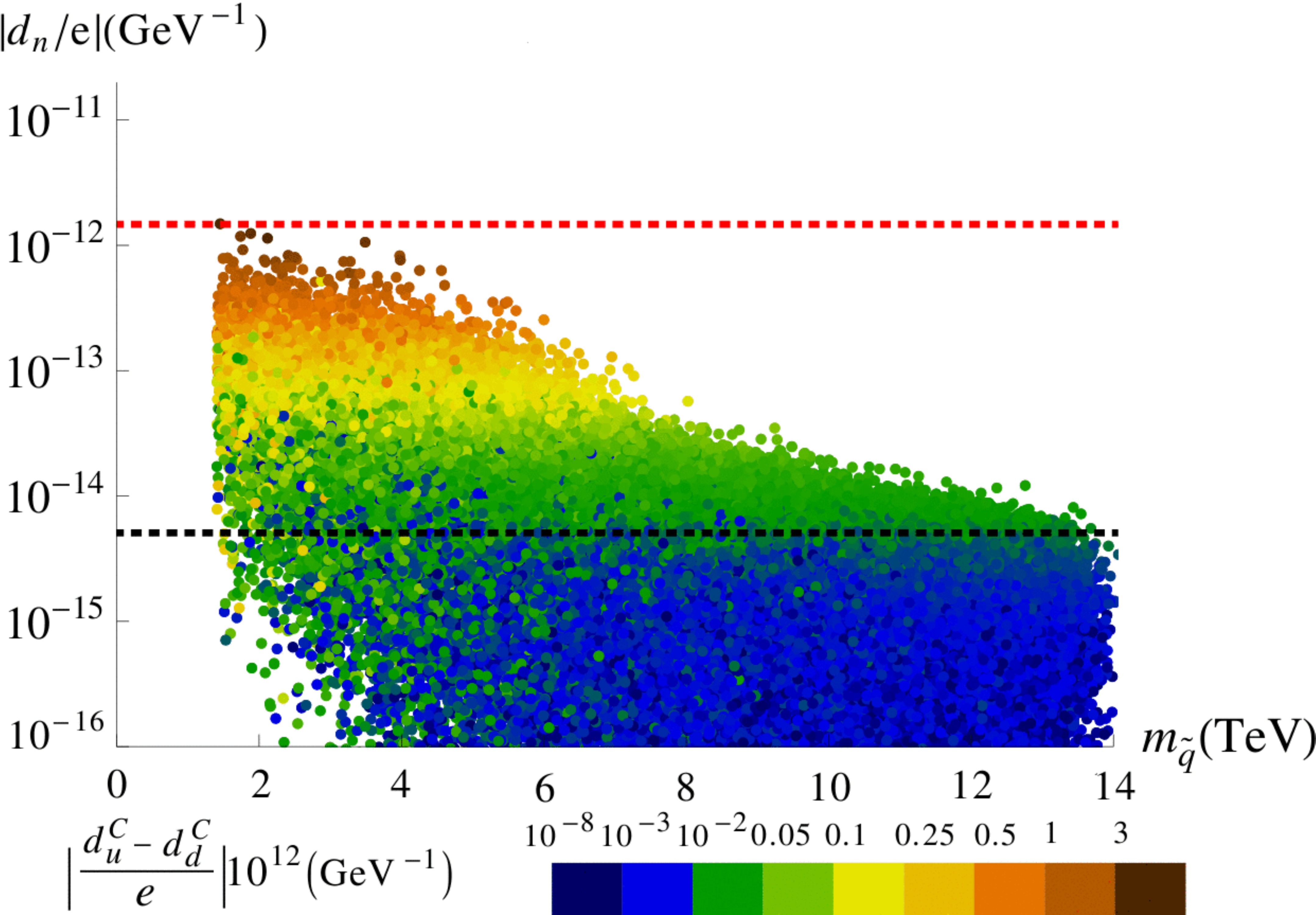}
\endminipage\hfill
\minipage{0.479 \textwidth}
  \includegraphics[width=\linewidth]{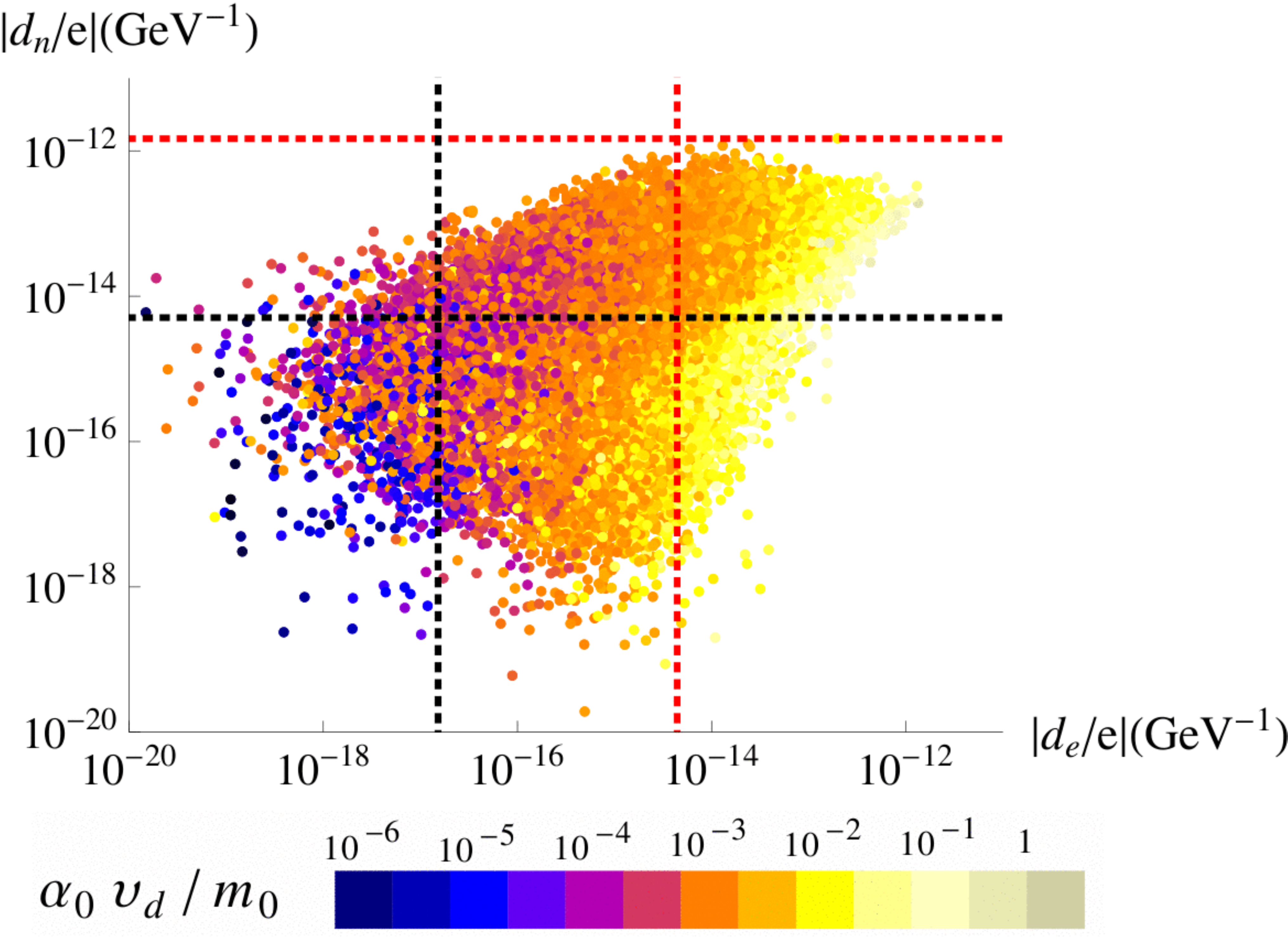}
\endminipage\hfill
\caption{The neutron EDM versus the average squark mass 
$m_{\tilde{q}}=\sqrt{m_{\tilde{u}}\,m_{\tilde{d}}}$, with 
$m_{\tilde{u}}$ and $m_{\tilde{d}}$ as defined in Eq.~\eqref{msqavdn} (left panel)
and versus the electron EDM (right~panel). 
The red dotted lines denote the current experimental limits as given in
Eqs.~(\ref{dn_limit},\ref{EDMlimit}) and the black dotted lines the future limits
 $|d_n/e|\lesssim 10^{-28}$\,cm~$\approx 
5\times 10^{-15}\,\text{GeV}^{-1}$ and 
$|d_e/e|\,\lesssim 3\times10^{-31}$\,cm\,$\approx 
1.52\times10^{-17} \,\text{GeV}^{-1}$~\cite{Hisano15}.}
\label{Fig:dn}
\end{figure}

In the left panel of Figure~\ref{Fig:dn}, we show our prediction for the
neutron EDM versus the average first generation squark mass
$m_{\tilde{q}}=\sqrt{m_{\tilde{u}}\,m_{\tilde{d}}}$.
For squark masses less than about 6~TeV, it lies just below the red line
denoting the experimental limit in Eq.~\eqref{dn_limit}. For heavier squarks it stays
below the limit by at least one order of magnitude. 
The colour coding corresponds to the predicted value of 
$|(d^C_u-d^C_d)/e|\times 10^{12}$\,GeV, which can also reach the
limit in Eq.~\eqref{Hd_limit} for large $|d_n/e|$ values. 
In the right panel of Figure~\ref{Fig:dn}, the neutron and electron EDMs are
plotted against each other. They are of the same order of magnitude, but it is
the current electron EDM limit that constrains our parameter space. When the
future experimental limits are reached, only the small part lying in the lower
left corner bounded by the black dotted lines will survive.




\section{Conclusions}
\label{conclusions}
\cleqn

In a recent paper we showed how MFV can emerge approximately from an $SU(5)$
SUSY GUT whose flavour structure is 
controlled by the family symmetry $S_4 \times U(1)$~\cite{companion}, providing 
a good description of all quark and lepton masses, mixings as
well as CP violation. We showed that the model leads to mass insertion parameters
in Eqs.~(\ref{eq:deltau},\ref{eq:deltad},\ref{eq:deltae}) which very closely
resemble the MFV forms, where $\delta^{u,d,e}_{LL,RR}$ are unit matrices and 
$\delta^{u,d,e}_{LR}$ are proportional to the Yukawa matrices.

Whereas in~\cite{companion} we focused on the similarity to MFV, here we
highlight the differences, which we do by considering the predictions for
electric dipole moments, lepton flavour violation, $B$ and $K$ meson
mixing as well as rare $B$ decays. As expected, many of the new physics contributions 
fall well below current limits. This is the case for example in $B$ physics observables, 
where deviations are negligible (at the 1\% level). Thus, our model would be
unable to explain any discrepancies between SM expectations and measurements
in $\Delta M_{B_{s,d}}$ or in the time dependent asymmetries $S_{J/\psi \phi}$
and $S_{J/\psi K_S}$. This is in marked contrast to the $SU(3)$ family
symmetry models previously studied, where large effects were expected in these 
observables. Thus, neutrino physics which led to $S_4 \times U(1)$, appears to
lead us towards models with small such deviations.

On the other hand there are observable effects which would distinguish the
$SU(5) \times S_4 \times U(1)$  SUSY GUT model 
from MFV. The most significant effects of the departure from MFV appear in the
(12) down-type quark and charged lepton sectors, related to Kaon mixing
observables and the branching ratio of $\mu \rightarrow e \gamma$. 
We find that $(\delta^{e}_{LL})_{12}$ provides the dominant contribution to
BR($\mu \rightarrow e \gamma$) and that our model requires rather heavy
sleptons, exceeding about 1 TeV, in order to satisfy the experimental
bound. Another important area where our model gives observable deviations from
MFV is CP violation, in particular the electron EDM, where again large (TeV
scale) slepton masses are required for compatibility with current bounds to be
achieved. The model therefore predicts that a signal should be observed in
both $\mu \rightarrow e \gamma$ and the electron EDM within the expected
future sensitivity of these experiments.

Turning to CP violation in the Kaon system, the model contributes
significantly to $\epsilon_K$ due to the phase of
$(\delta^{d}_{RR})_{12}$. The SM prediction for this observable depends  
sensitively on $|V_{cb}|$, which differs when considering inclusive or
exclusive decays, leading to a lower central value in the latter
case. However, even for inclusive values of $|V_{cb}|$, the SM expectation for 
$\epsilon_K$ is about 10\% below the measurement. 
Our model is capable of providing sufficient enhancement to explain the
experimentally observed value of $\epsilon_K$.

\begin{table}[t]\centerline{
\begin{tabular}{|c|c|c|c|c|c|c|c|c|}\hline
$d_e$ &
$\mu \rightarrow e \gamma$ & 
$\Delta M_{B_{s,d}}$ & 
$S_{J/\psi \phi}$ & 
$S_{J/\psi K_S}$ & 
$\Delta M_{K}$ & 
$\epsilon_K$& 
$B_{s,d} \rightarrow \mu^+\mu^-$ & 
$d_n$ 
\\\hline
\Red{$\bs{\star \star\star}$} &
\Red{$\bs{\star \star\star}$} &
$\bs{\star}$ & 
$\bs{\star}$ & 
$\bs{\star}$ & 
\Blue{$ \bs{\star\star}$} &
\Red{$\bs{\star \star\star}$} & 
$\bs{\star}$ &  
\Blue{$ \bs{\star\star}$} 
\\\hline
\end{tabular}}
\caption{\label{tab:DNA}The flavour ``DNA'' of our $SU(5) \times S_4\times
  U(1)$ SUSY GUT model following the labelling proposed
  in~\cite{AnatomyandPhenomenology}.
The predicted contributions to the various flavour observables are classified
into three categories: \Red{$\bs{\star \star\star}$} indicates large
observable effects while visible but small effects are marked by
\Blue{$\bs{\star\star}$}. The absence of sizable effects is shown by $\bs{\star}$.}
\end{table}
We collect our findings in Table~\ref{tab:DNA}, where we classify
various flavour observables according to the expected size of our model's
predictions. 
Large observable effects are indicated by 
{\Red{$\bs{\star \star\star}$}}, while visible but small effects are labelled by 
{\Blue{$ \bs{\star\star}$}}. A single star ${\bs{\star}}$ shows the absence of
sizable effects on a particular flavour observable. 
This classification, which was first suggested
in~\cite{AnatomyandPhenomenology}, is undoubtedly somewhat vague by nature and
therefore limited in its scope. Yet, it has proved to be a useful tool in
comparing characteristic predictions of various models of flavour. Table~8
of~\cite{AnatomyandPhenomenology} shows the expected predictions of a
selection of different models. Comparing this table with our model's DNA, see
Table~\ref{tab:DNA}, 
demonstrates the specific signatures of our $SU(5) \times S_4\times U(1)$ SUSY
GUT of flavour.  
According to the phenomenological study in~\cite{AnatomyandPhenomenology}, all
of the discussed models which predict large effects on $\epsilon_K$ also predict large
contributions to $S_{J/\psi \phi}$. In contrast, our model features large
contributions to $\epsilon_K$ in conjunction with negligible effects on
$S_{J/\psi \phi}$. 
Furthermore, all SUSY models in~\cite{AnatomyandPhenomenology} entail large
contributions to $B_s\rightarrow \mu^+\mu^-$ while such contributions are tiny
in our model.
Those models in~\cite{AnatomyandPhenomenology} which lead to a large electron
EDM ($d_e$) also predict a large neutron EDM ($d_n$). Again, our model differs
from this pattern by predicting large observable $d_e$ together with only small $d_n$.
Concerning $\mu\rightarrow e\gamma$ we observe that sizable effects are
expected for our model as well as all flavour models scrutinised
in~\cite{AnatomyandPhenomenology}.
This comparison illustrates that the phenomenological signatures of our
$SU(5) \times S_4\times U(1)$ SUSY GUT are indeed quite different from those of
previously discussed flavour models.

In summary, theories with discrete flavour symmetries such as the $SU(5)
\times S_4 \times U(1)$ SUSY GUT model, motivated by neutrino physics, seem to
lead to MFV-like flavour changing expectations, but with some important exceptions. 
This study shows that, while observable deviations in $B$ physics are generally not
expected to show up, departures from MFV are expected in both $\mu \rightarrow
e \gamma$ and the electron EDM within the foreseeable future sensitivity of
these experiments. CP violating effects may also be observed in $\epsilon_K$,
perhaps resolving some possible SM discrepancies.




\section*{Acknowledgements}

We thank Claudia Hagedorn for helpful discussions throughout this project. 
MD and SFK acknowledge partial support from the STFC Consolidated ST/J000396/1
grant and the European Union FP7 ITN-INVISIBLES 
(Marie Curie Actions, PITN-GA-2011-289442). 
CL is supported by the Deutsche Forschungsgemeinschaft (DFG) within
the Research Unit FOR 1873 ``Quark Flavour Physics and Effective Field Theories''.




\section*{Appendix}




\begin{appendix}


\section{Low energy mass insertion parameters}\label{App:LowMIs}
\cleqn

In this appendix, we show explicitly the full expressions of the low energy
mass insertion parameters used in our numerical analysis. They are given in
terms of the high energy order one coefficients introduced in
Section~\ref{HighScaleMatrices}. Performing the transformation to 
the SCKM basis, it is useful to define the corresponding GUT scale
parameters  
\begin{eqnarray}
 \tilde{b}_{12}&=&(b_{2}-b_{01}k_{2}),~~~~
\tilde{b}_{13}=-(b_4-b_{01}k_4),~~~~
\tilde{b}_{23}=-(b_{3}-b_{01}k_{3}),\label{Bt}\\
\tilde{B}_{12}&=&2\frac{\tilde{x}_2}{y_s}(b_{1}-b_{01}k_{1}),~~\:
\tilde{B}_{13}=\frac{\tilde{x}_2^2}{y_b\,y_s}(b_{01}-b_{02}),~~\:
\tilde{B}_{23}=\frac{y_s}{y_b}(b_{01}-b_{02}),~~\:
\tilde{R}_{12}=B_3-K_3,\nonumber
\end{eqnarray}
and
\begin{eqnarray}
\nonumber 
\tilde{a}^u_{11}&=&a_ue^{i(\theta^a_u-\theta^y_u)},~~~~
\tilde{a}^u_{22}=a_ce^{i(\theta^a_c-\theta^y_u)},~~~~
\tilde{a}^u_{33}=a_t,~~~~
\tilde{a}^u_{23}=z^u_2\left(\frac{a_t}{y_t}-e^{i(\theta^{z_{u_a}}_2-\theta^{z_u}_2)}\frac{z^{u_a}_2}{z^u_2}\right),\\
\nonumber \tilde{a}^d_{11}&=&\frac{\tilde{x}_2^2}{y_s}\left(2\frac{\tilde{x}^a_2}{\tilde{x}_2}e^{i(\theta^{\tilde{x}_a}_2-\theta^{\tilde{x}}_2)}-\frac{a_s}{y_s}e^{i(\theta^a_s-\theta^y_s)}\right),~~~~
\tilde{a}^d_{22}=a_se^{i(\theta^a_s-\theta^y_s)},~~~~
\tilde{a}^d_{33}=a_be^{i(\theta^a_b-\theta^y_b)},\\
\nonumber \tilde{a}^d_{12}&=&\tilde{x}_2\left(\frac{\tilde{x}^a_2}{\tilde{x}_2}
e^{i(\theta^{\tilde{x}_a}_2-\theta^{\tilde{x}}_2)}-\frac{a_s}{y_s}e^{i(\theta^a_s-\theta^y_s)}\right),~~~~
\tilde{a}^d_{23}=y_s\left(\frac{a_s}{y_s}e^{i(\theta^a_s-\theta^y_s)}-\frac{a_b}{y_b}
e^{i(\theta^a_b-\theta^y_b)}\right),\\
\nonumber 
\tilde{a}^d_{31}&=&z^d_3
\left(\frac{a_b}{y_b}e^{i(\theta^a_b-\theta^y_b)}-\frac{z^{d_a}_3}{z^{d}_3}e^{i(\theta^{z_{d_a}}_3-\theta^{z_d}_3)}\right),~~~~\\
\nonumber \tilde{a}^d_{32}&=&\frac{y_s^2}{y_b}\left(\frac{a_s}{y_s}e^{i(\theta^a_s-\theta^y_s)}-\frac{a_b}{y_b}e^{i(\theta^a_b-\theta^y_b)}\right)
+z^d_2\left(\frac{a_b}{y_b}e^{i(\theta^a_b-\theta^y_b)}-\frac{z^{d_a}_2}{z^d_2}e^{i(\theta^{z_{d_a}}_2-\theta^{z_d}_2)}\right),~~~~\\
\tilde{a}^e_{23}&=&9\frac{y_s^2}{y_b}\left(\frac{a_s}{y_s}e^{i(\theta^a_s-\theta^y_s)}-\frac{a_b}{y_b}e^{i(\theta^a_b-\theta^y_b)}\right)+z^d_2
\left(\frac{a_b}{y_b}e^{i(\theta^a_b-\theta^y_b)}-\frac{z^{d_a}_2}{z^{d}_2}e^{i(\theta^{z_{d_a}}_2-\theta^{z_d}_2)}
\right).~~~~ \label{adt}
\end{eqnarray}
Here, $z^u_2$ parameterises the (23) and (32) entries of the up-type quark
Yukawa matrix of order $\lambda^7$ before canonical normalisation; the
associated phase is given by $ \theta^{z_u}_2=3\theta^d_2+2\theta^d_3$. 
They become subdominant contributions to the (23) and (32) elements of
$Y^u_\text{GUT}$ in Eq.~\eqref{YuC}.
The parameter of the corresponding soft trilinear contribution is denoted by
$z^{u_a}_2$ with phase~$\theta^{z_{u_a}}_2$. 
In addition to $z^u_2$  we also need $z^d_4$ which parameterises a
subdominant contribution to the (22) and (23) elements of  $Y^d_\text{GUT}$ in
Eq.~\eqref{YdC} of order  $\lambda^5$. For the phase we have 
$ \theta^{z_d}_4=6\theta^d_2+4\theta^d_3$, and the corresponding parameters of
the $A$-terms are $z^{d_a}_4$ and~$\theta^{z_{d_a}}_4$. 
It is worth mentioning that all $\tilde{a}^f_{ij}$ become real in the limit
where the Yukawa and trilinear phase structures are aligned such that the
relation $\theta^{y}_f=\theta^a_f$ holds.

In order to describe the renormalisation group running from the GUT scale down
to low energies, we introduce the parameters in Eqs.~(\ref{Rysx},\ref{Rys}) as
well as
\begin{eqnarray}
 R^a_u&=&\eta\left(\frac{46}{5}g_U^2\frac{M_{1/2}}{A_0}+3a_t\,y_t\right)
+3\eta_N\,y_D\,\alpha_D\, ,~~~~~~~
R^a_t=R^a_u+3\,\eta\,a_t\,y_t\, ,~~~~\\
R^a_d&=&\eta\, \frac{44}{5}g_U^2\frac{M_{1/2}}{A_0}\, ,~~~~~~
R^a_b=R^a_d+\eta\,a_t\,y_t \, ,~~~~~~
R^a_e=\eta\frac{24}{5}g_U^2\frac{M_{1/2}}{A_0}+\eta_N\,y_D\alpha_D\, ,~~~~~~~~\\
R_\nu
&=&z^D_1-y_D(K_3+K^N_3)\,,~~~~~~
R^a_\nu=
z^{D_a}_1e^{i\theta^{z_{D_a}}_1}-\alpha_D(K_3+K^N_3)\,,
\end{eqnarray}
and
\begin{eqnarray}
R_\mu&=&4\eta\left(0.9\,g_U^2-\frac{3}{4}y_t^2\right)-3\eta_N\,y_D^2\,, \\
R_q&=&(2b_{02}+c_{H_u})\,y_t^2+\alpha_0^2\,a_t^2\,,\label{Rq}\\
R_l&=&(1+B^N_0+c_{H_u})y_D^2+\alpha_0^2\alpha_D^2\,, \label{Rl} \\
R_l'&=&(1+B^N_0+c_{H_u})y_D\,z^D_1+\alpha_0^2\alpha_D\,z^{D_a}_1e^{i\theta^{z_{D_a}}_1}\,.\label{Rlp}
\end{eqnarray}
In these expressions, $g_U\approx \sqrt{0.52}$ denotes the universal gauge coupling
constant at the GUT scale, $M_{1/2}$ is the universal gaugino mass parameter
and $A_0$ is the scale of the soft trilinear terms. Using the SUSY breaking mass
$m_0$, we have also introduced  $\alpha_0=A_0/m_0$, see Eq.~\eqref{x_alpha0}.
$\eta$ and $\eta_N$ have been defined in Eq.~\eqref{etas}, while $c_{H_u}$ is
given in Eq.~\eqref{cHud}. 

With these definitions, the $\mu$ parameter at the low energy scale can be
approximated by $\mu \approx\mu_{\text{GUT}}\left(1+R_{\mu}\right)$, and  the
low energy sfermion masses, whose GUT scale definitions are given in
Eq.~\eqref{fullmasses}, take the form 
\begin{eqnarray}
\nonumber m_{\tilde{u}_{LL}}&\approx&m_{\tilde{c}_{LL}}\approx m_0\,p^u_{L^{1G}}\,,~~~~\,~~m_{\tilde{t}_{LL}}\approx m_0\,p^u_{L^{3G}}\,,\\
m_{\tilde{u}_{RR}}&\approx&m_{\tilde{c}_{RR}}\approx
m_0\,p^u_{R^{1G}}\,,~~~~~~m_{\tilde{t}_{RR}}\approx
m_0\,p^u_{R^{3G}}\,,\label{uLR}\\[2mm]
\nonumber m_{\tilde{d}_{LL}}&\approx&m_{\tilde{s}_{LL}}\approx m_0\,p^d_{L^{1G}}\,,~~~~~~m_{\tilde{b}_{LL}}\approx m_0\,p^d_{L^{3G}}\,,\\
m_{\tilde{d}_{RR}}&\approx&m_{\tilde{s}_{RR}}\approx m_{\tilde{b}_{RR}}\approx
m_0\,p^d_{R}\,,\label{dLR}\\[2mm]
\nonumber m_{\tilde{e}_{LL}}&\approx&m_{\tilde{\mu}_{LL}}\approx m_{\tilde{\tau}_{LL}}\approx m_0\,p^e_{L}\,,\\
m_{\tilde{e}_{RR}}&\approx&m_{\tilde{\mu}_{RR}}\approx  m_0\,p^e_{R^{1G}}\,,~~~~~~m_{\tilde{\tau}_{RR}}\approx m_0\,p^e_{R^{3G}}\,,\label{eLR}
\end{eqnarray}
with
\begin{eqnarray}
\nonumber p^u_{L^{1G}}&=&\sqrt{b_{01}+6.5\,x},~~~~~~~\,
p^u_{L^{3G}}=\sqrt{b_{02}+6.5\,x-2\eta R_q+\frac{\upsilon_u^2}{m_0^2}y_t^2(1+R^y_t)^2}\,,\\
 p^u_{R^{1G}}&=&\sqrt{b_{01}+6.15\,x},~~~~~~p^u_{R^{3G}}=\sqrt{b_{02}+6.15\,x-4\eta R_q+\frac{\upsilon_u^2}{m^2_0}y_t^2(1+R^y_t)^2}\,,~~~~~~\label{pus}\\[2mm]
p^d_{L^{1G}}&=&\sqrt{b_{01}+6.5\,x},~~~~~p^d_{L^{3G}}=\sqrt{b_{02}+6.5\,x-4\eta
  R_q},~~~~~p^d_R=\sqrt{1+6.1x},\label{pds}~~~~~~~~\\[2mm]
 p^e_{R^{1G}}&=&\sqrt{b_{01}+0.15\,x},~~~~p^e_{R^{3G}}=\sqrt{b_{02}+0.15\,x},~~~~ 
p^e_L=\sqrt{1+0.5\,x-2\eta_N\,R_l}.~~~~~~~~\label{pes}
\end{eqnarray}
Here, $x=(M_{1/2}/m_0)^2$ as defined in Eqs.~\eqref{x_alpha0}. With these
definitions at hand, we can write the mass insertion parameters at the low
energy as follows.

\subsubsection*{Up-type quark sector:}

\begin{eqnarray}
\label{eq:uplowLL12}
(\delta^u_{LL})_{12}&=&\frac{1}{(p^u_{L^{1G}})^2}e^{-i\theta^d_2}\,\tilde{b}_{12}\,\lambda^4,\\
(\delta^u_{LL})_{13}&=&\frac{1}{p^u_{L^{1G}}p^u_{L^{3G}}}e^{-i(4\theta^d_2+\theta^d_3)}(1-\eta\,y_t^2)\,\tilde{b}_{13}\,\lambda^6,\\
(\delta^u_{LL})_{23}&=&\frac{1}{p^u_{L^{1G}}p^u_{L^{3G}}}e^{-i(7\theta^d_2+2\theta^d_3)}(1-\eta\,y_t^2)\,\tilde{b}_{23}\,\lambda^5,
\end{eqnarray}
\begin{eqnarray}
(\delta^u_{RR})_{12}&=&\frac{1}{(p^u_{R^{1G}})^2}e^{-i\theta^d_2}\,\tilde{b}_{12}\,\lambda^4,\\
(\delta^u_{RR})_{13}&=&\frac{1}{p^u_{R^{1G}}p^u_{R^{3G}}}(1-2\eta\,y_t^2)\,\tilde{b}_{13}\,\lambda^6,\\
(\delta^u_{RR})_{23}&=&\frac{1}{p^u_{R^{1G}}p^u_{R^{3G}}}e^{i(5\theta^d_2+\theta^d_3)}(1-2\eta\,y_t^2)\,\tilde{b}_{23}\,\lambda^5,
\end{eqnarray}
\begin{eqnarray}
(\delta^u_{LR})_{11}&=&\frac{\alpha_0\,\upsilon_u}{m_0\,p^u_{L^{1G}}\,p^u_{R^{1G}}}y_u
(1+R^y_u)\left(\frac{\tilde{a}^u_{11}}{y_u}-\frac{\mu(1+R_\mu)}{A_0\,t_\beta}-2\frac{R^a_u}{1+R^y_u}\right)\lambda^8,\\
(\delta^u_{LR})_{22}&=&\frac{\alpha_0\,\upsilon_u}{m_0\,p^u_{L^{1G}}\,p^u_{R^{1G}}}y_c
(1+R^y_u)\left(\frac{\tilde{a}^u_{22}}{y_c}-\frac{\mu(1+R_\mu)}{A_0\,t_\beta}-2\frac{R^a_u}{1+R^y_u}\right)\lambda^4,\\
(\delta^u_{LR})_{33}&=&\frac{\alpha_0\,\upsilon_u}{m_0\,p^u_{L^{3G}}\,p^u_{R^{3G}}}y_t
(1+R^y_t)\left(\frac{\tilde{a}^u_{33}}{y_t}-\frac{\mu(1+R_\mu)}{A_0\,t_\beta}-2\frac{R^a_t}{1+R^y_t}\right),
\end{eqnarray}
\begin{eqnarray}
(\delta^u_{LR})_{12}&=&(\delta^u_{LR})_{21}=(\delta^u_{LR})_{31}\label{deltauLR12Low}=0,\\
(\delta^u_{LR})_{13}&=&-\frac{\alpha_0\,\upsilon_u}{m_0\,p^u_{L^{1G}}\,p^u_{R^{3G}}}\tilde{x}_2\,y_b\,y_t
\left(\frac{\tilde{x}^a_2}{\tilde{x}_2}e^{i(\theta^{\tilde{x}_a}_2-\theta^{\tilde{x}}_2)}+\frac{R^a_t}{1+R^y_t}\right)2\eta\lambda^7,\label{deltauLR13Low}\\
\label{deltauLR23Low}
(\delta^u_{LR})_{23}&=&\frac{\alpha_0\,\upsilon_u}{m_0\,p^u_{L^{1G}}\,p^u_{R^{3G}}}\Bigg\{-y_s\,y_b\,y_t
\left(\frac{a_s}{y_s}e^{i(\theta^a_s-\theta^y_s)}+\frac{R^a_t}{1+R^y_t}\right)2\eta\lambda^6+\\
\nonumber&+&\lambda^7\Bigg[e^{i\theta^d_2}\tilde{a}^u_{23}(1+R^y_t-\eta\,y_t^2)+2\eta\,y_b\,y_t\Bigg(
e^{i\theta^d_2}\tilde{a}^d_{12}+\left(\frac{a_s}{y_s}e^{i(\theta^a_s-\theta^y_s)}+\frac{R^a_t}{1+R^y_t}\right)\times\\
\nonumber&\times&(\tilde{x}_2\cos(\theta^d_2)-z^d_4\cos(4\theta^d_2+\theta^d_3))+z^d_4e^{i(4\theta^d_2+\theta^d_3)}
\left(e^{i(\theta^a_s-\theta^y_s)}-\frac{z^{d_a}_4}{z^d_4}e^{i(\theta^{z_{d_a}}_4-\theta^{z_d}_4)}\right)\Bigg)\Bigg]\Bigg\},\\
\label{eq:uplowLR32}
 (\delta^u_{LR})_{32}&=&\frac{\alpha_0\,\upsilon_u}{m_0\,p^u_{L^{3G}}\,p^u_{R^{1G}}}(1+R^y_t-2\eta\,y_t^2)e^{i(3\theta^d_2+\theta^d_3)}\tilde{a}^u_{23}\,\lambda^7.
\end{eqnarray}
At the GUT scale, $(\delta^u_{LR})_{{13}}$ is zero up to the
order $\lambda^8$ where we truncate our expansion. The non-zero value in
Eq.~(\ref{deltauLR13Low}) is purely generated via the RG evolution. Similarly,
a term proportional to $\eta\,\lambda^6$ is generated in
$(\delta^u_{LR})_{23}$, which was of order $\lambda^7$ at the GUT scale. 
The $\lambda$-suppression of all other low energy mass insertion parameters 
$(\delta^f_{LL,RR,LR})_{ij}$ remains unaffected by the running, such that the
corresponding RG effects can simply be absorbed into new order one coefficients.

\subsubsection*{Down-type quark sector:}

\begin{eqnarray}
(\delta^d_{LL})_{12}&=&\frac{1}{(p^d_{L^{1G}})^2}\tilde{B}_{12}\,\lambda^3,\\[-1mm]
(\delta^d_{LL})_{13}&=&\frac{1}{p^d_{L^{1G}}p^d_{L^{13}}}e^{i\theta^d_2}\frac{\tilde{x}_2^2}{y_b\,y_s}\left(b_{01}-b_{02}+2\eta\,R_q\right)\left(1+\frac{\eta\,y_t^2}{1+R^y_b}\right)\,\lambda^4,\label{dLL13Low}\\[-1mm]
(\delta^d_{LL})_{23}&=&\frac{1}{p^d_{L^{1G}}p^d_{L^{13}}}\frac{y_s}{y_b}\left(b_{01}-b_{02}+2\eta\,R_q\right)\left(1+\frac{\eta\,y_t^2}{1+R^y_b}\right)\,\lambda^2,\label{dLL23Low}
\end{eqnarray}
\vspace{-2mm}
\begin{eqnarray}
(\delta^d_{RR})_{12}&=&
-(\delta^d_{RR})_{13}=\frac{1}{(p^d_{R})^2}e^{i\theta^d_2}\,\tilde{R}_{12}\,\lambda^4,\label{dRR13Low}
\\[-1mm]
(\delta^d_{RR})_{23}&=&-\frac{1}{(p^d_{R})^2}\tilde{R}_{12}\,\lambda^4,\label{dRR23Low}
\end{eqnarray}
\vspace{-2mm}
\begin{eqnarray}
(\delta^d_{LR})_{11}&=&\frac{\alpha_0\,\upsilon_d}{m_0\,p^d_{L^{1G}}\,p^d_{R}}\frac{\tilde{x}_2^2}{y_s}
(1+R^y_d)\left(\frac{\tilde{a}^d_{11}}{\tilde{x}_2^2/y_s}-\frac{\mu\,t_\beta(1+R_\mu)}{A_0}-2\frac{R^a_d}{1+R^y_d}\right)\lambda^6,\\[-1mm]
(\delta^d_{LR})_{22}&=&\frac{\alpha_0\,\upsilon_d}{m_0\,p^d_{L^{1G}}\,p^d_{R}}y_s
(1+R^y_d)\left(\frac{\tilde{a}^d_{22}}{y_s}-\frac{\mu\,t_\beta(1+R_\mu)}{A_0}-2\frac{R^a_d}{1+R^y_d}\right)\lambda^4,\\[-1mm]
(\delta^d_{LR})_{33}&=&\frac{\alpha_0\,\upsilon_d}{m_0\,p^d_{L^{3G}}\,p^d_{R}}y_b
(1+R^y_b)\left(\frac{\tilde{a}^d_{33}}{y_b}-\frac{\mu\,t_\beta(1+R_\mu)}{A_0}-2\frac{R^a_b}{1+R^y_b}\right)\lambda^2,
\end{eqnarray}
\vspace{-1mm}
\begin{eqnarray}
(\delta^d_{LR})_{12}&=&-(\delta^d_{LR})_{21}=(\delta^d_{LR})_{13}=\frac{\alpha_0\,\upsilon_d}{m_0\,p^d_{L^{1G}}\,p^d_{R}}(1+R^y_d)\tilde{a}^d_{12}\,\lambda^5,\\[-1mm]
(\delta^d_{LR})_{23}&=&\frac{\alpha_0\,\upsilon_d}{m_0\,p^d_{L^{1G}}\,p^d_{R}}y_s(1+R^y_d)
\left(\frac{\tilde{a}^d_{23}}{y_s}+2\frac{\eta\,y_t^2}{1+R^y_b}\left(\frac{a_t}{y_t}+\frac{R^a_d}{1+R^y_d}\right)\right)\lambda^4,~~~~~~~~~~\label{dLR23Low}\\[-1mm]
(\delta^d_{LR})_{31}&=&\frac{\alpha_0\,\upsilon_d}{m_0\,p^d_{L^{3G}}\,p^d_{R}}e^{-i\theta^d_2}(1+R^y_b)\tilde{a}^d_{31}\,\lambda^6,\\[-1mm]
\nonumber\label{dLR32Low}(\delta^d_{LR})_{32}&=&\frac{\alpha_0\,\upsilon_d}{m_0\,p^d_{L^{3G}}\,p^d_{R}}(1+R^y_b)y_b\Bigg(\frac{\tilde{a}^d_{32}}{y_b}+2\eta
y_t^2\frac{y_s^2}{y_b^2}\Bigg[\frac{2(1+R^y_b)+\eta
    y_t^2}{2(1+R^y_b)^2}\frac{\tilde{a}^d_{23}}{y_s}\\[-3mm]
&+&\left(\frac{a_t}{y_t}+\frac{R^a_d}{1+R^y_d}\right)\frac{(1+R^y_d)^2}{(1+R^y_b)^3}\Bigg]\Bigg)\lambda^6.
\end{eqnarray}

\subsubsection*{Charged lepton sector:}

\begin{eqnarray}
 (\delta^e_{LL})_{12}&=&-(\delta^e_{LL})_{23}=\frac{1}{(p^e_{L})^2}\left(\tilde{R}_{12}-2\eta_N\tilde{E}_{12}\right)\lambda^4,\label{eLL12Low}\\[-1mm]
\label{eLL12Low-13}
 (\delta^e_{LL})_{13}&=&-\frac{1}{(p^e_{L})^2}\left(\tilde{R}_{12}-2\eta_N\tilde{E}^*_{12}\right)\lambda^4,
\end{eqnarray}
\vspace{-2mm}
\begin{eqnarray}
 (\delta^e_{RR})_{12}&=&-\frac{1}{(p^e_{R^{1G}})^2}e^{i\theta^d_2}\frac{\tilde{B}_{12}}{3}\,\lambda^3,\\[-1mm]
(\delta^e_{RR})_{13}&=&\frac{1}{p^e_{R^{1G}}\,p^e_{R^{3G}}}\frac{\tilde{B}_{13}}{3}\,\lambda^4,\label{eRR13Low}\\[-1mm]
(\delta^e_{RR})_{23}&=&\frac{1}{p^e_{R^{1G}}\,p^e_{R^{3G}}}3\tilde{B}_{23}\,\lambda^2,
\end{eqnarray}
\vspace{-2mm}
\begin{eqnarray}
(\delta^e_{LR})_{11}&=&\frac{1}{p^e_{L}\,p^e_{R^{1G}}}\frac{\upsilon_d\,\alpha_0}{m_0}
\frac{\tilde{x}_2^2}{3\,y_s}(1+R^y_e)\left(\frac{y_s}{\tilde{x}_2^2}\tilde{a}^d_{11}-\frac{\mu\,t_\beta}{A_0}(1+R_\mu)-2\frac{R^a_e}{1+R^y_e}\right)\lambda^6,~~~~~~\\[-1mm]
 (\delta^e_{LR})_{22}&=&\frac{1}{p^e_{L}\,p^e_{R^{1G}}}\frac{\upsilon_d\,\alpha_0}{m_0}
3\,y_s(1+R^y_e)\left(\frac{\tilde{a}^d_{22}}{y_s}-\frac{\mu\,t_\beta}{A_0}(1+R_\mu)-2\frac{R^a_e}{1+R^y_e}\right)\lambda^4,\\[-1mm]
 (\delta^e_{LR})_{33}&=&\frac{1}{p^e_{L}\,p^e_{R^{3G}}}\frac{\upsilon_d\,\alpha_0}{m_0}
y_b(1+R^y_e)\left(\frac{\tilde{a}^d_{33}}{y_b}-\frac{\mu\,t_\beta}{A_0}(1+R_\mu)-2\frac{R^a_e}{1+R^y_e}\right)\lambda^2,
\end{eqnarray}
\vspace{0mm}
\begin{eqnarray}
 (\delta^e_{LR})_{12}&=&\frac{1}{p^e_{L}\,p^e_{R^{1G}}}\frac{\upsilon_d\,\alpha_0}{m_0}(1+R^y_e)e^{i\theta^d_2}\tilde{a}^d_{12}\,\lambda^5,\\[-1mm]
 (\delta^e_{LR})_{13}&=&\frac{1}{p^e_{L}\,p^e_{R^{3G}}}\frac{\upsilon_d\,\alpha_0}{m_0}\left((1+R^y_e)\tilde{a}^d_{31}
+2
\eta_N\,y_D\,R_\nu\,y_b\left(\frac{\alpha_D}{y_D}+\frac{R^a_e}{1+R^y_e}\right)\right)\lambda^6,~~~~~~~~\\[-1mm]
(\delta^e_{LR})_{21}&=&(\delta^e_{LR})_{31}=-\frac{1}{p^e_{L}\,p^e_{R^{1G}}}\frac{\upsilon_d\,\alpha_0}{m_0}(1+R^y_e)e^{-i\theta^d_2}\tilde{a}^d_{12}\,\lambda^5,\\[-1mm]
 (\delta^e_{LR})_{23}&=&\frac{1}{p^e_{L}\,p^e_{R^{3G}}}\frac{\upsilon_d\,\alpha_0}{m_0}\left((1+R^y_e)\tilde{a}^e_{23}
+2 \eta_N\,y_D\,R_\nu\,y_b\left(\frac{R^a_\nu}{R_\nu}+\frac{R^a_e}{1+R^y_e}\right)\right)\lambda^6,\label{eLR23Low}\\[-1mm]
 (\delta^e_{LR})_{32}&=&\frac{1}{p^e_{L}\,p^e_{R^{1G}}}\frac{\upsilon_d\,\alpha_0}{m_0}(1+R^y_e)3\,\tilde{a}^d_{23}\,\lambda^4.
\end{eqnarray}
Here we have additionally introduced $\tilde E_{12}$ which parameterises the
off-diagonal entries of $(\delta^e)_{LL}$ in
Eqs.~(\ref{eLL12Low},\ref{eLL12Low-13}) induced by the RG running. It is defined as
\begin{eqnarray}
\tilde{E}_{12}&=&y_D^2\left(\tilde R_{12}+B^N_3-K^N_3B^N_0\right)+R_l'-(K_3+K^N_3)R_l\,.\label{E12}
\end{eqnarray}


\section{Loop functions}
\label{Loop Functions}
\cleqn

The dimensionless functions $C_{B}$, $C'_{L}$, $C'_{R}$, $C'_2$, $C'_{B,R}$,
$C'_{B,L}$ and $C''_B$ which appear in the expressions for the EDM of the electron in
Section~\ref{Electron-EDM} and the branching ratio of $\mu \to e \gamma$ in
Section~\ref{mutoegamma} are defined as~\cite{Sleptonarium}
\begin{eqnarray}
C_{i}=\frac{m_0^4}{\mu^2}I_{i} \ ,
\end{eqnarray}
where
\begin{eqnarray}
I_B (M_1^2,\, m_{L}^2,\, m_{R}^2) & =  & 
\frac{1}{m_{R}^2 - m_{L}^2} \left[ y_L \, g_1 \left( x_L \right) 
- y_R \, g_1 \left( x_R \right) \right]  ,\\
I'_L (m_L^2,\, M_1^2,\, \mu^2 ) & = & \frac{1}{m_L^2}
\frac{y_L}{y_L - x_L}\left[\, h_1 \left( x_L\right)
- \, h_1 \left( y_L\right)\right] ,\\ 
I'_R (m_R^2,\, M_1^2,\, \mu^2 ) & = & \frac{1}{m_R^2}
\frac{y_R}{y_R - x_R}\left[\, h_1 \left( x_R\right) 
- \, h_1 \left( y_R\right)\right] , 
\end{eqnarray}
\begin{eqnarray}
I'_2 (m_L^2,\, M_2^2,\, \mu^2 ) & = & \frac{M_2\cot ^2\theta_W}{M_1 m_L^2}
\frac{y_L}{y_L - x'_L} \left[\, h_2 \left( x'_L\right) 
- \, h_2 \left( y_L\right)\right] , \\ 
I'_{B,R} (M_1^2,\, m_L^2,\, m_{R}^2) & = &- \frac{1}{m_R^2 - m_{L}^2}
\left( y_R \, h_1\left( x_R\right)  
- m_R^2 I_B \right) , \\
I'_{B,L} (M_1^2,\, m_{L}^2,\, m_{R}^2) & = & \frac{1}{m_R^2 - m_L^2} 
\left(  y_L \, h_1\left( x_L\right)  
- m_L^2 I_B \right) ,\\
I''_B (M_1^2,\, m_{L}^2,\, m_{R}^2) & = & \frac{m_L^2\,m_{R}^2}{m_R^2 - m_L^2}{\frac{1}{\mu^2}}
\left( y_R I'_{B,R} - y_L I'_{B,L} \right),
\label{Iapp}
\end{eqnarray}
with
\begin{eqnarray}
x_L= \frac{M_1^2}{m_L^2},~~\quad x_R= \frac{M_1^2}{m_R^2} ,~~\quad x'_L=
\frac{M_2^2}{m_L^2} ,~~ \quad y_L= \frac{\mu^2}{m_L^2},~~\quad y_R= \frac{\mu^2}{m_R^2},
\end{eqnarray}
and
\begin{eqnarray}
 \nonumber g_1 (y) &=& \frac{1 - y^2 + 2y \ln (y) }{(1 - y)^3}  ,\\
 \nonumber  h_1 (y) &=& \frac{1+4y-5y^2 + (2y^2 + 4y)\ln (y)}{(1 - y)^4},\\
 h_2 (y) &=& \frac{7y^2+4y-11 - 2(y^2 +6y+2) \ln (y) }{2(y - 1)^4} .
\end{eqnarray}
Note that we assume real and positive values for $M_i$ and $\mu^2$.

The loop functions appearing in the meson mixing amplitudes of
Section~\ref{sec:meson} as well as the branching ratios of $B_{s,d} \to
\mu^+\mu^-$ in Section~\ref{sec:Bsdtomumu} read~\cite{AnatomyandPhenomenology} 
\begin{eqnarray}
f_6(y)&=&\frac{6(1+3y)\ln(y)+y^3-9y^2-9y+17}{6(y-1)^5},\\
\tilde{f}_6(y)&=&\frac{6y(1+y)\ln(y)-y^3-9y^2+9y+1}{3(y-1)^5},\\
f_1(y)&=&\frac{1}{1-y}+\frac{y}{(1-y)^2}\ln(y),\\
f_3(y)&=&-\frac{1+y}{2(1-y)^2}-\frac{y}{(1-y)^3}\ln(y),\\
f_4(x,y)&=&-\frac{x\ln(x)}{(1-x)^2(y-x)}-\frac{y\ln(y)}{(1-y)^2(x-y)}+\frac{1}{(1-x)(1-y)},\\
f_5(y)&=&\frac{2+5y-y^2}{6(1-y)^3}+\frac{y}{(1-y)^4}\ln(y).
\end{eqnarray}

The relevant functions for the branching ratio of $b\to s\gamma$ in
Section~\ref{sec:bsgamma}  are given by~\cite{Gabbiani:1996hi}
\begin{eqnarray}
M_1(y)&=&\frac{1+4y-5y^2+4y\ln(y)+2y^2\ln(y)}{2(1-y)^4},\\
M_3(y)&=&\frac{-1+9y+9y^2-17y^3+18y^2\ln(y)+6y^3\ln(y)}{12(y-1)^5}.
\end{eqnarray}

\newpage

Finally, the loop functions entering the hadronic EDM expressions in
Section~\ref{sec:hadrEdM} are~\cite{Hisano04}
\begin{eqnarray}
N_1(y)&=&\frac{3+44y-36y^2-12y^3+y^4+12y(2+3y)\ln(y)}{6(y-1)^6},\\
N_2(y)&=&-\frac{10+9y-18y^2-y^3+3(1+6y+3y^2)\ln(y)}{3(y-1)^6}.
\end{eqnarray}




\end{appendix}






\end{document}